\documentclass{Config/aa} 

\usepackage[svgnames]{xcolor}
\usepackage{graphicx}
\usepackage{subfigure}
\usepackage{cancel}
\usepackage{txfonts}
\usepackage{multirow}
\usepackage{nicefrac}
\usepackage[d]{esvect}
\usepackage{lipsum}
\usepackage{diagbox}
\usepackage{comment}
\usepackage[colorlinks=True,
            linkcolor=blue,
            citecolor= blue]{hyperref}

\begin{document} 

   \title{Axisymmetric investigation of differential rotation in contracting stellar radiative zones}

   \author{B. Gouhier,
           \inst{1}
           F. Lignières
          \inst{1}
          \and
          L. Jouve
          \inst{1}
          }
          

   \institute{Institut de Recherche en Astrophysique et Planétologie (IRAP), Université de Toulouse,
              14 Avenue Edouard Belin, 31400 Toulouse\\
              \email{[bgouhier, flignieres, ljouve]@irap.omp.eu}
             }
   \date{Received August 25, 2020 ; accepted November 11, 2020}

\abstract
{Stars experience rapid contraction or expansion at different phases of their evolution. Modelling the angular momentum and chemical elements transport occurring during these phases remains an unsolved problem.}
{We study a stellar radiative zone undergoing radial contraction and investigate the induced differential rotation and meridional circulation.}
{We consider a rotating spherical layer crossed by an imposed radial velocity field that mimics the contraction and solve numerically the axisymmetric hydrodynamical equations in both the Boussinesq and anelastic approximations. An extensive parametric study is conducted to cover regimes of contraction, rotation, stable stratification and density stratification that are relevant for stars.}  
{The differential rotation and the meridional circulation result from a competition between the contraction-driven inward transport of angular momentum and an outward transport dominated by either viscosity or an Eddington-Sweet type circulation, depending on the value of the $P_r \left( N_0/\Omega_0 \right)^2 $ parameter, where $P_r$ is the Prandtl number, $N_0$ the Brunt-Väisäilä frequency and $\Omega_0$ the rotation rate. 
Taking the density stratification into account is important to study more realistic radial contraction fields but also because the resulting flow is less affected by unwanted effects of the boundary conditions.
In these different regimes and for a weak differential rotation, we derive scaling laws that relate the amplitude of the differential rotation to the contraction timescale.}
{}


   \maketitle

\section{Introduction}

The transport of angular momentum (thereafter AM) and chemical elements in stars strongly affects their evolution, from pre-main sequence (PMS) to evolved stages. These processes are particularly crucial in the stellar radiative zones but their modelling remains an open question. In standard evolution models, these stably stratified zones have been assumed to be motionless despite early works pointing out the lack of a static solution in uniformly rotating stars \citep{von1924radiative} and a related large-scale meridional circulation driven by the centrifugal acceleration \citep{eddington1925circulating,sweet1950importance}. A more complete formalism was then introduced by \cite{zahn1992circulation} including a self-consistent meridional flow and models of the turbulent transport driven by hydrodynamical instabilities. The importance of additional processes, internal waves \cite[e.g.][]{talon2005hydrodynamical} and magnetic fields \citep{spruit2002dynamo}, has also been investigated. Zahn's formalism has been successful at explaining a number of observed stellar properties, like the nitrogen abundances at the surface of red supergiants or the observed blue to red supergiants ratio in the Small Magellanic Cloud \citep{maeder2001stellar}. The dynamics of internal gravity waves could possibly explain the flat rotation profile of the solar radiative zone (\cite{talon2002angular} and \cite{charbonnel2005influence}) inferred through helioseismology \citep{schou1998helioseismic}, as well as the lithium dip in solar-like stars \citep{charbonnel2005influence}. Despite these early encouraging results, many stellar observations remain unexplained, especially the internal rotation rates revealed by asteroseismology in evolved stars and in
intermediate-mass main-sequence stars (see \cite{aerts2019angular} for a review). Actually, the theoretical understanding of the structure of the differential rotation in stellar radiative zones, crucial for the development of instabilities and thus for the related turbulent transport, is still largely incomplete. In particular, the assumption that the differential rotation is mostly radial, rather than radial and latitudinal, is at the base of Zahn's formalism, but the validity of this assumption has never been thoroughly tested so far. In this paper, we will be particularly interested in the differential rotation and the large-scale meridional flows generated in periods of the stellar life when either contraction or expansion or both processes occur. This is the case for example for PMS stars which are gravitationally contracting before starting their core nuclear reactions, or for subgiant and giant stars which undergo contraction of their core and expansion of their envelope. Within these stars, a strong redistribution of AM is expected to happen, producing differential rotation and thus potentially unstable shear flows.\\

Solar-like PMS stars such as T-Tauri stars contract during their evolution up to the main sequence (MS). The increase of the rotation rate due to this contraction can be estimated using AM conservation and assuming uniform rotation. For example, taking typical rotation rates from \cite{gallet2013improved}, a star rotating at $8\%$ of the break-up velocity at $1$ Myr that experiences a decrease in radius from $2.58$ \(\textup{R}_\odot\) to $0.9$ \(\textup{R}_\odot\) (while the moment of inertia goes from $19.6$ \(\textup{I}_\odot\) to $1.03$ \(\textup{I}_\odot\) where \(\textup{I}_\odot\) $= 6.41 \cdot 10^{\hspace*{0.02cm} 53}$ g$\cdot$cm$^{2}$ is the moment of inertia of the present Sun) would rotate at $32 \%$ of the break-up velocity at the zero age main sequence (ZAMS). The observation of stars in young open clusters rather indicates that their surface typically rotates at only $10$\% of their break-up velocity, showing the need to add other processes like AM exchanges with a circumstellar disk and AM extraction through a magnetised stellar wind \citep{bouvier1986rotation}. Interestingly, simple models of the internal AM transport, where a coupling time between a uniformly rotating outer convective zone and a uniformly rotating radiative core determines the surface rotation, strongly suggest that a certain level of radial differential rotation is present in the interior of these contracting young solar-type stars \citep{gallet2013improved}.

Another stellar class of interest for our purpose are the more massive Herbig Ae/Be stars. Just like the T-Tauri stars, these objects experience a contraction during the PMS phase which should lead them to spin-up up to the MS. Using typical values  from \cite{bohm1995rotation} and assuming uniform rotation, we can estimate that a $2$ \(\textup{M}_\odot\) star rotates beyond its break-up velocity at the ZAMS while a $3$ \(\textup{M}_\odot\) star should rotate at around $65\%$ of its break-up velocity and a $5$ \(\textup{M}_\odot\) at around $52\%$ break-up. Again, observations indicate lower surface rotation rates at the ZAMS, corresponding to $30\%$ break-up for a $2$ \(\textup{M}_\odot\) star and around $40\%$ for the $3$ and $5$ \(\textup{M}_\odot\) stars \citep{alecian2013high2}. These data would be compatible either with a $ \propto 1/r^2$ inner differential rotation and no AM losses or with wind-driven AM losses and smoother radial gradients. An additional interest of the Herbig Ae/Be stellar class is the small fraction of Herbig Ae/Be stars that are known to be strongly magnetic \citep{wade2005discovery,alecian2013high} because of the contraction-driven AM transport in their radiative envelope would surely be affected by these magnetic fields. 

When studying AM transport, red giant stars and particularly early red giants and subgiants constitute objects of particular interest. Indeed, asteroseismology puts constraints on the rotation rates of both the core and the envelope of these stars and was able to show for the first time that a significant differential rotation exists in a stellar radiative zone \citep{deheuvels2012seismic,beck2012fast}. By studying a sample of six Kepler stars at different evolutionary stages, \cite{deheuvels2014seismic} also showed that the level of differential rotation increases with the stellar age. During the subgiant phase, the combination of core contraction and envelope expansion indeed tends to produce a strong radial differential rotation. However, up-to-date hydrodynamical models based on Zahn's formalism predict core rotation rates two or three orders of magnitude higher than observed \citep{eggenberger2012angular, marques2013seismic, ceillier2013understanding}, thus showing that some model assumptions are not valid or that additional processes extract the AM from the core. Internal gravity waves \citep{pinccon2017can} or magnetic fields \citep{fuller2019slowing,eggenberger2019asteroseismology2, den2020asteroseismology} have been invoked as possible candidates. 

In this work, we propose to investigate the large-scale flows induced by a steady mass flux through a stably stratified radiative zone, thus modelling the contraction of the star either during the PMS or the subgiant and giant phases. More specifically, we intend to describe the structure of the steady axisymmetric solutions, that is the differential rotation and the meridional circulation, in the different physical regimes relevant for stars undergoing contraction. As far as we know, such studies are scarce. In \cite{hypolite2014dynamics}, the effect of the gravitational contraction was studied in a linear regime without stratification or assuming an approximate treatment of the stable stratification proposed by \cite{rieutord2006dynamics}. The expansion was considered with a similar approach by \cite{rieutord2014dynamics}. The differential rotation induced by the contraction can in principle give rise to various kinds of instabilities that would in turn affect the AM distribution. In this context, our present work on steady axisymmetric solutions can be viewed as a first step towards a future study on the instabilities of contraction-driven flows and their role on the AM transport in stars.

The paper is organised as follows: in Sect. \ref{mathematical_formulation} the mathematical model is presented. In Sect. \ref{timescales_physical_processes} we discuss the different physical timescales. After shedding light on the values of the relevant parameters in stars in Sect. \ref{relevance_stellar_context}, the numerical method is presented in Sect. \ref{numerical_method}. The results under the Boussinesq and anelastic approximations are then respectively described in Sects. \ref{description_results} and \ref{effect_of_the_density_stratification}. Finally, a summary and a conclusion are given in Sect. \ref{summary_and_conclusions}.

\section{Mathematical Formulation}
\label{mathematical_formulation}

In this work, we intend to model the contraction of a portion of stellar radiative zone. To do so, the fluid dynamics equations are solved numerically, using two different approaches: either the Boussinesq approximation or the anelastic approximation. These two approximations are based on a sound-proof approach where the acoustic waves are filtered out. The Boussinesq case is studied to analyse the response to contraction of an incompressible fluid, without loosing the effects of buoyancy. Then, the anelastic approximation is introduced to study the effects of the density stratification. In this section, we first present the dimensionless Boussinesq and anelastic equations then discuss the reference state and the choice of the forcing velocity field that mimics the contraction of the stellar radiative zone.

\subsection{Governing Equations}
\label{governing_equations}

To model a stellar radiative zone we consider a spherical shell of inner radius $r_i$ and outer radius $r_0$ filled with a self-gravitating ideal gas subject to a forced radial velocity field intended to mimic the stellar contraction. In the following, the non-dimensional form of the governing equations in a rotating frame is obtained using the radius of the outer sphere $r_0$ as the reference length, the value of the contraction velocity field at the outer sphere $V_0$ as a characteristic velocity and the contraction timescale $\tau_{\text{c}} = r_0 / V_0$ as the reference timescale. The frame rotates at $\Omega_0$, the rotation rate of the outer sphere. All the thermodynamics variables are expanded as background plus fluctuations, the background variables being identified with an overbar and the fluctuations with a prime. The background density and temperature field are non-dimensionalised respectively by the outer sphere density $\rho_0$ and temperature $T_0$, while the background gradients of temperature and entropy fields are adimensionalised using the temperature and entropy difference $\Delta \overline{T}$ and $\Delta \overline{S}$ between the two spheres. The pressure fluctuations are non-dimensionalised by $\rho_0 r_0 \Omega_0 V_0$ and the entropy fluctuations by $C_p \Omega_0 V_0/g_0$, where $C_p$ is the heat capacity and $g_0$ the gravity at the outer sphere. Thereafter, all dimensionless variables are identified with a tilde. Under these conditions, the dimensionless form of the governing equations for conservation of mass, momentum and entropy is given by (\cite{jones2011anelastic}, \cite{gastine2012effects}, \cite{wicht2018modeling} and \cite{dietrich2018penetrative}):

\begin{equation}
\begin{array}{lll}
\vv{\nabla} \cdot \left \lbrack \tilde{\overline{\rho}} \left( \vv{\tilde{U}} + \tilde{V}_f(r) \vv{e}_r \right) \right \rbrack = 0
\label{anel_continuity_adim}
\end{array}
\end{equation}

\begin{equation}
\begin{array}{lll}
R_o \left \lbrack \displaystyle \frac{\partial \vv{\tilde{U}}}{\partial t} + \left( \left( \vv{\tilde{U}} + \tilde{V}_f(r) \vv{e}_r \right) \cdot \vv{\nabla} \right) \left( \vv{\tilde{U}} + \tilde{V}_f(r) \vv{e}_r \right) \right \rbrack ~ + \\\\ 2 \vv{e}_z \times \left( \vv{\tilde{U}} + \tilde{V}_f(r) \vv{e}_r \right) = - \vv{\nabla} \left( \displaystyle \frac{\tilde{\Pi}^{'}}{\tilde{\overline{\rho}}} \right) +
 \displaystyle \frac{\tilde{S}^{'}}{\tilde{r}^2} \hspace*{0.05cm} \vv{e}_r ~ + \displaystyle \frac{E}{\tilde{\overline{\rho}}} ~ \vv{\nabla} \cdot \vv{\vv{\tilde{\sigma}}}
 \label{anel_momentum_adim}
 \end{array}
\end{equation}

\begin{equation}
\begin{array}{lll}
\tilde{\overline{\rho}} \tilde{\overline{T}} \left \lbrack P_r R_0 \left( \displaystyle \frac{\partial \tilde{S}^{'}}{\partial t} \right.  + \left( \left( \vv{\tilde{U}} + \tilde{V}_f(r) \vv{e}_r \right) \cdot \vv{\nabla} \right) \tilde{S}^{'} \right) ~ + \\\\ P_r \left( \displaystyle \frac{N_0}{\Omega_0} \right)^2 \left( \tilde{U}_r + \tilde{V}_f(r) \vv{e}_r \right) \left. \displaystyle \frac{\text{d} \tilde{\overline{S}}}{\text{d} r} \right \rbrack = E ~ \vv{\nabla} \cdot \left( \tilde{\overline{\rho}} \tilde{\overline{T}} \vv{\nabla} \tilde{S}^{'} \right) + D_i Pe_c E^2 \tilde{\mathcal{Q}}{_{\nu}}
\end{array}
\label{anel_entropie_adim}
\end{equation}

\noindent where $V_f \hspace*{0.02cm} (r)$ is the mass-conserving contraction velocity field whose expression is given in Sect. \ref{velocity_forcing}. The dimensionless momentum equation Eq. \eqref{anel_momentum_adim} was derived using the Lantz-Braginsky-Roberts (LBR) approximation (\cite{lantz1992dynamical}, \cite{braginsky1995equations}) and assuming that the kinematic viscosity $\nu$ is uniform. The dimensionless equation of entropy evolution Eq. \eqref{anel_entropie_adim} was obtained using the formalism described by \cite{braginsky1995equations} and \cite{clune1999computational} to introduce the diffusion of entropy instead of the diffusion of temperature. The thermal diffusivity $\kappa$ is here assumed to be uniform. In the above equations $\tilde{\sigma}_{ij}$ is the dimensionless stress tensor and $\tilde{\mathcal{Q}}_{\nu}$ is the dimensionless viscous heating. Furthermore, the gravity model assumes that the mass of the star is concentrated in the centre:

\begin{equation}
\begin{array}{lll}
\vv{g} = - \displaystyle \frac{G M_c}{r^2} \hspace*{0.05cm} \vv{e}_r = - g_0 \displaystyle \frac{r_0^2}{r^2} \vv{e}_r
\label{anel_gravity}
\end{array}
\end{equation}

\noindent where $G$ is the gravitational constant and $M_c$ is the mass at the centre.

In Eqs. \eqref{anel_continuity_adim}, \eqref{anel_momentum_adim} and \eqref{anel_entropie_adim}, $R_o$ is a Rossby number based on the amplitude of the contraction velocity field $R_o = V_0 / \Omega_0 r_0$, $E$ is the Ekman number $E = \nu / \Omega_0 r_0^2$, $P_r$ is the Prandtl number $P_r = \nu / \kappa$ and $N_0$ is a reference value of the Brunt-Väisälä frequency defined in Sect. \ref{background_state}. It gives a mean measure of the magnitude of the stable stratification. The ratio between the Rossby number and the Ekman number allows us to define the contraction Reynolds number $Re_c$ from which we define the Péclet number associated with the contraction $Pe_c = P_r \hspace*{0.05cm} Re_c$. In Eq. \eqref{anel_entropie_adim}, a parameter related to the density stratification appears, namely the dissipation number:

\begin{equation}
D_i = \displaystyle \frac{g_0 r_0}{T_0 C_p}
\end{equation}

The Boussinesq approximation is then found in the limit where the effects of compressibility are neglected, except in the buoyancy term of the momentum equation. In this case, the temperature fluctuations are adimensionalised using $\Omega_0 V_0 T_0 / g_0$ as temperature scale. The dimensionless governing equations for continuity, momentum evolution and temperature evolution become:

\begin{equation}
\begin{array}{lll}
\vv{\nabla} \cdot \left( \vv{\tilde{U}} +  \tilde{V}_f(r) \vv{e}_r \right) = 0
\label{bouss_continuity_adim}
\end{array}
\end{equation}

\begin{equation}
\begin{array}{lll}
R_o \left \lbrack \displaystyle \frac{\partial \vv{\tilde{U}}}{\partial t} + \left( \left( \vv{\tilde{U}} +  \tilde{V}_f(r) \vv{e}_r \right) \cdot \vv{\nabla} \right) \left( \vv{\tilde{U}} +  \tilde{V}_f(r) \vv{e}_r \right)\right \rbrack ~ + \\\\ 2 \vv{e}_z \times \left( \vv{\tilde{U}} +  \tilde{V}_f(r) \vv{e}_r \right) = - \vv{\nabla} \tilde{\Pi}^{'} + \tilde{\Theta}^{'} \hspace*{0.02cm} \vv{\tilde{r}} ~ + E ~ \vv{\nabla}^2 \left( \vv{\tilde{U}} +  \tilde{V}_f(r) \vv{e}_r \right)
 \label{bouss_momentum_adim}
\end{array}
\end{equation}

\vspace*{-0.4cm}

\begin{equation}
\begin{array}{lll}
P_r R_0 \left \lbrack \displaystyle \frac{\partial \tilde{\Theta}^{'}}{\partial t} + \left( \left( \vv{\tilde{U}} +  \tilde{V}_f(r) \vv{e}_r \right) \cdot \vv{\nabla} \right) \tilde{\Theta}^{'} \right \rbrack ~ + \\\\ P_r \left( \displaystyle \frac{N_0}{\Omega_0} \right)^2 \left( \tilde{U}_r +  \tilde{V}_f(r) \vv{e}_r \right) \displaystyle \frac{\text{d} \tilde{\overline{T}}}{\text{d} r} = E ~ \vv{\nabla}^2 \tilde{\Theta}^{'}
\label{bouss_entropie_adim}
\end{array}
\end{equation}

\vspace*{0.2cm}

In Eq. \eqref{bouss_momentum_adim}, we use a gravity profile corresponding to a uniform fluid density $\rho_c$:

\begin{equation}
\vv{g} = - \displaystyle \frac{4 }{3} \hspace*{0.04cm} \pi \hspace*{0.04cm} r \hspace*{0.04cm} \rho_c \hspace*{0.04cm} G \hspace*{0.04cm} \hspace*{0.02cm} \vv{e}_r = - g_0 \displaystyle \frac{r}{r_0} \vv{e}_r
\label{bouss_gravity}
\end{equation}

To conclude this section, we note that when the anelastic approximation is used, the parameter space is defined by six dimensionless numbers, namely: $\epsilon_s$ (defined in Sect. \ref{background_state}), $D_i$, $Re_c$, $E$, $P_r$ and $N_0^2 / \Omega_0^2$ whereas only the last four are necessary in the Boussinesq approximation.

\subsection{Background State}
\label{background_state}

In the anelastic approximation, we choose to use a non-adiabatic reference state which enables us to prescribe a profile for the background entropy gradient. We here simply use a uniform positive entropy gradient to produce a stable stratification. The uniform background entropy gradient $\text{d} \overline{S}/\text{d} r$ is related to the Brunt-Väisälä frequency through:

\begin{equation}
N(r) = \sqrt{\displaystyle \frac{g}{C_p} \displaystyle \frac{\text{d} \overline{S}}{\text{d} r}}
\end{equation}

\noindent from which we define a reference Brunt-Väisälä frequency:

\begin{equation}
N_0 = \sqrt{\displaystyle \frac{g_0}{C_p} \displaystyle \frac{\Delta \overline{S}}{r_0}}
\label{Anel_Vaisala}
\end{equation}

For the anelastic approximation to remain valid, we nonetheless need to ensure that the deviation to the isentropic state is small. The magnitude of this deviation is controlled by the parameter $\epsilon_s$:

\begin{equation}
\epsilon_s = \displaystyle \frac{\Delta \overline{S}}{C_p}
\end{equation}

This parameter, together with the dissipation number, sets the temperature profile and the magnitude of the density stratification. 

Finally, the prescribed background gradient of entropy $\text{d} \overline{S} / \text{d} r$, the dissipation number $D_i$ and the $\epsilon_s$ parameter give the background temperature profile through the following equation: 

\begin{equation}
\displaystyle \frac{\text{d} \tilde{\overline{T}}}{\text{d} r} = \epsilon_s ~ \tilde{\overline{T}} ~ \displaystyle \frac{\text{d} \tilde{\overline{S}}}{\text{d} r} - \displaystyle \frac{D_i}{\tilde{r}^2}
\label{background_temperature_profile}
\end{equation}

\noindent as well as the background density profile:

\begin{equation}
\displaystyle \frac{\text{d} \ln{\tilde{\overline{\rho}}}}{\text{d} r} = \left(\displaystyle \frac{1}{\gamma - 1}\right) \displaystyle \frac{\text{d} \ln \tilde{\overline{T}}}{\text{d} r} - \epsilon_s \left(\displaystyle \frac{\gamma}{\gamma - 1}\right) \displaystyle \frac{\text{d} \tilde{\overline{S}}}{\text{d} r}
\label{background_density_profile}
\end{equation}

\noindent where $\gamma$ is the adiabatic index that is equal to $5/3$  for a monatomic gas. When the Boussinesq approximation is used, the background temperature $\overline{T}$ is the solution of the conduction equation:

\begin{equation}
\vv{\nabla}^2 ~ \overline{T} = 0
\end{equation}

\noindent for prescribed temperatures at the spherical boundaries, that is:

\begin{equation}
\overline{T} = \overline{T}(r=r_0) - \Delta \overline{T} ~ \displaystyle \frac{r_i}{r} \left( \displaystyle \frac{r-r_0}{r_i-r_0} \right)
\end{equation}

\noindent where $\Delta \overline{T} = \overline{T}(r=r_0) - \overline{T}(r=r_i) > 0$, again ensuring stable stratification. The background temperature gradient $\text{d} \overline{T} / \text{d} r$ is then related to the Brunt-Väisälä frequency through the following relationship:

\begin{equation}
N(r) = \sqrt{\displaystyle \frac{g}{\overline{T}} \displaystyle \frac{\text{d} \overline{T}}{\text{d} r}} 
\end{equation}

\noindent from which the reference Brunt-Väisälä frequency is defined as:

\begin{equation}
N_0 = \sqrt{\displaystyle \frac{g_0}{T_0} \displaystyle \frac{\Delta \overline{T}}{r_0}}
\label{Bouss_Vaisala}
\end{equation}

\subsection{Contraction Velocity Field}
\label{velocity_forcing}

The specificity of our model is to introduce a contraction process which will force large-scale flows inside the stably stratified stellar layer. We assume here that the contraction is produced by a steady mass flux oriented towards the centre of the star. In other words, a fixed radial velocity field $\vv{V_f}$ is added to the flow velocity and this forcing velocity field is chosen to be radial and directed towards the centre of the star. In order for this additional velocity field to conserve mass, we need to ensure that $\vv{\nabla} \cdot \left( \overline{\rho} \hspace*{0.05cm}  \vv{V_f} \right) =0$. In the anelastic case, the contraction velocity field thus simply reads:

\begin{equation}
\vv{V_f} = V_f \hspace*{0.02cm} (r) \hspace*{0.02cm} \vv{e}_r = - \displaystyle \frac{V_0 \hspace*{0.04cm} \rho_0 \hspace*{0.02cm} r{_0^2}}{\overline{\rho} r^2} \hspace*{0.02cm} \vv{e}_r
\label{Anelcontraction}
\end{equation}

Under the Boussinesq approximation, since the background density is uniform, this velocity field simplifies into:

\begin{equation}
\vv{V_f} = V_f \hspace*{0.02cm} (r) \hspace*{0.02cm} \vv{e}_r = - \displaystyle \frac{V_0 \hspace*{0.02cm} \hspace*{0.02cm} r{_0^2}}{r^2} \hspace*{0.02cm} \vv{e}_r
\label{Bousscontraction}
\end{equation}

We would like to point out that to model stellar expansion, this forcing velocity field $V_f \hspace*{0.02cm} (r)$ can be chosen to be positive.

\section{Timescales of physical processes}
\label{timescales_physical_processes}

In this section we use the equation of AM conservation to derive the timescales of the physical processes acting on the differential rotation. This allows us to distinguish a priori the parameter regimes where each process dominates. In its dimensional form, the equation of AM conservation reads:

\begin{equation}
\hspace*{-0.1cm}
\displaystyle \frac{\partial U_{\phi}}{\partial t} + \text{NL} + \underbrace{2 \Omega_0 U_s}_{\text{Coriolis term}} -\underbrace{\nu \hspace*{0.05cm} D^2 U_{\phi}}_{\text{Viscous term}} =
\underbrace{\displaystyle \frac{V_0 r{_0^2}}{r^3} \displaystyle \frac{\partial}{\partial r} \left(r U_{\phi} + r^2 \sin{\theta} \hspace*{0.05cm} \Omega_0 \right)}_{\text{Contraction}}
\label{angular_momentum_evolution}
\end{equation}
\normalsize

\noindent where $D^2 = \left( \vv{\nabla}^2 - 1/r^2 \sin^2 \theta \right) $ is the azimuthal component of the vector Laplacian operator, $U_s = \cos \theta \hspace*{0.05cm} U_{\theta} \hspace*{0.05cm} + \hspace*{0.05cm} \sin \theta \hspace*{0.05cm} U_r $ is the cylindrical velocity field and NL denotes the non-linear advection term. By considering the balance between the partial time derivative of azimuthal velocity field and the contraction term, we recover the contraction timescale defined in Sect. \ref{governing_equations}, namely $\tau_{\text{c}}=r_0/V_0$, when $\Delta \Omega/\Omega_0 \gtrsim 1$, or its linear version $\tau_{\text{c}}^{~\text{L}}=(r_0/V_0)(\Delta \Omega/\Omega_0)$, when $\Delta \Omega/\Omega_0 \ll 1$ so that $r  U_{\phi} \ll r^2 \sin{\theta} \hspace*{0.05cm} \Omega_0$. From the balance with the viscous and Coriolis terms, we now define three additional timescales for the AM transport. The first one is the viscous timescale which characterises a redistribution of AM by viscous processes:

\begin{equation}
\tau_{\nu} = \displaystyle \frac{r_0^2}{\nu}
\label{viscous_timescale}
\end{equation}

The two other timescales are related to the AM transport by meridional flows. In the linear regime, the timescale of AM redistribution by the meridional motions is:

\begin{equation}
\tau = \displaystyle \frac{r_0}{U_s} \displaystyle \frac{\Delta \Omega}{\Omega_0}
\label{circulation_timescale}
\end{equation}

\noindent and thus depends on the nature of the circulation through its velocity scale $U_s$. We now consider two well-known types of circulation, the Eddington-Sweet and the Ekman circulations, that will be relevant to analyse our simulation results. We note that the steady-state Eddington-Sweet circulation debated by \cite{busse1981eddington} or \cite{busse1982problem}, is here consistent with the conservation of AM (see Sect. \ref{summary_and_conclusions}).

In a stably stratified medium, large-scale meridional motions are efficiently prevented as the increasing buoyancy content of rising (or sinking) fluid elements produces strong restoring buoyancy forces. However, by diminishing the amplitude of temperature fluctuations and thus of the fluid element buoyancy, thermal diffusion enables the so-called Eddington-Sweet type circulation. To derive its characteristic timescale, we use the thermal wind balance, obtained by taking the curl of Eq. \eqref{bouss_momentum_adim_without_contraction_pressure} projected in the azimuthal direction. It gives: 

\begin{equation}
2 \displaystyle \frac{\partial U_{\phi}}{\partial z} = \displaystyle \frac{g_0}{\Omega_0 r_0 \overline{T}(r)} \displaystyle \frac{\partial \delta \Theta}{\partial \theta}
\label{boussthermalwind}
\end{equation}

We then use the steady linear dimensional form of the thermal balance Eq. \eqref{final_form_of_entropy_equation}:

\begin{equation}
U_r \frac{\text{d} T_{\text{m}}}{\text{d} r} = \kappa \hspace*{0.05cm} \vv{\nabla}^2 \delta \Theta
\label{thermal_balance}
\end{equation}

\noindent where $T_{\text{m}}(r) = \overline{T}(r) + T^{'}(r)$ and the temperature deviations $\delta \Theta(r,\theta) = \Theta^{'}(r,\theta) - T^{'}(r)$ are respectively, the mean spherically symmetric temperature field and the associated fluctuations obtained by taking the additional radial thermal background $T^{'}(r)$ induced by the contraction field into account (details on this are given in Appendix \ref{second_hydrostatic_state}). 

Accordingly, the characteristic amplitudes of the differential rotation $\Delta \Omega$, the temperature deviations $\delta \Theta$ and the radial velocity field $U_r$ are related as follows:

\begin{equation}
\delta \Theta \approx \displaystyle \frac{\Omega_0^2}{N_0^2} \displaystyle \frac{\Delta \Omega}{\Omega_0} \Delta \overline{T}
\label{thermal_fluct}
\end{equation}

\begin{equation}
U_r \approx \displaystyle \frac{\kappa}{r_0} \displaystyle \frac{\delta \Theta}{\Delta \overline{T}}
\label{relation_Ur_Theta}
\end{equation}

\noindent from which we get the following relationship between the radial velocity field and the level of differential rotation:

\begin{equation}
U_r \approx \displaystyle \frac{\kappa}{r_0} \displaystyle \frac{\Omega_0^2}{N_0^2} \displaystyle \frac{\Delta \Omega}{\Omega_0}
\label{ur_estimation}
\end{equation}

Then, from the continuity equation, we have $U_{\theta} \approx U_r \approx U_s$, and by injecting the expected order of magnitude of $U_s$ in Eq. \eqref{circulation_timescale}, we obtain the Eddington-Sweet timescale:

\begin{equation}
\tau_{\text{ED}} = \displaystyle \frac{r_0^2}{\kappa} \displaystyle \frac{N_0^2}{\Omega_0^2}
\label{edd_timescale}
\end{equation}

In an unstratified medium, the Ekman pumping mechanism produces a large-scale circulation controlled by the boundary layers. Indeed, the mismatch between boundary conditions applied at spherical surfaces and the inviscid interior flow solutions leads to thin viscous boundary layers that in practice modify the boundary conditions seen by the interior flow. For example, the no-slip boundary conditions at the outer sphere induce an Ekman layer that imposes the following relationship between the azimuthal and the radial velocity fields: 

\begin{equation}
U_r(r_0,\theta) = \displaystyle \frac{- \sqrt{E}}{\sin \theta} \displaystyle \frac{\partial}{\partial \theta} \left( \displaystyle \frac{\text{sgn} \left( \cos{\theta} \right) \hspace*{0.02cm} \sin \theta \hspace*{0.02cm} U_{\phi}^{I}\left(r_0,\theta \right)}{2 \sqrt{\left | \cos \theta \right |}} \right)
\label{ekman_pumping}
\end{equation}

\noindent as the boundary condition satisfied by the interior flow at the outer sphere. A well-known example of Ekman circulation is the classical unstratified spherical Couette flow where the necessary jumps in rotation rate between the interior and the spherical boundaries, denoted $\Delta \Omega_J$, force a meridional circulation of the order of $U_r \sim \sqrt{E} \hspace*{0.05cm} L \hspace*{0.05cm} \Delta \Omega_J$ that extends over the whole domain ($L$ is the characteristic length in the Couette problem). From Eq. \eqref{circulation_timescale}, the associated timescale for the AM transport, also called the Ekman spin-up timescale, is:  

\begin{equation}
\tau_{E} = \sqrt{\displaystyle \frac{r_0^2}{\Omega_0 \nu}}
\end{equation}

For a stably stratified medium, Ekman layers can still exist at the boundaries because the buoyancy force has a negligible effect on motions of very small radial extent. But they are not expected to drive a large-scale global circulation \citep{barcilon1967linear}. As we shall see later, an Ekman layer is present at the outer sphere of our numerical simulations to connect the interior circulation to the prescribed boundary condition and because the Ekman number is small but finite this will play a non-negligible role on the resulting differential rotation.

The ratios which compare the relative importance of these three AM transport processes:

\begin{equation}
\displaystyle \frac{\tau_{\text{ED}}}{\tau_{\nu}} = P_r \left( \displaystyle \frac{N_0}{\Omega_0}\right)^2 \text{;} \quad \displaystyle \frac{\tau_{\text{E}}}{\tau_{\nu}} = \sqrt{E} ~ \text{;} \quad \displaystyle \frac{\tau_{\text{E}}}{\tau_{\text{ED}}} = \displaystyle \frac{\sqrt{E}}{ P_r \left( \displaystyle \frac{N_0}{\Omega_0}\right)^2}
\end{equation}

\noindent are controlled by two dimensionless numbers, the Ekman number $E$ and the $P_r \left( N_0/\Omega_0\right )^2$ parameter. As we are interested in low Ekman number cases, $\sqrt{E}<<1$, we can thus define three regimes of interest, namely:

\begin{equation}
\begin{array}{lll}
P_r \left( \displaystyle \frac{N_0}{\Omega_0}\right)^2 \ll \sqrt{E} \ll 1 ~ \text{;} \qquad \sqrt{E} \ll P_r \left( \displaystyle \frac{N_0}{\Omega_0}\right)^2 \ll 1 ~ \text{;} \\\\ \hspace*{2cm} \sqrt{E} \ll 1 \ll P_r \left( \displaystyle \frac{N_0}{\Omega_0}\right)^2
\end{array}
\end{equation}

The actual differential rotation will result from a competition between the contraction forcing and these three AM transport processes. The corresponding timescale ratios are:

\begin{equation}
\displaystyle \frac{\tau_{\nu}}{\tau_{\text{c}}}  = \frac{R_o}{E} = Re_c ~ \text{;} \quad \displaystyle \frac{\tau_{\text{ED}}}{\tau_{\text{c}}} =  P_r \left( \displaystyle \frac{N_0}{\Omega_0}\right)^2 \displaystyle \frac{R_o}{E} ~ \text{;} \quad \displaystyle \frac{\tau_{\text{E}}}{\tau_{\text{c}}} = \frac{R_o}{\sqrt{E}}
\label{ratios_with_contraction}
\end{equation}

This introduces an additional parameter, either $R_o$ or $Re_c$ that will have a direct impact on the level of differential rotation. In our numerical study, this parameter will be varied to explore the linear regime, $\Delta \Omega/ \Omega_0 \ll 1$, and the beginning of the non-linear one, $\Delta \Omega/ \Omega_0 \gtrsim 1$.

A non-linear regime is indeed expected if the contraction is the dominant transport process. In this case, the steady state of the AM conservation Eq. \eqref{angular_momentum_evolution} reads:

\begin{equation}
\displaystyle \frac{V_0 r_0^2}{r^3} \displaystyle \frac{\partial}{\partial r} \left( r^2 \delta \Omega(r) \right) + 2 \Omega_0 V_0 \displaystyle \frac{r_0^2}{r^2} = 0
\label{forcing_equation}
\end{equation}

\noindent which can be solved to get:

\begin{equation}
\displaystyle \frac{\delta \Omega(r)}{\Omega_0} = \displaystyle \frac{r_0^2}{r^2} - 1
\label{estimation_max_rot_diff}
\end{equation}

Thus, for our set-up with $r_i= 0.3 ~ r_0$, the maximum value of the differential rotation is:

\begin{equation}
\displaystyle \frac{\Delta \Omega}{\Omega_0} = \displaystyle \frac{\delta \Omega(r_i)}{\Omega_0} = \frac{\Omega(r_i) - \Omega(r_0)}{\Omega_0} \approx 10.1
\label{max_rot_diff}
\end{equation}

\noindent which is clearly in the non-linear regime. We thus expect to reach this regime when the timescale ratios Eq. \eqref{ratios_with_contraction} are larger than one, while a linear regime $\Delta \Omega / \Omega_0 \ll 1$ should exist when these timescale ratios are small enough.

\section{Estimate of the relevant physical parameters in stars}
\label{relevance_stellar_context}

In this section, we estimate the Ekman number $E$, the $P_r \left( N_0/\Omega_0\right)^2$ parameter and the contraction Reynolds number $Re_c$ in stars undergoing contraction.

\subsection{Pre-main-sequence stars}

Among PMS stars, low mass stars are characterised by a bimodal distribution of their rotational period (P$_{\text{rot}} \approx 8 $ days for those experiencing a disk-locking phase and P$_{\text{rot}} \approx 1 $ day for the others \citep{maeder2008physics}), while intermediate and massive mass stars are rather rapid rotators (P$_{\text{rot}} \approx 1 $ day) except when they are magnetised (P$_{\text{rot}} \approx 1.6 ~ \text{to} ~ 4.3 $ days) \citep{alecian2013high2}. 

Stellar models of a 3 \(\textup{M}_\odot\) Population I PMS star \citep{talon2008angular} show that the thermal part of the Brunt-Väisälä frequency varies from $N_{\text{T}}^2 = 10^{-7.5} ~ \text{s}^{-2}$ at the beginning of the PMS to $N_{\text{T}}^2 = 10^{-5.8} ~ \text{s}^{-2}$ at the ZAMS. The Prandtl number is very low in non-degenerate stellar interiors, typically $P_r \approx 10^{-6}$ \citep{garaud2015excitation}. We thus obtain the following orders of magnitude of $P_r \left(N_0/\Omega_0\right)^2$ in PMS stars: 

\begin{table}[h]
\begin{center}
\begin{tabular}{c|c|c}
\hline
\diagbox[width=12em]{Status}{Period} & $1$ day & $8$ days \\
\hline
 \rule[-0.1cm]{0.cm}{0.6cm} $N_{\text{T}}^2 = 10^{-7.5} ~ \text{s}^{-2}$ & \multirow{2}{*}{$5.98 \cdot 10^{-6}$} & \multirow{2}{*}{$3.83 \cdot 10^{-4}$} \\
 \rule[-0.2cm]{0.cm}{0.5cm} (Beginning of the PMS) & & \\
\hline
 \rule[-0.1cm]{0.cm}{0.6cm} $N_{\text{T}}^2 = 10^{-5.8} ~ \text{s}^{-2}$ & \multirow{2}{*}{$3.00 \cdot 10^{-4}$} & \multirow{2}{*}{$1.91 \cdot 10^{-2} $} \\
\rule[-0.2cm]{0.cm}{0.5cm} (End of the PMS) & & \\ 
\hline
\end{tabular}
\vspace*{0.2cm}
\caption{Estimates of $P_r \left(N_0/\Omega_0\right)^2$ in PMS stars.}
\label{prn2o2_pms}
\end{center}
\end{table}

Another relevant parameter is the contraction Reynolds number $Re_c$. Since this number can be expressed as the ratio of the contraction timescale and the viscous timescale ($Re_c = \tau_{\nu}/\tau_{\text{c}}$), a rough estimate can be derived by replacing the contraction timescale by the Kelvin-Helmholtz timescale $\tau_{\kappa} = r_0^2 / \kappa$. In this case $Re_c \approx P_r^{-1}$, thus $\tau_{\text{ED}}/\tau_{\text{c}} = Re_c \hspace*{0.05cm} P_r\left(N_0/\Omega_0\right)^2 \approx \left(N_0/\Omega_0\right)^2$ is always larger than one for PMS stars, regardless of their evolutionary status or rotation period.

The Ekman number is always very small in stellar interiors. According to \cite{lara2013self}, at the ZAMS, the radiative viscosity in stellar radiative zones ranges from $\nu \approx 8\cdot 10^{\hspace*{0.04cm} 1} ~ \text{cm}^{2} \cdot \text{s}^{-1}$ in the core to $\nu \approx 5\cdot 10^{\hspace*{0.04cm} 2} ~ \text{cm}^{2} \cdot \text{s}^{-1}$ in the middle of the envelope for a $3$ \(\textup{M}_\odot\) star, while its value stays approximately uniform between the core and the middle of the envelope for a $7$ \(\textup{M}_\odot\) or a $10$ \(\textup{M}_\odot\) star (around $\nu \approx 10^{\hspace*{0.04cm} 3} ~ \text{cm}^2\cdot \text{s}^{-1}$). During the PMS, the radius of a star varies a lot. As an example, for a solar-mass star, the radius ranges from $2.6$ \(\textup{R}_\odot\) during the early PMS phase, to $0.9$ \(\textup{R}_\odot\) for more advanced stages of the PMS \citep{maeder2008physics,gallet2013improved}. During the PMS evolution of more massive stars, the radius can decrease from $8$ \(\textup{R}_\odot\) to $3.4$ \(\textup{R}_\odot\) for a $7$ \(\textup{M}_\odot\) star and diminishes from $4.5$ \(\textup{R}_\odot\) to $2.3$ \(\textup{R}_\odot\) for a $3$ \(\textup{M}_\odot\) star \citep{alecian2013high}. For such values, the Ekman number is indeed always very small, varying from $E = 10^{-15}$ to $E = 10^{-17}$. We thus conclude that the dominant parameter regime for PMS stars is the non-linear Eddington-Sweet regime defined by: 

\begin{equation}
\sqrt{E} \ll P_r \left(\displaystyle \frac{N_0}{\Omega_0}\right)^2 \ll 1 \quad \text{and} \quad Re_c \hspace*{0.05cm} P_r \left(\displaystyle \frac{N_0}{\Omega_0}\right)^2 \gg 1
\end{equation}

\subsection{Subgiants}

In the degenerate cores of subgiants, the Prandtl number increases to $\sim 10^{-3}$ due to the degeneracy of electrons \citep{garaud2015excitation}. Meanwhile, the Brunt-Väisälä frequency in these dense cores is quite large, its value varying between $N_\text{T}^2 = 10^{-4.5} ~ \text{s}^{-2}$ at the beginning of the subgiant phase to $N_\text{T}^2 = 10^{-3} ~ \text{s}^{-2}$ at a more advanced stage \citep{talon2008angular}. Asteroseismology puts strong constraints on the core rotation rate of these stars \citep{deheuvels2014seismic}. Considering a typical value of $\Omega_{\hspace*{0.035cm} \text{core}} = 3.34 \cdot 10^{-6} ~ \text{rad} \cdot \text{s}^{-1}$, we find that the degenerate core is characterised by very large values of the parameter $P_r \left(N_0/\Omega_0\right)^2$ up to $\sim 10^{5}$. 

Outside the degenerate core, typical $P_r \left(N_0/\Omega_0\right)^2$ are $10^{-3}$, that is much smaller than one, as in PMS stars. Given the small Ekman numbers, we find that the non-linear viscous regime is the relevant one inside the degenerate core of the subgiants:

\begin{equation}
\sqrt{E} \ll 1 \ll P_r \left(\displaystyle \frac{N_0}{\Omega_0}\right)^2 \quad \text{and,} \quad Re_c \gg 1
\end{equation}

\noindent whereas outside the degenerate core, the non-linear Eddington-Sweet regime dominates:

\begin{equation}
\sqrt{E} \ll P_r \left(\displaystyle \frac{N_0}{\Omega_0}\right)^2 \ll 1 \quad \text{and,} \quad Re_c \hspace*{0.05cm} P_r \left(\displaystyle \frac{N_0}{\Omega_0}\right)^2 \gg 1
\end{equation}

\section{Numerical method}
\label{numerical_method}

In order to study numerically the hydrodynamical steady states obtained in the various regimes described above, we carry out numerical simulations using the pseudo-spectral code MagIC. MagIC is based on a Chebyshev discretisation in the radial direction and spherical harmonic discretisation for the latitudinal and azimuthal directions. MagIC is a fully documented, publicly available code (\url{https://github.com/magic-sph/magic}) which solves the (magneto)-hydrodynamical equations in a spherical shell. An axisymmetric version of \cite{wicht2002inner} for the Boussinesq approximation and of \cite{gastine2012effects} for the anelastic approximation is used to solve Eqs. \eqref{anel_momentum_adim} and \eqref{anel_entropie_adim} as well as Eqs. \eqref{bouss_momentum_adim} and \eqref{bouss_entropie_adim} numerically. The solenoidal condition of Eqs. \eqref{anel_continuity_adim} and \eqref{bouss_continuity_adim} is ensured by a poloidal/toroidal decomposition of the mass flux. Our domain extent is $r \in \left \lbrack r_i = 0.3 \hspace*{0.03cm} \text{;} \hspace*{0.1cm} r_0 = 1.0 \right \rbrack$ and $\theta \in \left \lbrack 0 \hspace*{0.025cm} \text{;} \hspace*{0.1cm} \pi \right \rbrack$.

For all simulations we impose stress-free conditions at the inner sphere for the latitudinal and azimuthal velocity fields, and impermeability condition for the radial velocity field:

\begin{equation}
U_r = \displaystyle \frac{\partial}{\partial r} \left( \displaystyle \frac{U_{\phi}}{r} \right) = \displaystyle \frac{\partial}{\partial r} \left( \displaystyle \frac{U_{\theta}}{r} \right) = 0 \quad \text{at} \quad r = r_i
\end{equation}

At the outer sphere, we impose impermeability condition on the radial velocity field, a no-slip condition on the latitudinal velocity field and uniform rotation:

\begin{equation}
U_r = U_{\theta} = 0 \quad \text{and} \quad \Omega = \Omega_0 \quad \text{at} \quad r = r_0
\end{equation}

The boundary conditions for $\Theta^{'}$ (respectively $S^{'}$) are set to zero at the top and bottom of the domain in the Boussinesq (respectively anelastic) case.

We initially impose a uniform rotation in the whole domain $\Omega(r,\theta,t=0) = \Omega_0$ in the Boussinesq case and a fixed level of differential rotation

\begin{equation}
\Omega(r,\theta,t=0) = \overline{\rho}(r) ~ \Omega_0 \hspace*{0.05cm} \exp{\left( \displaystyle \frac{-\left(r - r_0 \right)}{\sigma}\right)}
\end{equation}

\noindent in the anelastic case, where $\sigma$ controls the magnitude of the rotation contrast.

For most of the simulations, the grid resolution is $N_r \times N_{\theta} = 193 \times 256$. The large number of points in the radial direction is because we also aim at characterising the different boundary layers and their potential role. When the Ekman number becomes very small (typically $10^{-7}$), the resolution is sometimes increased to $N_r \times N_{\theta} = 217 \times 288$. We always ensure that the number of radial grid cells in the Ekman layer is non-negligible, from five grid cells in the $E=10^{-7}$ case to twelve grid cells in the $E=10^{-5}$ case, for our best resolved cases.

\section{Results in the Boussinesq case}
\label{description_results}

The analysis of the different transport mechanisms performed in Sect. \ref{timescales_physical_processes} led to define three different regimes depending on the Ekman number and the $P_r \left( N_0 / \Omega_0 \right)^2$ parameter. A third parameter, $Re_c$ or $R_o$, appears to control the level of differential rotation and thus whether the non-linear terms play a role. The numerical investigation presented below has been carried out by varying these parameters and the results are analysed with the guidance of the timescale analysis of Sect. \ref{timescales_physical_processes}.

\subsection{Taylor-Proudman regime}
\label{taylor_proudman_regime}

We find that the differential rotation is near cylindrical when $P_r \left( N_0/\Omega_0 \right)^2 \ll \sqrt{E} \ll 1 $. According to the thermal wind balance Eq. \eqref{boussthermalwind}, such a regime is possible if the buoyancy force is weak enough. In the following of the paper, the regimes of cylindrical rotation will be called Taylor-Proudman regimes although strictly speaking the Taylor-Proudman theorem tells that all the velocity components must be cylindrical. 

\vspace*{0.25cm}

\begin{table}[h]
\begin{center}
\begin{tabular}{c|c|c|c}
\hline
\hspace*{0.05cm} \textbf{Simulation} \hspace*{0.05cm} & \hspace*{0.5cm} $\boldsymbol{E}$ \hspace*{0.5cm} & \hspace*{0.1cm} $\boldsymbol{P_r \left( N_0 / \Omega_0 \right)^2}$ \hspace*{0.1cm} & \hspace*{0.5cm} $\boldsymbol{Re_c}$ \hspace*{0.5cm} \\ 
\hline
1.1 & $10^{-2}$ & $10^{-4}$ & $10^{-2}$ \\
1.2 & $10^{-3}$ & $10^{-4}$ & $10^{-2}$ \\
1.3 & $10^{-4}$ & $10^{-4}$ & $10^{-2}$ \\
1.4 & $10^{-5}$ & $10^{-4}$ & $10^{-2}$ \\
1.5 & $10^{-6}$ & $10^{-4}$ & $10^{-2}$ \\
\hline
2.1 & $10^{-5}$ & $10^{-4}$ & $10^{-1}$ \\
2.2 & $10^{-5}$ & $10^{-4}$ & $1$ \\
2.3 & $10^{-5}$ & $10^{-4}$ & $10$ \\
\hline
3.1 & $10^{-6}$ & $10^{-4}$ & $10^{-1}$ \\
3.2 & $10^{-6}$ & $10^{-4}$ & $1$ \\
3.3 & $10^{-6}$ & $10^{-4}$ & $10$ \\
\hline
\end{tabular}
\vspace*{0.2cm}
\caption{Parameters of the simulations carried out in the Taylor-Proudman regime.}
\label{parameters_TP}
\end{center}
\end{table}

\begin{figure}[h]
\begin{center}
\includegraphics[width=4.4cm]{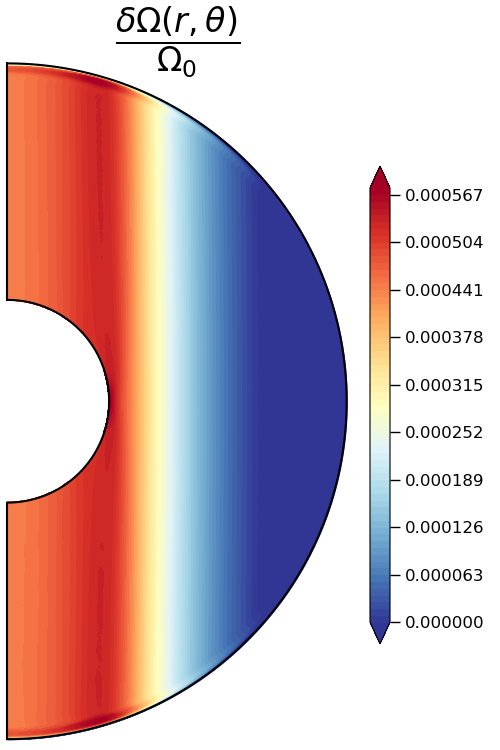}
\includegraphics[width=4.4cm]{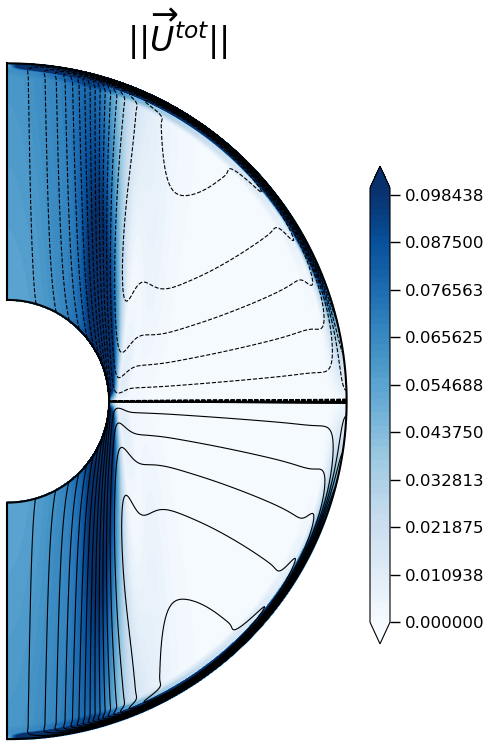}
\caption{Steady axisymmetric flow in the Taylor-Proudman regime. In the left panel, the coloured contours represent the differential rotation $\delta \Omega(r,\theta) = \Omega(r,\theta) - \Omega_0$ normalised to the rotation rate of the outer sphere $\Omega_0$ while in the right panel, we show the norm of the total meridional velocity field $\protect\vv{U}^{\hspace*{0.05cm} \text{tot}} = \left(U_r + V_f \hspace*{0.02cm} (r) \right) \protect\vv{e}_r + U_{\theta} \protect\vv{e}_{\theta}$. The parameters of the simulation are $E=10^{-4}$, $P_r \left(N_0/\Omega_0\right)^2 = 10^{-4}$ and $Re_c = 10^{-2}$ (simulation $1.3$ of Table \ref{parameters_TP}). The black lines show the streamlines of the total meridional circulation i.e. taking the effect of the contraction into account. Outside the tangent cylinder, the dashed lines correspond to a counterclockwise circulation and the full lines to a clockwise one while inside the tangent cylinder, the dashed lines represent a downward circulation and the full ones, an upward circulation.}
\label{Rot_Diff_TP}
\end{center}
\end{figure}

The meridional cut in the left panel of Fig. \ref{Rot_Diff_TP} shows the structure of the flow for a simulation in this regime ($E=10^{-4}$, $P_r \left(N_0/\Omega_0\right)^2 = 10^{-4}$, $Re_c = 10^{-2}$). The differential rotation exhibits a cylindrical profile, the rotation rate being maximum near the tangent cylinder. For this particular simulation, the maximal amplitude of the normalised differential rotation $ \left (\Omega(r,\theta)-\Omega_0\right)/\Omega_0 $ is barely $0.001$ which thus corresponds to a linear case. The right panel of the same figure displays the norm and the streamlines of the total meridional flow $\protect\vv{U}^{\hspace*{0.05cm} \text{tot}} = \left(U_r + V_f \hspace*{0.02cm} (r) \right) \protect\vv{e}_r + U_{\theta} \protect\vv{e}_{\theta}$. We see that a fluid particle initially located at the outer sphere is deflected into the Ekman layer towards the pole before it enters into the interior flow. From then on, if the particle is outside the tangent cylinder, it goes back towards the outer sphere thus forming a cell. On the contrary, if the particle is inside the tangent cylinder, it goes towards the inner sphere. A remarkable feature is the strong vertical jet directed towards the equator of the inner sphere. 

In order to characterise this regime, we have performed several simulations for a range of Ekman and contraction Reynolds numbers, $E = 10^{-2} - 10^{-6}$, $Re_c = 10^{-2} - 10$, as listed in Table \ref{parameters_TP}. These numerical results will be compared with an analytical solution that can be obtained in the linear and low Ekman number regime. Indeed, considering the balance between the contraction term and the Coriolis term in the linear inviscid steady equation of AM evolution Eq. \eqref{angular_momentum_evolution}, we first get the cylindrical radial component of the velocity field:

\begin{equation}
U_s = \displaystyle V_0 \frac{r_0^2}{r^2} \sin{\theta}
\label{cylindrical_radial_velocity_TP}
\end{equation}

The full inviscid meridional flow is then determined using the $U_r=0$ condition at the inner sphere. This solution does not fulfil the boundary conditions at the outer sphere and this is done through an Ekman boundary layer which comes with a jump in the rotation rate. Then, using the Ekman pumping condition Eq. \eqref{ekman_pumping} as a boundary condition for the interior flow, it is possible to express a mass budget inside a cylinder of radius $s = r \sin{\theta}$ which involves the inward flow from the Ekman layer and the outward flow through the cylinder. The inward and the outward mass-fluxes are obtained by integrating Eqs. \eqref{ekman_pumping} and \eqref{cylindrical_radial_velocity_TP} respectively. After some algebra (see details in Appendix \ref{demonstration_rot_diff_TP}) we find \footnote{We noted that, outside the tangent cylinder, that is for $s> r_i$, this analytical solution is similar at the expression derived by \cite{hypolite2014dynamics} up to a factor related to the aspect ratio.}:

\begin{equation}
\begin{array}{lll} 
\displaystyle \frac{\delta \Omega_{\hspace*{0.02cm} \text{TP}}(s \leq r_i)}{\Omega_0} = \displaystyle \frac{R_o}{\sqrt{E}} \left \lbrack \displaystyle \frac{2r_0^2}{s^2} \left( \cos \theta_0 - \cos \theta_i \right) \sqrt{\left|\cos \theta_0\right|} \right \rbrack \\\\
\displaystyle \frac{\delta \Omega_{\hspace*{0.02cm} \text{TP}}(s > r_i)}{\Omega_0} = \displaystyle \frac{R_o}{\sqrt{E}} \left \lbrack \displaystyle \frac{2r_0^2}{s^2} \cos \theta_0 \sqrt{\left|\cos \theta_0\right|} \right \rbrack
\end{array}
\label{differential_rotation_TP}
\end{equation}

\begin{figure*}[h]
\begin{center}
\hspace*{-0.7cm}
\includegraphics[width=8.5cm]{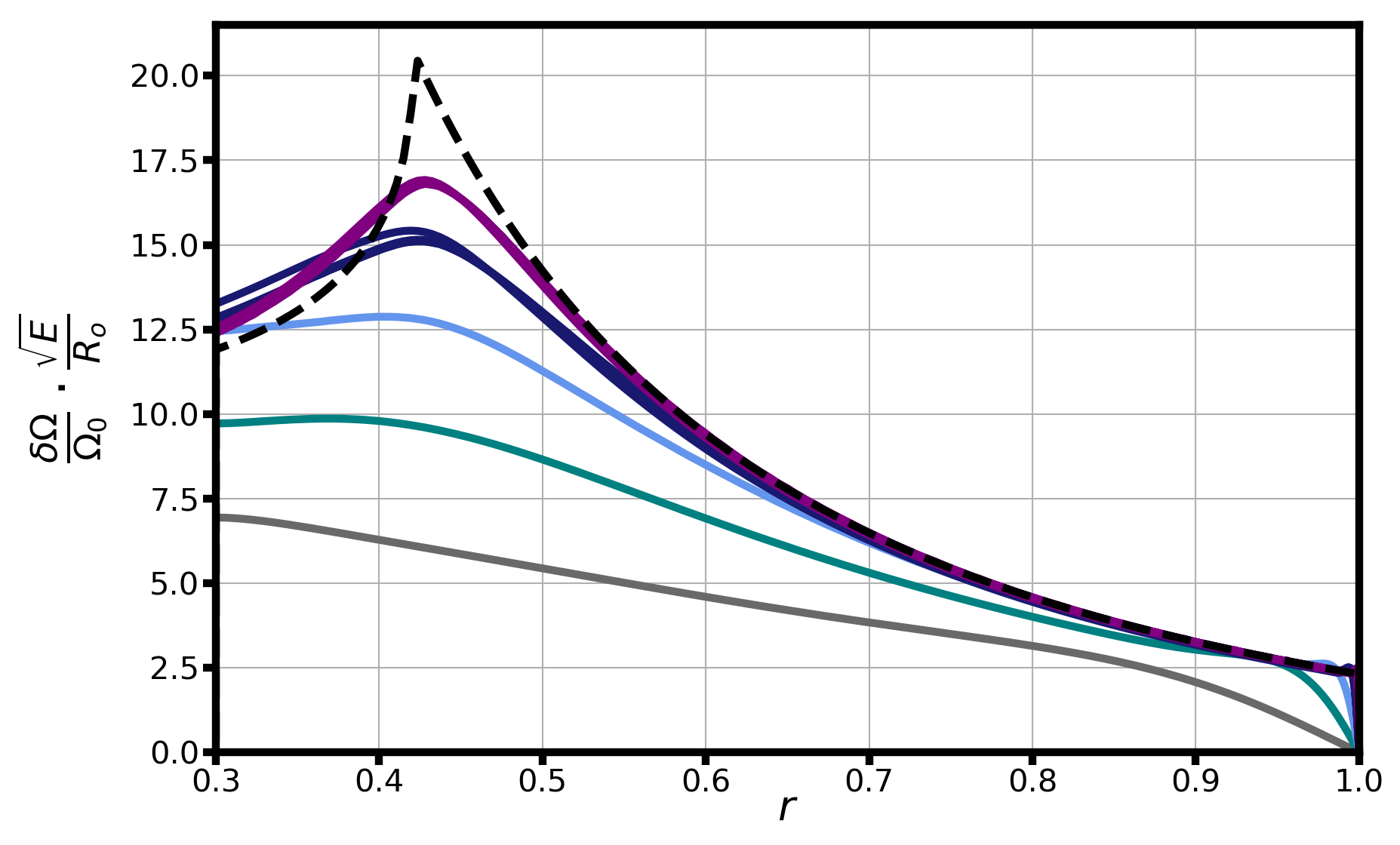}
\includegraphics[width=8.5cm]{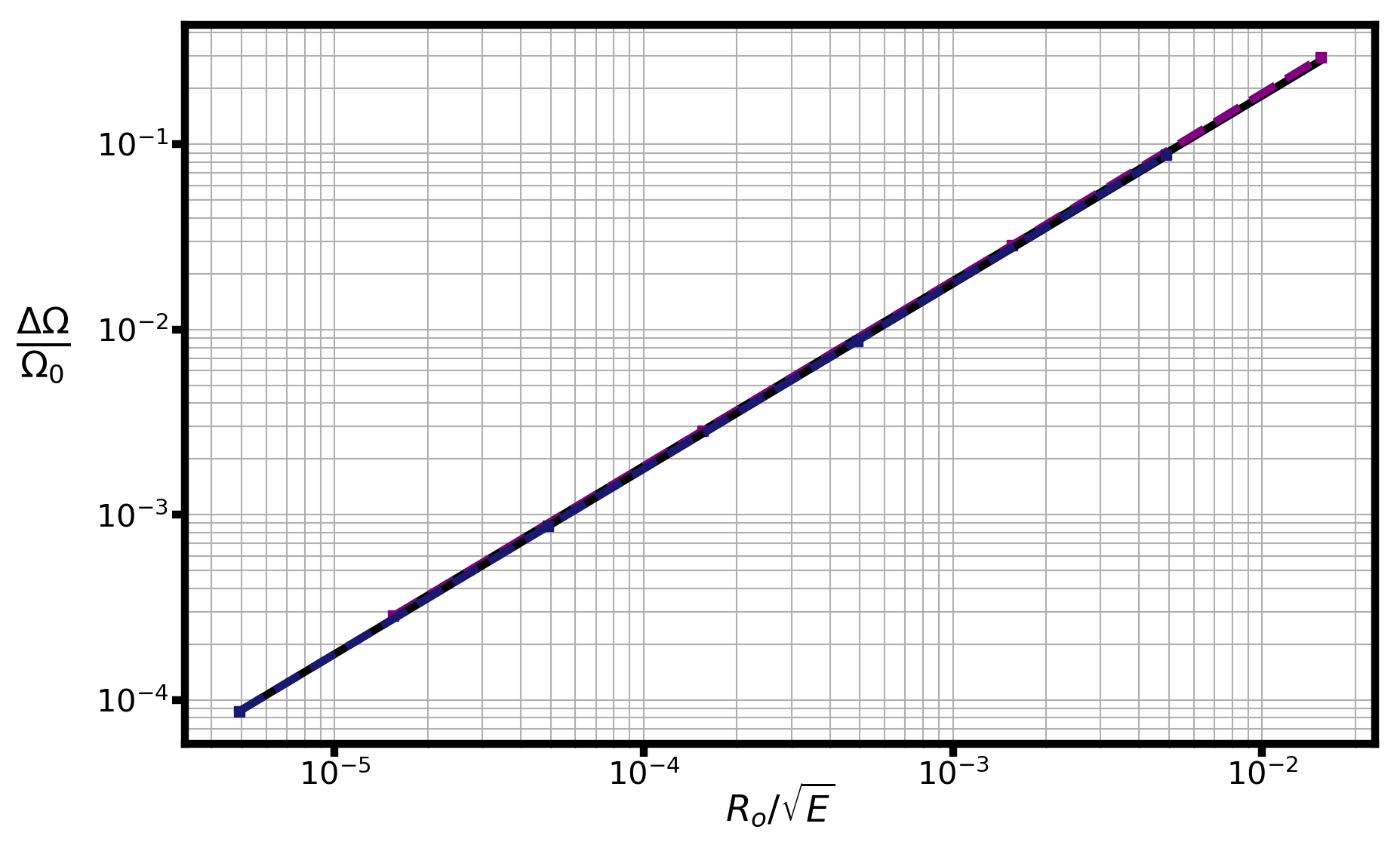}
\caption{Axisymmetric steady solutions of the differential rotation as a function of radius at the latitude $\theta = \pi/4$ in the Taylor-Proudman regime. [Left panel]: the analytical solution Eq. \eqref{differential_rotation_TP} is plotted in black dashed lines and compared with the numerical simulations obtained for fixed $P_r \left(N_0/\Omega_0\right)^2$ parameter ($10^{-4}$) and contraction Reynolds number $Re_c$ ($10^{-2})$ with various Ekman numbers namely, $10^{-2}$ in grey, $10^{-3}$ in cyan, $10^{-4}$ in light blue, $10^{-5}$ in dark blue and $10^{-6}$ in purple (runs $1.1$ to $1.5$ of Table \ref{parameters_TP} respectively). For $E=10^{-5}$ and $E=10^{-6}$, additional contraction Reynolds numbers are studied, namely $10^{-1}$ (runs $2.1$ and $3.1$ of Table \ref{parameters_TP}), $1$ (runs $2.2$ and $3.2$ of Table \ref{parameters_TP}) and $10$ (runs $2.3$ and $3.3$ of Table \ref{parameters_TP}). All curves are rescaled by $\sqrt{E}/R_o$. [Right panel]: the differential rotation between the inner and outer spheres $\Delta \Omega = \Omega(r_i) - \Omega(r_0)$ at the same latitude and normalised with the rotation rate taken at the outer sphere $\Omega_0$, is now plotted as a function of $R_o / \sqrt{E}$. The previous numerical solutions obtained for $E = 10^{-5}$ (runs $1.4$ and $2.1$ to $2.3$ of Table \ref{parameters_TP}) are plotted in dark blue while the simulations performed for $E=10^{-6}$ (runs $1.5$ and $3.1$ to $3.3$ of Table \ref{parameters_TP}) are plotted in purple.}
\label{Rot_Diff_Taylor_Proudman}
\end{center}
\end{figure*}

This solution can be interpreted as follows: at leading order, the contraction forces a global meridional circulation that is unable to match the boundary conditions at the outer sphere. This is done in an Ekman boundary layer which in turn produces a certain level of differential rotation determined by the Ekman pumping relation for the prescribed interior radial field Eq. \eqref{ekman_pumping}. This differential rotation is communicated to the bulk of the flow through the Taylor-Proudman constraint $\partial U_{\phi} / \partial z = 0$.

The analytical solution scaled by $R_o/\sqrt{E}$ is plotted with a black dashed line in the left panel of Fig. \ref{Rot_Diff_Taylor_Proudman} and compared with numerical solutions obtained for different Ekman and Rossby numbers. It shows that the flow behaves differently inside and outside the tangent cylinder. The rotation rate is maximum on the tangent cylinder. At this location the analytical inviscid solution is not differentiable and the vertical velocity is also discontinuous because it tends to minus infinity when the tangent cylinder is approached from the left. In low Ekman number flows, such discontinuities are expected to produce boundary layers whose thickness decreases with the Ekman number. This is visible in Fig. \ref{Rot_Diff_Taylor_Proudman} which also shows that, as expected, the agreement with the analytical solution improves for smaller Ekman numbers. The right panel further shows that for low enough Ekman number and in the linear regime
the global differential rotation between the inner and the outer spheres taken along a given radius does scale as:

\begin{equation}
\displaystyle \frac{\Delta \Omega}{\Omega_0} \propto \displaystyle \frac{R_o}{\sqrt{E}}.
\end{equation}

Finally, the existence of the Taylor-Proudman regime in a stably stratified atmosphere where $P_r \left(N_0/\Omega_0\right)^2 \ll \sqrt{E}$ can be explained as follows:
from Eq. \eqref{cylindrical_radial_velocity_TP}, we see that $U_r \approx V_0$. Thus, according to the thermal balance Eq. \eqref{thermal_balance}, $\delta \Theta \approx P_r \left(N_0/\Omega_0\right)^2 \hspace*{0.05cm} E^{-1} \hspace*{0.05cm} V_0$ while according to Eq. \eqref{ekman_pumping}, $U_{\phi} \approx E^{-1/2} \hspace*{0.05cm} V_0$. The ratio of these two quantities is $U_{\phi} / \delta \Theta \approx \sqrt{E} / P_r \left(N_0/\Omega_0\right)^2$. As a consequence, in the regime $P_r \left(N_0/\Omega_0\right)^2 \ll \sqrt{E}$, the effect of the buoyancy forces
in the thermal wind equation Eq. \eqref{boussthermalwind} is too weak and the Taylor-Proudman constraint $\partial U_{\phi} / \partial z =  0$ holds.

\subsection{Viscous regime}
\label{viscous_regime}

According to the timescale analysis of Sect. \ref{timescales_physical_processes}, the viscous transport of AM dominates when $1 \ll  P_r \left( N_0/\Omega_0 \right)^2$. To investigate this regime, numerical simulations have been performed for values of $P_r \left( N_0/\Omega_0 \right)^2$ varying between $10^2$ and $10^{\hspace*{0.02cm} 4}$. The different runs characterised by the three parameters, $P_r \left( N_0/\Omega_0 \right)^2$, $E$, $Re_c$, are listed in Table \ref{parameters_viscous}.

\begin{table}[h]
\begin{center}
\begin{tabular}{c|c|c|c}
\hline
\hspace*{0.05cm} \textbf{$\text{Simulation}$} \hspace*{0.05cm} & \hspace*{0.5cm} $\boldsymbol{E}$ \hspace*{0.5cm} & \hspace*{0.05cm} $\boldsymbol{P_r \left( N_0 / \Omega_0 \right)^2}$ \hspace*{0.05cm} & \hspace*{0.5cm} $\boldsymbol{Re_c}$ \hspace*{0.5cm} \\ 
\hline
1.1 & $10^{-2}$ & $10^{2}$ & $10^{-2}$ \\
1.2 & $10^{-3}$ & $10^{2}$ & $10^{-2}$ \\
1.3 & $10^{-4}$ & $10^{2}$ & $10^{-2}$ \\
1.4 & $10^{-5}$ & $10^{2}$ & $10^{-2}$ \\
\hline
2.1 & $10^{-2}$ & $10^{3}$ & $10^{-2}$ \\
2.2 & $10^{-3}$ & $10^{3}$ & $10^{-2}$ \\
2.3 & $10^{-4}$ & $10^{3}$ & $10^{-2}$ \\
2.4 & $10^{-5}$ & $10^{3}$ & $10^{-2}$ \\
\hline
3.1 & $10^{-2}$ & $10^{4}$ & $10^{-2}$ \\
3.2 & $10^{-3}$ & $10^{4}$ & $10^{-2}$ \\
3.3 & $10^{-4}$ & $10^{4}$ & $10^{-2}$ \\
3.4 & $10^{-5}$ & $10^{4}$ & $10^{-2}$ \\
\hline
4.1 & $10^{-4}$ & $10^{4}$ & $10^{-1}$ \\
4.2 & $10^{-4}$ & $10^{4}$ & $1$ \\
4.3 & $10^{-4}$ & $10^{4}$ & $10$ \\
\hline
5.1 & $10^{-5}$ & $10^{4}$ & $1$ \\
\hline
\end{tabular}
\vspace*{0.2cm}
\caption{Parameters of performed simulations in the viscous regime.}
\label{parameters_viscous}
\end{center}
\end{table}

\begin{figure*}[h]
\begin{center}
\includegraphics[width=4.4cm]{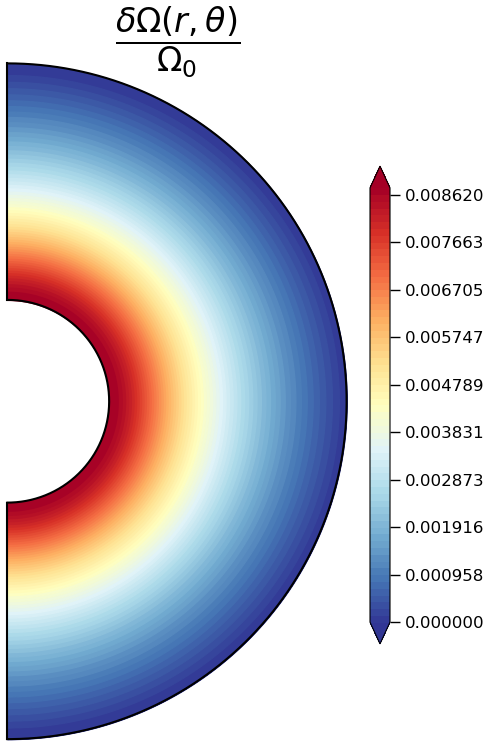}
\includegraphics[width=4.4cm]{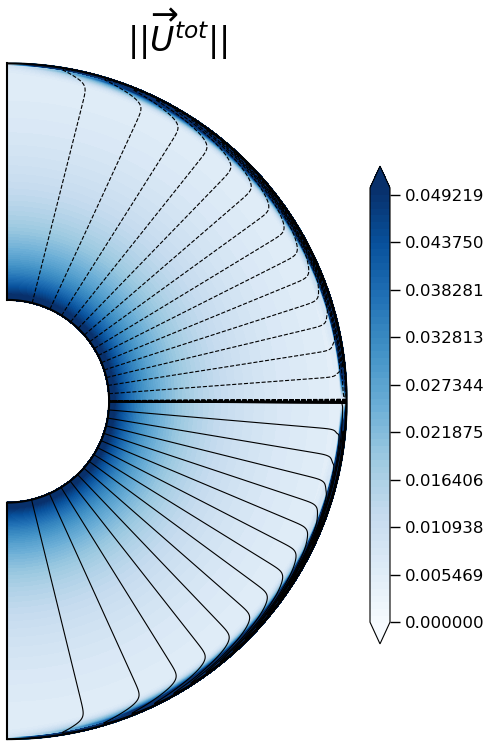}
\includegraphics[width=4.4cm]{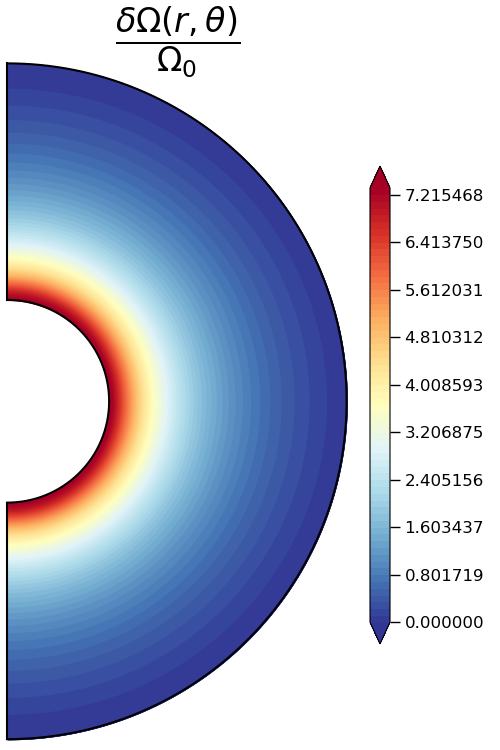}
\includegraphics[width=4.4cm]{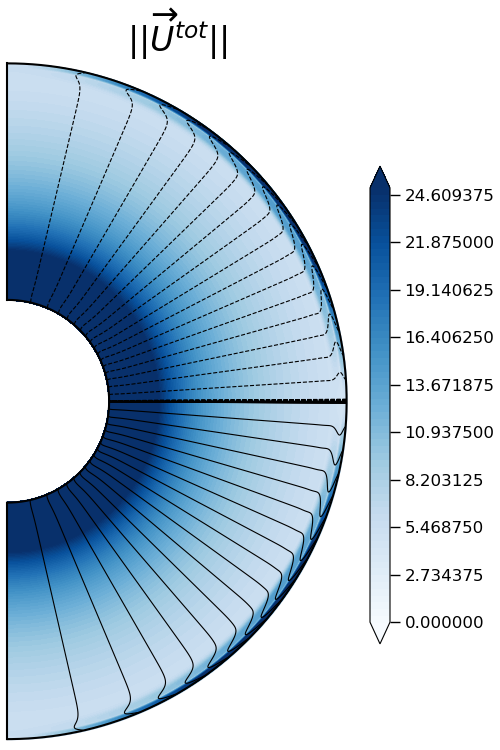}
\caption{Differential rotation normalised to the rotation rate of the outer sphere $\delta \Omega(r,\theta)/\Omega_0$ in the viscous regime (first and third panels) and norm of the total meridional velocity field $\protect\vv{U}^{\hspace*{0.05cm} \text{tot}} = \left(U_r + V_f \hspace*{0.02cm} (r) \right) \protect\vv{e}_r \hspace*{0.02cm} + \hspace*{0.02cm} U_{\theta} \protect\vv{e}_{\theta}$ (second and fourth panels) with the streamlines of the associated total meridional circulation (in black). In the two left panels, the parameters of the simulation are $P_r \left( N_0 / \Omega_0 \right)^2 = 10^{\hspace*{0.02cm} 4}$, $Re_c = 10^{-2}$, and $E =10^{-5}$ (run $3.4$ of Table \ref{parameters_viscous}) and in the two right panels, the parameters are $P_r \left( N_0 / \Omega_0 \right)^2 = 10^{\hspace*{0.02cm} 4}$, $Re_c = 10$, and $E =10^{-4}$ (run $4.3$ of Table \ref{parameters_viscous}). In the second and fourth panels, the black full lines represent the streamlines of an ascending and inwardly directed circulation while the black dashed lines correspond to a descending and inwardly directed one.}
\label{Rot_Diff_Viscous}
\end{center}
\end{figure*}

In this parameter range, we find that the stationary differential rotation is mostly radial and that the level of differential rotation increases with $Re_c$. This is illustrated by the first and third panels in Fig. \ref{Rot_Diff_Viscous} where the differential rotation is shown for two runs with equally high $P_r \left( N_0 / \Omega_0 \right)^2 = 10^{\hspace*{0.02cm} 4}$ and different $Re_c$. The differential rotation is closely radial in both cases while the rotation contrast between the outer and the inner spheres goes from $\sim 10^{-2}$ for $ Re_c = 10^{-2}$ to $\sim8$ as the contraction rate is increased to $ Re_c=10 $. The other two panels display the total meridional velocity field, through its norm and streamlines, showing that it is clearly dominated by the imposed radial contraction flow. Subtracting this contribution, we would see a meridional circulation confined into boundary layers at the outer and inner spheres, the confinement being stronger for smaller Ekman numbers.

The results of the numerical simulations are now compared with the analytical solution that can be derived by neglecting the Coriolis and the non-linear advection terms in the AM conservation Eq. \eqref{angular_momentum_evolution}. The steady balance between the contraction and viscous terms indeed reads: 

\begin{equation}
-\nu ~ \left( \vv{\nabla}^2 
- \displaystyle \frac{1}{r^2 \sin^2 \theta} \right) U_{\phi} = \displaystyle \frac{V_0 r{_0^2}}{r^3} \left( \frac{\partial}{\partial r} \left( r U_{\phi} + r^2 \sin{\theta} ~ \Omega_0 \right) \right)
\label{viscous_balance}
\end{equation}

\noindent whose solution is given by the following expression:

\begin{equation}
\begin{array}{lll}
\displaystyle \frac{\delta \Omega_{\hspace*{0.02cm} \nu}(r)}{\Omega_0} = Re_c \Bigg{[} 2 \left( Re_c + \displaystyle \frac{r_i}{r_0} \right) \left( \exp{\left(Re_c\right)} - \exp{\left(Re_c\displaystyle \frac{r_0}{r}\right)} \right) ~ + \\\\ \displaystyle \frac{Re_c r_i}{r} \exp{ \left(Re_c\displaystyle \frac{r_0}{r_i}\right)} \left( Re_c \left( 1 - \displaystyle \frac{r^2}{r_0^2} \right) \displaystyle \frac{r_0}{r} + 2 \left( 1 - \displaystyle \frac{r}{r_0} \right) \right) \Bigg{]} ~ ~ \Bigg{/} \\\\ \Bigg{[} \displaystyle \frac{r_i}{r_0} \hspace*{0.02cm} \left( Re_c^2 + 2 Re_c + 2 \right) \exp{\left(Re_c\displaystyle \frac{r_0}{r_i}\right)} - 2 \hspace*{0.02cm} \left( Re_c + 1 \right) \exp{\left(Re_c\right)} \Bigg{]}
\end{array}
\label{full_solution_viscous}
\end{equation}

If the differential rotation $\delta \Omega_{\hspace*{0.02cm} \nu}(r)/\Omega_0$ is smaller than one, the forcing term in Eq. \eqref{viscous_balance} is reduced to $\left(2 \Omega_0 \sin{\theta} V_0 r{_0^2}\right)/r^2$, and the analytical solution for the differential rotation simplifies into:

\begin{equation}
\displaystyle \frac{\delta \Omega_{\hspace*{0.02cm} \nu}^{\hspace*{0.02cm} \text{L}}(r)}{\Omega_0} = Re_c \left \lbrack \displaystyle \frac{r_i^2}{3 r_0^2} \left( 1 - \displaystyle \frac{r_0^3}{r^3} \right) + \left( \displaystyle \frac{r_0}{r} - 1 \right) \right \rbrack
\label{simple_sol}
\end{equation}

\noindent which implies that this approximate solution should be valid for low Reynolds numbers $Re_c$.

\begin{figure}[h]
\includegraphics[width=8.5cm]{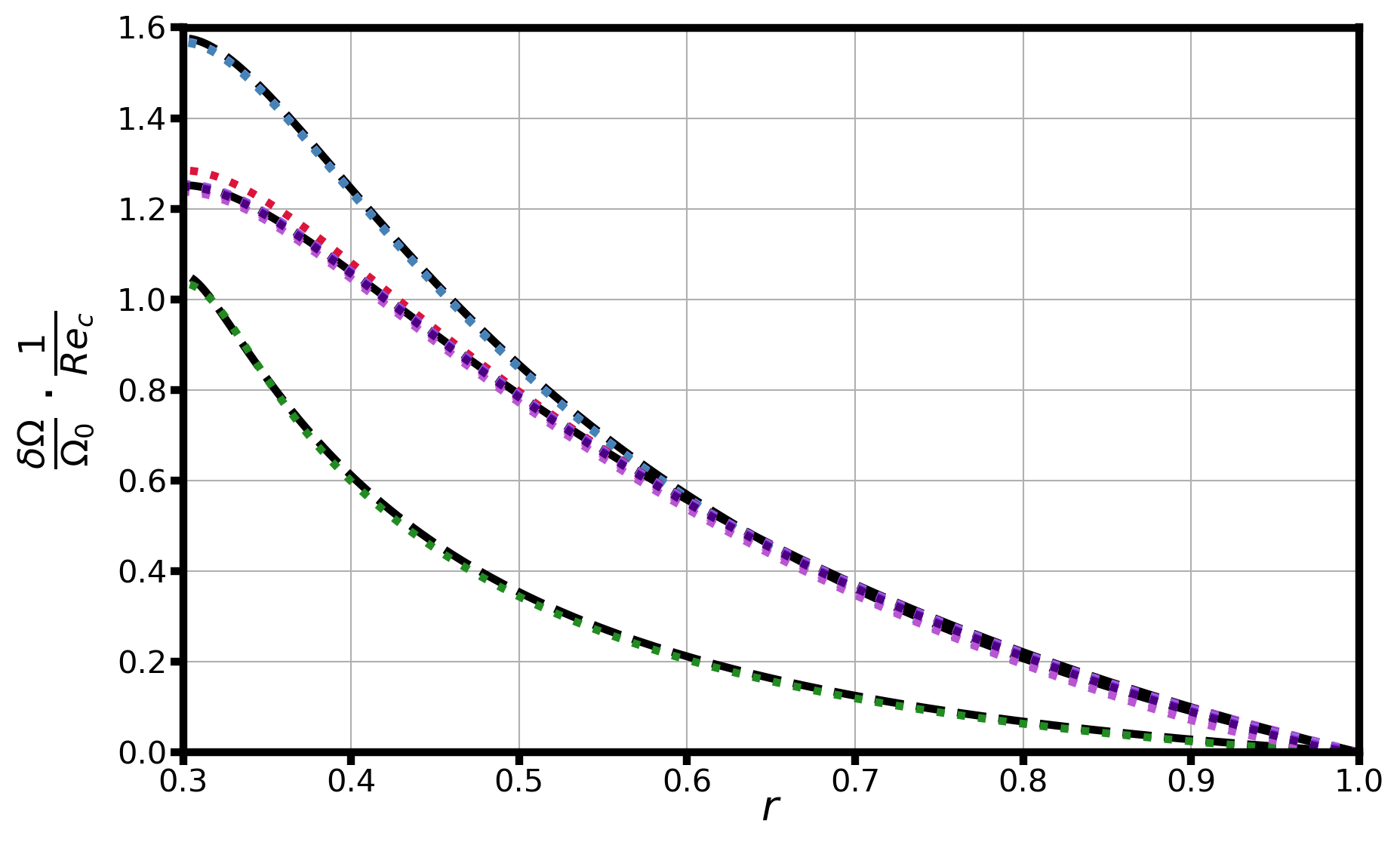}
\caption{Differential rotation in the viscous regime scaled by $Re_c$ as a function of radius at the latitude $\theta = \pi/8$. The analytical solution Eq. \eqref{full_solution_viscous} (black dashed lines) is compared with the numerical results. Low-Reynolds number runs ($1.1$ to $4.1$ of Table \ref{parameters_viscous}, with $Re_c=10^{-2} - 10^{-1}$) are plotted with violet dotted lines and are all very close to the low Reynolds number version of the analytical solution, Eq. \eqref{simple_sol}. Simulations at $Re_c= 1$ (runs $4.2$ and $5.3$ of Table \ref{parameters_viscous}) and $Re_c=10$ (run $4.3$ Table \ref{parameters_viscous}) are also in agreement with the analytical solution.}
\label{Viscous_Scaling_Bouss}
\end{figure}

The analytical solution is compared with the numerical ones in Fig. \ref{Viscous_Scaling_Bouss}. The agreement is very good showing that the viscous term indeed dominates the transport in the regime $P_r \left( N_0/\Omega_0 \right)^2 \ge 10^2$ considered. We also find that the linear relation $\Delta \Omega/\Omega_0 \propto Re_c = \tau_{\nu}/\tau_{\text{c}}$ valid for low $Re_c$ still provides a good order of magnitude estimate of the differential rotation between the inner and the outer spheres up to the largest $Re_c$ considered. 

In this sub-section we have shown that for high enough $P_r \left( N_0 / \Omega_0 \right)^2 \gtrsim 10$ stable stratification is so strong that the meridional circulation is inhibited and the effect of the contraction is balanced by the viscous transport of AM. An analytical solution of the resulting radial differential rotation is available.

\subsection{Eddington-Sweet regime}
\label{eddington_sweet_regime}

When the stratification is not too strong $\sqrt{E} \ll P_r \left(  N_0/\Omega_0 \right)^2 \ll 1 $, we expect that the transport of AM is dominated by an Eddington-Sweet type meridional circulation.

\begin{table}[h]
\begin{center}
\begin{tabular}{c|c|c|c}
\hline
\hspace*{0.1cm} \textbf{$\text{Simulation}$} \hspace*{0.1cm} & \hspace*{0.4cm} $\boldsymbol{E}$ \hspace*{0.4cm} & \hspace*{0.2cm} $\boldsymbol{P_r \left( N_0 / \Omega_0 \right)^2}$ \hspace*{0.2cm} & \hspace*{0.4cm} $\boldsymbol{Re_c}$ \hspace*{0.4cm} \\ 
\hline
1.1 & $10^{-3}$ & $10^{-1}$ & $10^{-2}$ \\
1.2 & $10^{-4}$ & $10^{-1}$ & $10^{-2}$ \\
1.3 & $10^{-5}$ & $10^{-1}$ & $10^{-2}$ \\
1.4 & $10^{-6}$ & $10^{-1}$ & $10^{-2}$ \\
\hline
2.1 & $10^{-5}$ & $10^{-2}$ & $10^{-2}$ \\
2.2 & $10^{-6}$ & $10^{-2}$ & $10^{-2}$ \\
2.3 & $10^{-7}$ & $10^{-2}$ & $10^{-2}$ \\
\hline
3.1 & $10^{-6}$ & $10^{-1}$ & $10^{-1}$ \\
3.2 & $10^{-6}$ & $10^{-2}$ & $10^{-1}$ \\
\hline
4.1 & $10^{-6}$ & $10^{-1}$ & $1$ \\
4.2 & $10^{-6}$ & $10^{-2}$ & $1$ \\
\hline
5.1 & $10^{-6}$ & $10^{-1}$ & $5$ \\
5.2 & $10^{-6}$ & $10^{-2}$ & $5$ \\
\hline
6.1 & $10^{-6}$ & $10^{-1}$ & $10$ \\
\hline
\end{tabular}
\vspace*{0.2cm}
\caption{Parameters of the simulations in the Eddington-Sweet regime.}
\label{parameters_edd}
\end{center}
\end{table}

\begin{figure}[h]
\begin{center}
\includegraphics[width=4.4cm]{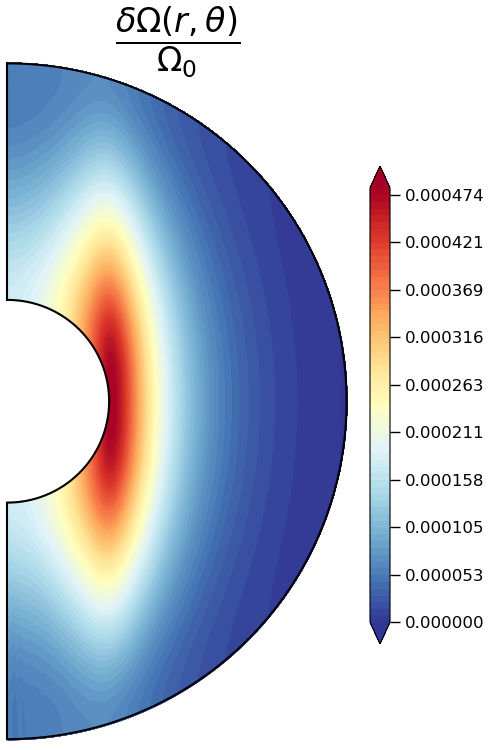}
\includegraphics[width=4.4cm]{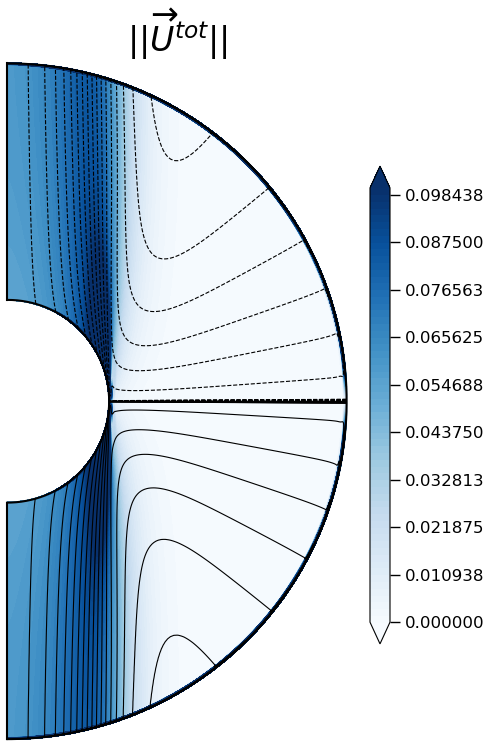}
\caption{Steady differential rotation (left panel) normalised to the top value as well as, norm and streamlines of the total meridional velocity field $\protect\vv{U}^{\hspace*{0.05cm} \text{tot}} = \left(U_r + V_f \hspace*{0.02cm} (r) \right) \protect\vv{e}_r + U_{\theta} \protect\vv{e}_{\theta}$ (right panel). The parameters of the simulation (run $1.4$ of Table \ref{parameters_edd}) are $E=10^{-6}$, $P_r \left( N_0/\Omega_0 \right)^2 = 10^{-1}$ and $Re_c=10^{-2}$. Outside the tangent cylinder, the black dashed lines represent the streamlines of a counterclockwise circulation and the full lines, the ones of a clockwise circulation. Inside the tangent cylinder the dashed lines simply correspond to the streamlines of a downward circulation while the full lines are for an upward one.}
\label{Rot_Diff_Edd_Sweet}
\end{center}
\end{figure}

The meridional cut displayed in the left panel of Fig. \ref{Rot_Diff_Edd_Sweet} shows the typical flow structure obtained in this regime. The differential rotation exhibits a dependence in both latitude and radius. Thus, contrary to the Taylor-Proudman and viscous regimes, it is neither cylindrical nor spherical. Meanwhile the total meridional circulation (right panel)
is similar to the one already described in the Taylor-Proudman case with an inward flow concentrated inside the tangent cylinder and more particularly in 
a vertical jet towards the equator of the inner sphere. To understand what determines the level and distribution of the differential rotation in the Eddington-Sweet regime, we performed a parametric study where $E$, $P_r \left(N_0/\Omega_0\right)^2$ and $Re_c$ were varied, the values of these parameters being listed in Table \ref{parameters_edd}.

Owing to the flow complexity, we were not able to find an analytical solution for the azimuthal velocity field. We shall instead compare our numerical results with a scaling relation for the amplitude of differential rotation, that we derive as follows. Assuming a linear regime, the balance between the Coriolis term and the contraction term in the AM evolution equation Eq. \eqref{angular_momentum_evolution} implies that the circulation velocities scale as the imposed contraction radial velocity field: 

\begin{equation}
U_r \approx U_\theta \approx V_0
\label{order_ur_edd}
\end{equation}

Then, the thermal wind balance \eqref{boussthermalwind} and the thermal balance \eqref{thermal_balance} provide an estimate for the radial velocity amplitude given by Eq. \eqref{ur_estimation}. Injecting Eq. \eqref{order_ur_edd} in Eq. \eqref{ur_estimation}, we obtain:

\begin{equation}
V_0 \approx \displaystyle \frac{\kappa}{r_0} \displaystyle \frac{\Omega_0^2}{N_0^2} \displaystyle \frac{\Delta \Omega}{\Omega_0}
\label{new_order}
\end{equation}

\noindent which is equivalent to :

\begin{equation}
\displaystyle \frac{\Delta \Omega}{\Omega_0}  \approx \displaystyle \frac{\tau_{\text{ED}}}{\tau_{\text{c}}} = P_r \left( \displaystyle \frac{N_0}{\Omega_0} \right)^2 \displaystyle \frac{R_o}{E}
\label{edd_estimation}
\end{equation}

Physically, it says that the steady differential rotation is determined by the balance between the inward transport of AM forced by the contraction with timescale $\tau_{\text{c}}^{~\text{L}}$, and the outward transport of AM by the Eddington-Sweet circulation with timescale $\tau_{\text{ED}}$.

\begin{figure}[h]
\begin{center}
\hspace*{-0.4cm}
\includegraphics[width=8.5cm]{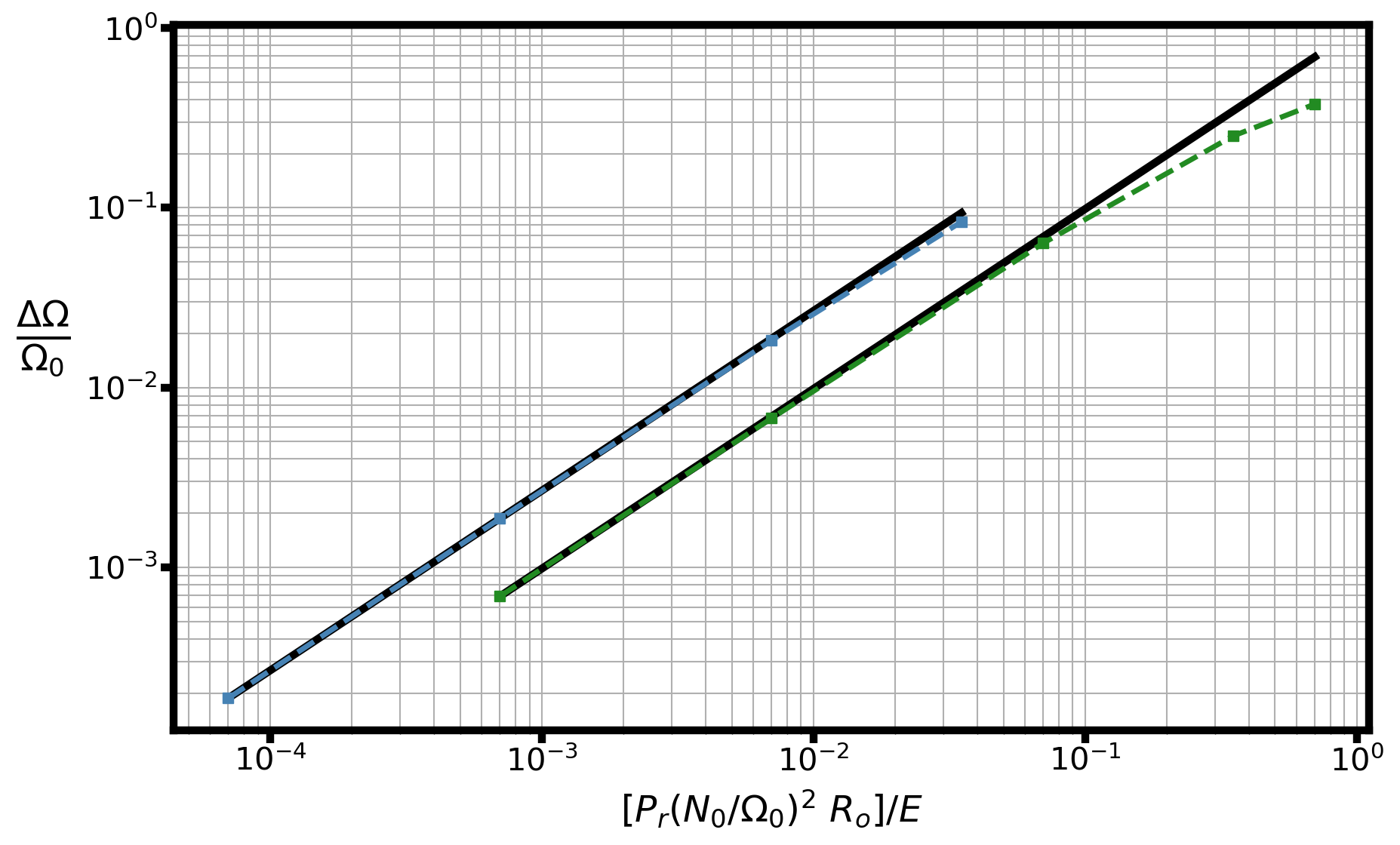}
\caption{Differential rotation between the inner equator and the outer sphere normalised to the outer value of the rotation rate $\Omega_0$ in the Eddington-Sweet regime, plotted as a function of $P_r\left(N_0/\Omega_0\right)^2 R_o/ E$. The Rossby number $R_o$ is varied to encompass the linear and non-linear regimes. The Ekman number is fixed to $E= 10^{-6}$.  The green curve corresponds to simulations with $P_r \left(N_0/\Omega_0\right)^2 = 10^{-1}$ (runs $1.4$, $3.1$, $4.1$, $5.1$ and $6.1$ of the Table \ref{parameters_edd}) and the blue one to simulations
with $P_r \left(N_0/\Omega_0\right)^2 = 10^{-2}$ (runs $2.2$, $3.2$, $4.2$ and $5.2$ of the same table).}
\label{Rot_Diff_Scaling_Edd_Sweet}
\end{center}
\end{figure}

Coming back to the numerical results, the left panel in Fig. \ref{Rot_Diff_Edd_Sweet} shows that the differential rotation taken between the inner and the outer spheres depends on the latitude considered at the inner sphere and that it is maximum at the equator. Figure \ref{Rot_Diff_Scaling_Edd_Sweet} displays the differential rotation between the equator of the inner sphere and the outer sphere for simulations performed at various $R_o$, considering two different $P_r \left(N_0/\Omega_0\right)^2 = 10^{-1}$ and $10^{-2}$, and a fixed low Ekman number $E= 10^{-6}$.

This figure first shows that a linear regime exists as long as $\Delta \Omega / \Omega_0$ is small enough. Indeed, the differential rotation is clearly linearly dependent on $R_o$ until $\Delta \Omega / \Omega_0 \gtrsim 10^{-1}$, where it starts to deviate. The differential rotation then falls below the linear relation because the non-linear advection terms, both in the momentum and temperature equations, increase the efficiency of the AM redistribution.

This figure also shows that the scaling Eq. \eqref{edd_estimation} does not reproduce the differential rotation amplitude because the curves corresponding to the two different $P_r \left(N_0/\Omega_0\right)^2$ do not collapse. The numerical results would be better reproduced by a relation of the form $\Delta \Omega / \Omega_0 \propto \sqrt{P_r \left( N_0/\Omega_0 \right)^2} \hspace*{0.02cm} R_o/E$. 

However, we now argue that Eq. \eqref{edd_estimation} should still be approximately valid for the very low stellar Ekman numbers, by showing that the observed discrepancy is due to the relatively high Ekman numbers of our simulations. Indeed, we find that our results are best described by a model where the differential rotation is the superposition of two terms: 

\begin{equation}
\frac{\delta \Omega_{\hspace*{0.02cm} \text{tot}}(r,\theta)}{\Omega_0} = \displaystyle \frac{R_o}{\sqrt{E}} \left( \frac{\delta \Omega_{~ \text{TP}}^{(0)}(r,\theta)}{\Omega_0} + \displaystyle \frac{P_r \left(N_0/\Omega_0\right)^2}{\sqrt{E}} \frac{\delta \Omega_{~ \text{ED}}^{(0)}(r,\theta)}{\Omega_0} \right)
\label{combination}
\end{equation}

\noindent the first term being a by-product of the Ekman layer induced by the mismatch between the inviscid contraction-driven interior circulation (Eqs. \eqref{radial_Edd} and \eqref{latitudinal_Edd}) and  the boundary condition $U_r = 0$ at the outer sphere. This term is actually close to the unstratified Taylor-Proudman solution described in Sect. \ref{taylor_proudman_regime} that shows a $\mathcal{O}\left(1/\sqrt{E}\right)$ jump of differential rotation across the $\mathcal{O}\left(\sqrt{E}\right)$ outer Ekman layer. The second term scales with $P_r \left(N_0/\Omega_0\right)^2/E$ and is of the Eddington-Sweet type. It connects smoothly to the outer sphere boundary condition, that is $\delta \Omega_{\hspace*{0.02cm} \text{ED}}(r_0,\theta) =0$. Therefore, whereas the Eddington-Sweet term is dominant in the interior for small enough $\sqrt{E}/P_r \left(N_0/\Omega_0\right)^2$, the first term is never fully negligible close to the outer sphere. In this model the interior meridional circulation velocities and temperature fluctuations scale as Eqs. \eqref{order_ur_edd} and \eqref{thermal_fluct}, respectively (details are given in Appendix \eqref{superposition}).

\begin{figure*}
\begin{center}
\includegraphics[width=4.4cm]{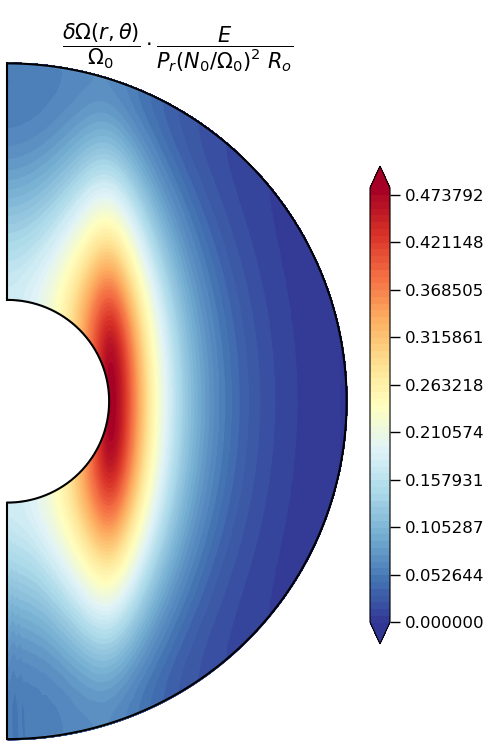}
\hspace*{1.1cm}
\includegraphics[width=4.4cm]{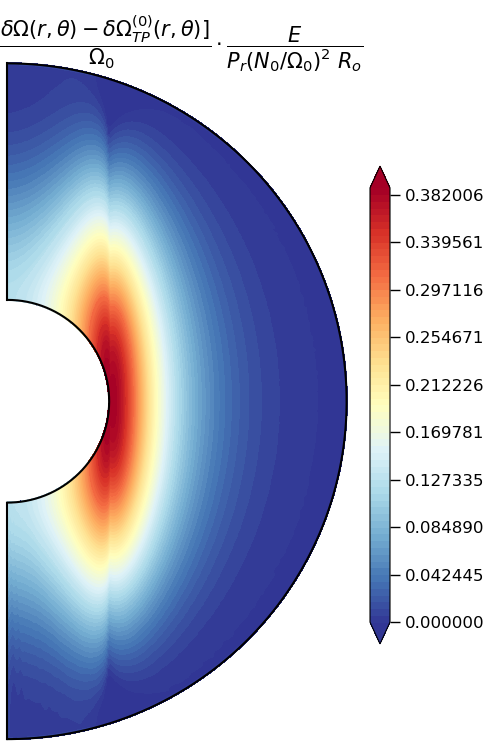}
\hspace*{1cm}
\includegraphics[width=4.4cm]{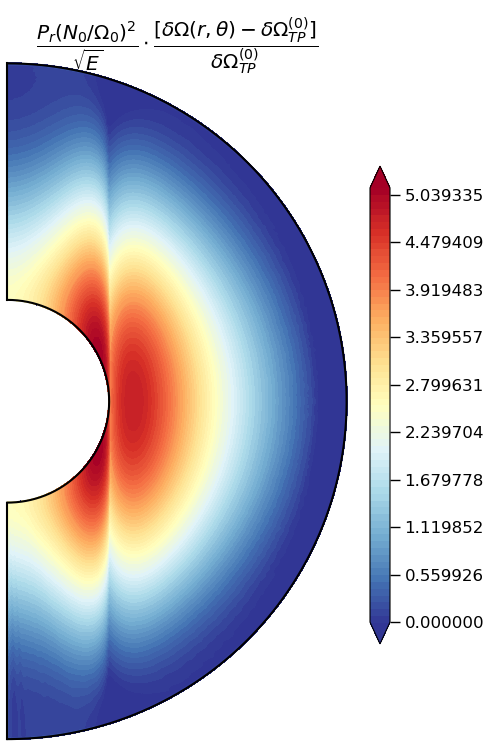}
\end{center}
\end{figure*}
\begin{figure*}
\begin{center}
\includegraphics[width=4.4cm]{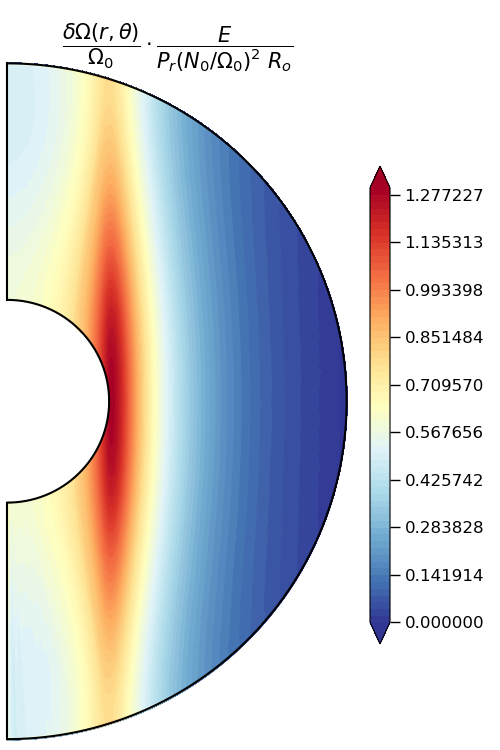}
\hspace*{1.1cm}
\includegraphics[width=4.4cm]{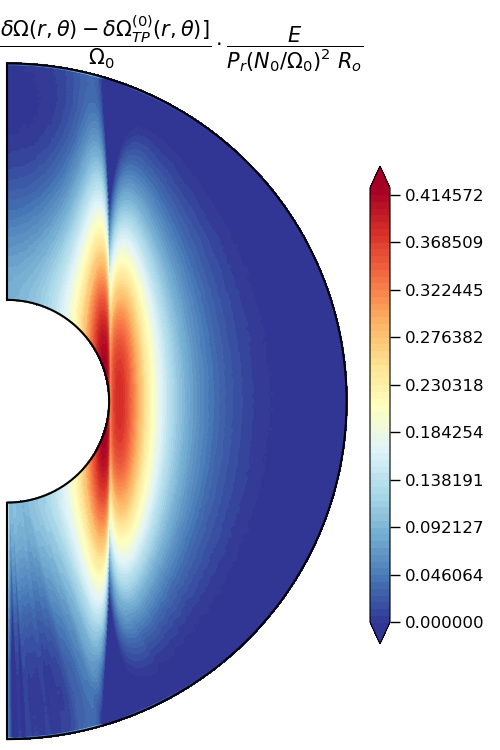}
\hspace*{1cm}
\includegraphics[width=4.4cm]{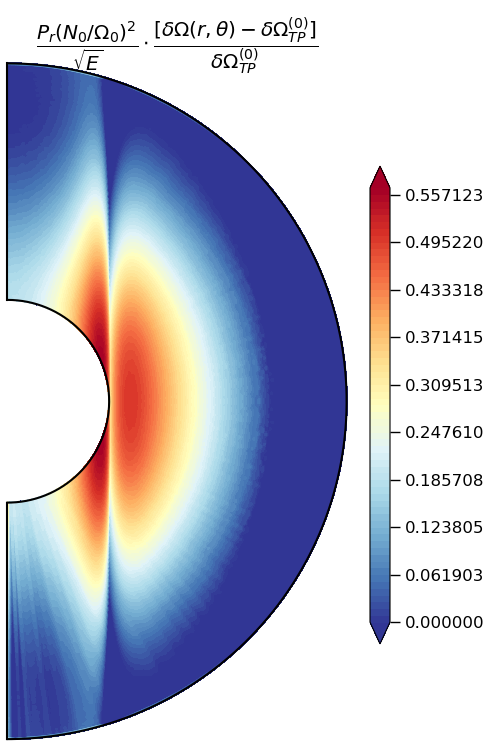}
\end{center}
\caption{Differential rotation in the Eddington-Sweet linear regime for $P_r \left(N_0/\Omega_0\right)^2 = 10^{-1}$ (top panel) and $10^{-2}$ (bottom panel), the other two parameters $E =10^{-6}$ and $Re_c = 10^{-2}$ being identical (runs $1.4$ and $2.2$ of Table \ref{parameters_edd}). The coloured contours are normalised to the rotation rate of the outer sphere $\Omega_0$. [Left column]: differential rotation $\delta \Omega(r,\theta)/\Omega_0$ rescaled by $E ~ \left(P_r \left(N_0/\Omega_0\right)^2 R_o \right)^{-1}$. [Middle column]: differential rotation $\left(\delta \Omega - \delta \Omega_{\text{TP}}^{(0)}\right) / \Omega_0$ obtained after subtracting the unstratified Taylor-Proudman solution given by Eq. \eqref{differential_rotation_TP} and rescaling by $E ~ \left(P_r \left(N_0/\Omega_0\right)^2 R_o \right)^{-1}$ (see Eq. \eqref{combination}). [Right column]: ratio $ \left( P_r \left(N_0/\Omega_0\right)/\sqrt{E} \right) \cdot \left \lbrack \left(\delta \Omega - \delta \Omega_{\text{TP}}^{(0)}\right) / \delta \Omega_{\hspace*{0.02cm} \text{TP}}^{(0)} \right \rbrack$.}
\label{Full_Rot_Diff_Rescaled_Edd_Sweet}
\end{figure*}

As illustrated in Fig. \ref{Full_Rot_Diff_Rescaled_Edd_Sweet}, our numerical results are consistent with this model.  Considering flows in the linear regime for two different $P_r \left(N_0/\Omega_0\right)^2 = 10^{-1}$ and $10^{-2}$, and the same Ekman number $E=10^{-6}$, we test the relevance of the Eddington-Sweet scaling given by Eq. \eqref{edd_estimation}, when applied to the full differential rotation on one hand, and to the field obtained after subtracting the unstratified solution Eq. \eqref{differential_rotation_TP} on the other hand. We observe that, in this last case, the distribution and amplitude of the rescaled differential rotations are very similar, whereas significant amplitude differences persist when the unstratified solution is not subtracted before rescaling. The fact that the scaling Eq. \eqref{edd_estimation} is not accurate for the full differential rotational thus appears to be due to the non-negligible contribution of the $\propto 1/\sqrt{E}$ term. This is confirmed by the comparison of the two terms of Eq. \eqref{combination} displayed in the right column of Fig. \ref{Full_Rot_Diff_Rescaled_Edd_Sweet}. This comparison clearly shows that $\sqrt{E}/P_r \left(N_0/\Omega_0\right)^2$ has to be as low as $10^{-2}$ and that one has to consider layers away from the outer sphere for the Eddington-Sweet term to clearly dominate over the Taylor-Proudman term.

Runs performed at other Ekman numbers confirm the relevance of this description. In particular, as expected from the expression Eq. \eqref{combination}, simulations with the same ratio $\sqrt{E}/P_r \left(N_0/\Omega_0\right)^2$ (such as runs 1.3 and 2.3 of Table \ref{parameters_edd}) have similar scaled differential rotation $\bigg{(} \delta \Omega_{\hspace*{0.02cm} \text{ED}} / \Omega_0 \bigg{)} \bigg{(} E \big{/} P_r \left(N_0/\Omega_0\right)^2 \bigg{)}$.

While the Eddington-Sweet scaling and its $\mathcal{O}\left(\sqrt{E}/P_r \left(N_0/\Omega_0\right)^2\right)$ Taylor-Proudman correction provide a global description of the differential rotation, additional physics is required to describe regions close to boundary layers. This concerns the $\mathcal{O}\left(E^{1/2}\right)$ Ekman layer at the outer sphere, the free layer that smooths the vertical velocity jump across the tangent cylinder, and also the Ekman layer at the inner sphere induced by the mismatch between the interior flow and the boundary conditions there. As for the classical spherical Couette flow, the Ekman layer at the inner sphere is stronger (that is, more extended and with higher amplitudes) near the equator than at higher latitudes. As can be observed in Fig. \ref{Full_Rot_Diff_Rescaled_Edd_Sweet}, we find that the thicknesses of the layer along the tangent cylinder and of the equatorial layer decrease with $E$ and $P_r \left(N_0/\Omega_0\right)^2$, although determining their exact scaling with these non-dimensional numbers might be tricky and is beyond the scope of the present study.

\section{Effect of the density stratification : the anelastic case}
\label{effect_of_the_density_stratification}

\begin{table}[h]
\begin{center}
\begin{tabular}{c|c|c}
\hline
$\hspace*{0.5cm} \boldsymbol{\epsilon_s} \hspace*{0.5cm}$ & \hspace*{0.4cm} $\boldsymbol{D_i}$ \hspace*{0.4cm} & \hspace*{0.2cm} $\boldsymbol{\rho_i} \mathbf{/} \boldsymbol{\rho_0}$ \hspace*{0.2cm} \\ 
\hline
$0.1$ & $0.1$ & $1.75$\\
$0.1$ & $1.0$ & $10.8$\\
$0.1$ & $1.7$ & $20.9$\\
$1.0$ & $1.0$ & $60.2$\\
\hline
\end{tabular}
\vspace*{0.2cm}
\caption{Density contrasts between the inner and outer spheres investigated in the anelastic approximation, and their associated parameters.}
\label{density_ratio_viscous_anel}
\end{center}
\end{table}

In this section, we analyse the effects of the density stratification using the anelastic approximation. To do so, the magnitude of the background density contrast between the inner and the outer spheres, $\rho_i/\rho_0$, is varied from $1.75$ to $60.2$. Table \ref{density_ratio_viscous_anel} lists the different density contrasts considered together with the associated parameters of the background state, the dissipation number $D_i$ and the $\epsilon_s$ (see Eqs. \eqref{background_temperature_profile} and \eqref{background_density_profile}).

\begin{figure}[h]
\begin{center}
\hspace*{-0.7cm}
\includegraphics[width=8.5cm]{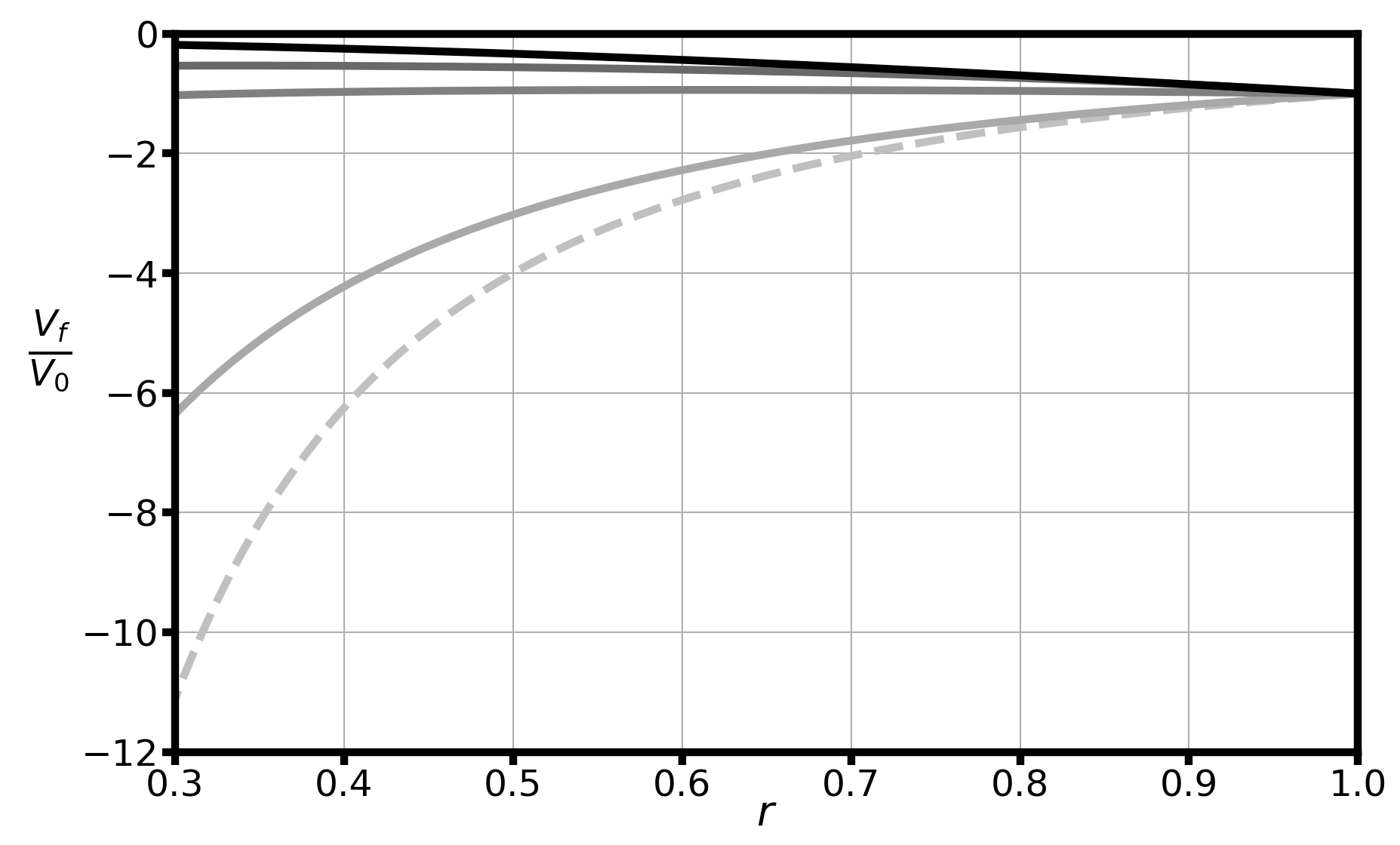}
\caption{Contraction radial velocity field $V_f$, Eq. \eqref{Anelcontraction}, normalised to the top value $V_0$ and plotted as a function of radius $r$. From the lightest to the darkest, correspond the weakest to the strongest density contrasts between the inner and outer spheres which are respectively $ \rho_i/\rho_0 = 1.75$, $10.8$, $20.9$ and $60.2$. The contraction radial velocity field in the Boussinesq case given by Eq. \eqref{Bousscontraction} (i.e. corresponding to a density ratio of $1$) is plotted in dashed lines as a reference.}
\label{comparison_contraction_radial_velocity_field}
\end{center}
\end{figure}

The first effect of the density stratification is to modify the profile of the contraction velocity field (see Eqs. \eqref{Anelcontraction} and \eqref{Bousscontraction}). Figure \ref{comparison_contraction_radial_velocity_field} shows that for a density contrast of $1.75$ the absolute value of the contraction velocity at the inner sphere is twice as small as in the Boussinesq case. For stronger contrasts, that is when $\rho_i/\rho_0 > \left( r_0/r_i\right)^2 \approx 11.11$, the absolute value of the contraction velocity field even decreases with depth.

\subsection{Anelastic Taylor-Proudman regime}
\label{anelastic_taylor_proudman_regime}

We first study the effects of the density stratification in the regime $P_r \left(N_0/\Omega_0\right)^2 \ll \sqrt{E} \ll 1$ where the Boussinesq simulations have shown that the rotation is cylindrical because the buoyancy force has a negligible effect on the dynamics. The Ekman number, the $P_r \left(N_0/\Omega_0\right)^2$ parameter, and the contraction Reynolds number are fixed to $10^{-5}$, $10^{-4}$, $10^{-2}$ respectively, while the density contrast between the inner and outer spheres is varied from $1.75$ to $60.2$. Two additional simulations have been also performed at $E=10^{-4}$ and $E=10^{-6}$, the other parameters being fixed. The different runs and their associated parameters are summarised in Table \ref{parameters_tp_anel}.

\begin{table}[h]
\begin{center}
\begin{tabular}{c|c|c|c|c}
\hline
\hspace*{0.01cm} \textbf{$\text{Simulation}$} \hspace*{0.01cm} & \hspace*{0.3cm} $\boldsymbol{E}$ \hspace*{0.3cm} &  $\boldsymbol{P_r \left( N_0 / \Omega_0 \right)^2}$ & \hspace*{0.1cm} $\boldsymbol{Re_c}$ \hspace*{0.1cm} & \hspace*{0.1cm} $\boldsymbol{\rho_i} \mathbf{/} \boldsymbol{\rho_0}$ \hspace*{0.1cm} \\ 
\hline
1.1 & $10^{-4}$ & $10^{-4}$ & $10^{-2}$ & $20.9$\\
\hline
2.1 & $10^{-5}$ & $10^{-4}$ & $10^{-2}$ & $1.75$\\
2.2 & $10^{-5}$ & $10^{-4}$ & $10^{-2}$ & $10.8$\\
2.3 & $10^{-5}$ & $10^{-4}$ & $10^{-2}$ & $20.9$\\
2.4 & $10^{-5}$ & $10^{-4}$ & $10^{-2}$ & $60.2$\\
\hline
3.1 & $10^{-6}$ & $10^{-4}$ & $10^{-2}$ & $20.9$\\
\hline
\end{tabular}
\vspace*{0.2cm}
\caption{Parameters of the simulations in the anelastic Taylor-Proudman regime.}
\label{parameters_tp_anel}
\end{center}
\end{table}

\begin{figure}[h]
\begin{center}
\includegraphics[width=4.4cm]{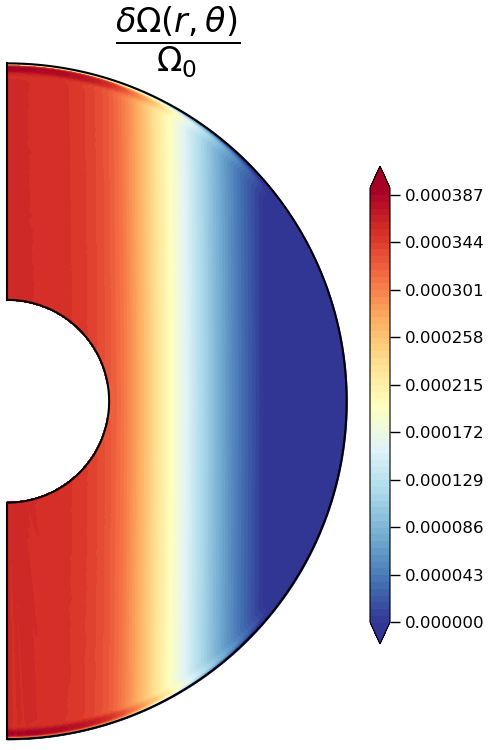}
\includegraphics[width=4.4cm]{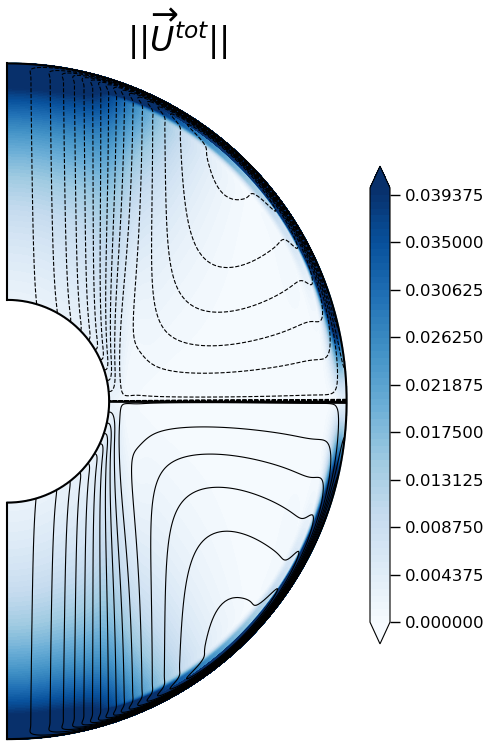}
\caption{Meridional cuts showing the coloured contours of the steady differential rotation normalised to the top value in the left panel, and the norm of the total meridional velocity field $\protect\vv{U}^{\hspace*{0.05cm} \text{tot}} = \left(U_r + V_f \hspace*{0.02cm} (r) \right) \protect\vv{e}_r + U_{\theta} \protect\vv{e}_{\theta}$ with its associated streamlines (in black) in the right panel, in the Taylor-Proudman regime. As previously, outside the tangent cylinder the full lines describe a clockwise circulation and the dashed ones, a counterclockwise circulation whereas inside the tangent cylinder, the full lines represent an upward meridional flow and the dashed lines a downward one. The parameters of the simulation are $E=10^{-4}$, $P_r \left(N_0/\Omega_0\right)^2 = 10^{-4}$, $Re_c = 10^{-2}$ and $\rho_i/\rho_0 = 20.9$ (run $1.1$ of Table $\ref{parameters_viscous_anel}$).}
\label{Rot_Diff_TP_Anel}
\end{center}
\end{figure}

Figure \ref{Rot_Diff_TP_Anel} shows the steady flow obtained for a density contrast $\rho_i/\rho_0 = 20.9$.
It can be compared to the Boussinesq simulation displayed in Fig. \ref{Rot_Diff_TP}, a simulation performed with the same non-dimensional numbers but without density stratification. As we can see, the differential rotation is still cylindrical and its amplitude is similar to the Boussinesq case. This property is confirmed at other density ratios (not shown). The main difference in the AM distribution concerns the local maximum of the rotation rate along the tangent cylinder which is present in the Boussinesq case but absent in the anelastic case. The differences in the meridional circulation are more striking because, in the density stratified case, the circulation is no longer characterised by a strong vertical jet towards the inner equator.

We shall first interpret the fact that the amplitude of the differential rotation appears to be independent of the density contrast between the two spheres. As in the Boussinesq case, we expect the AM conservation to drive a meridional circulation that balances the effect of the imposed contraction field. In Appendix \ref{demonstration_rot_diff_TP_anelastic}, a linear and inviscid solution of such meridional flow is given. As in the Boussinesq case, this circulation does not satisfy the boundary conditions at the outer sphere and an Ekman boundary layer takes charge of it. This induces a  differential rotation across the Ekman
layer which is determined by the relation, Eq. \eqref{ekman_pumping}, between the azimuthal velocity jump and the interior radial velocity field. As the meridional velocity field near the outer sphere should not depend on the density contrast, the jump in rotation rate is not affected either. This differential rotation is then communicated to the rest of the flow through the Taylor-Proudman constraint $\partial U_{\phi} / \partial z = 0$. This reasoning explains why the amplitude of the differential rotation is not affected by the density stratification and why its value is similar to the Boussinesq case.

We further show in Appendix \ref{demonstration_rot_diff_TP_anelastic} that the analytical solution found in the Boussinesq case for the differential rotation is also a solution in the anelastic case. Away from the tangent cylinder, this solution indeed provides a good approximation to the numerical results. However, the lack of a local maximum along the tangent cylinder observed in the anelastic numerical simulations makes it less relevant than in the Boussinesq case.

\begin{figure*}[h]
\begin{center}
\hspace*{-0.5cm}
\includegraphics[width=6cm]{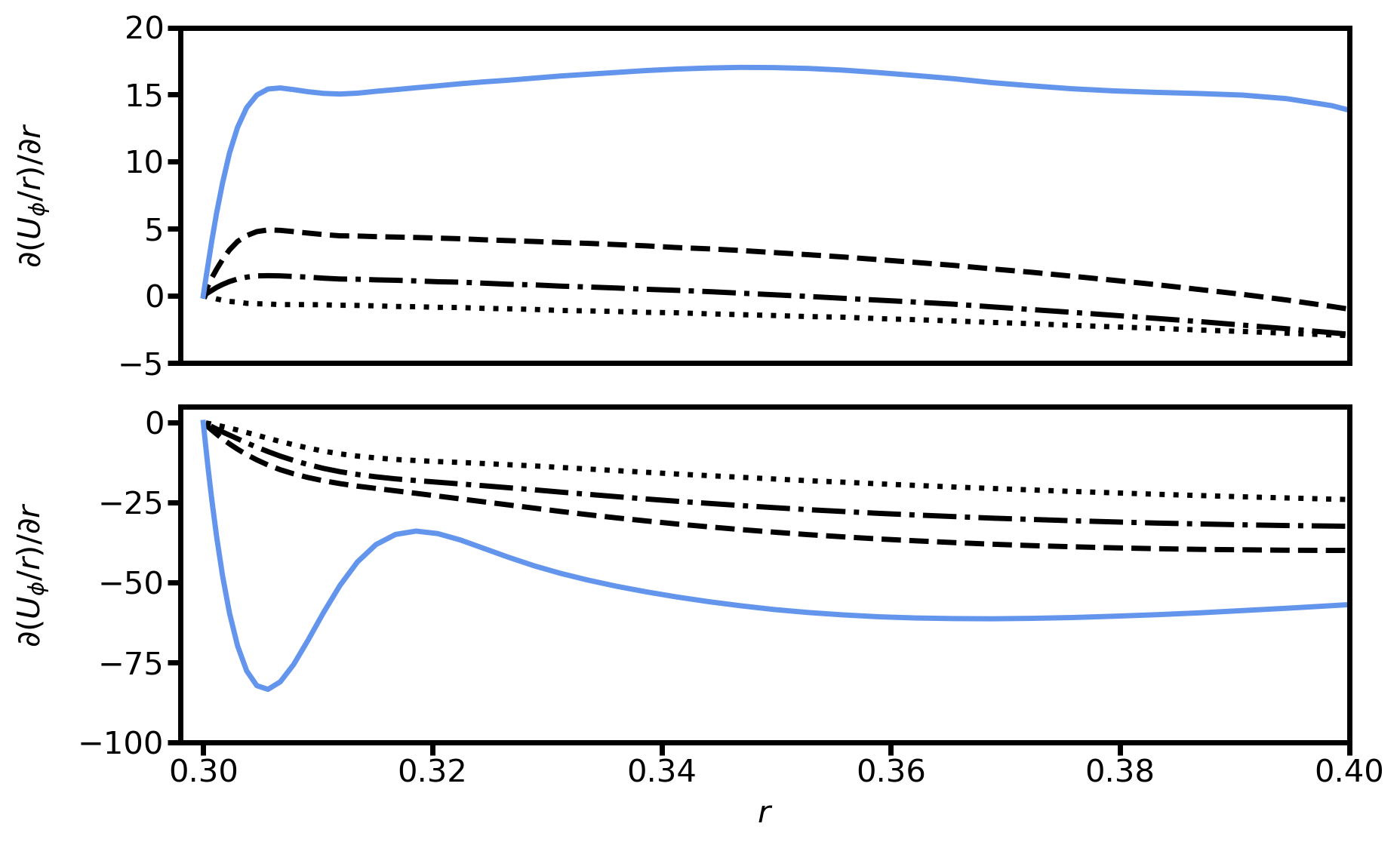}
\includegraphics[width=6cm]{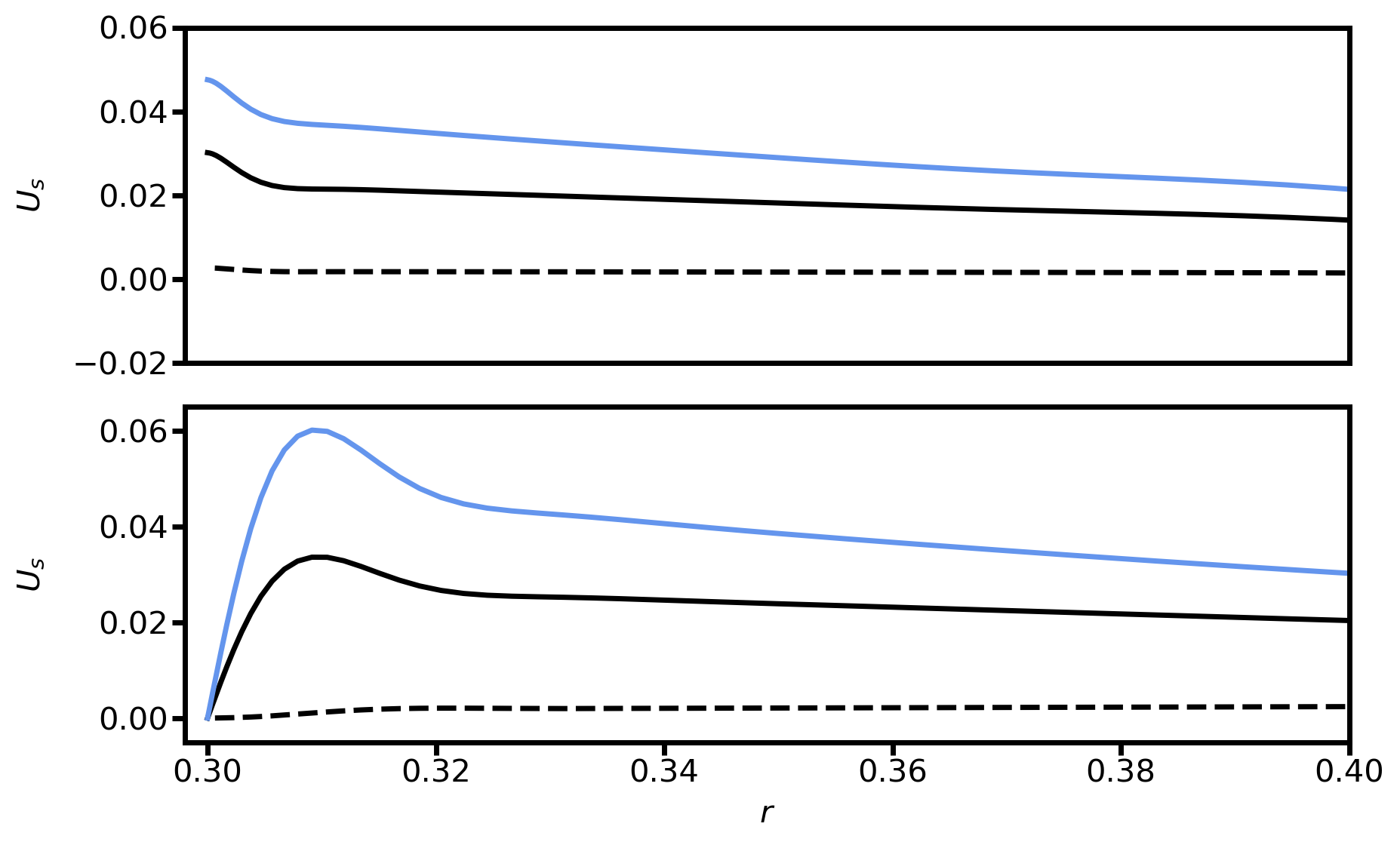}
\includegraphics[width=6cm]{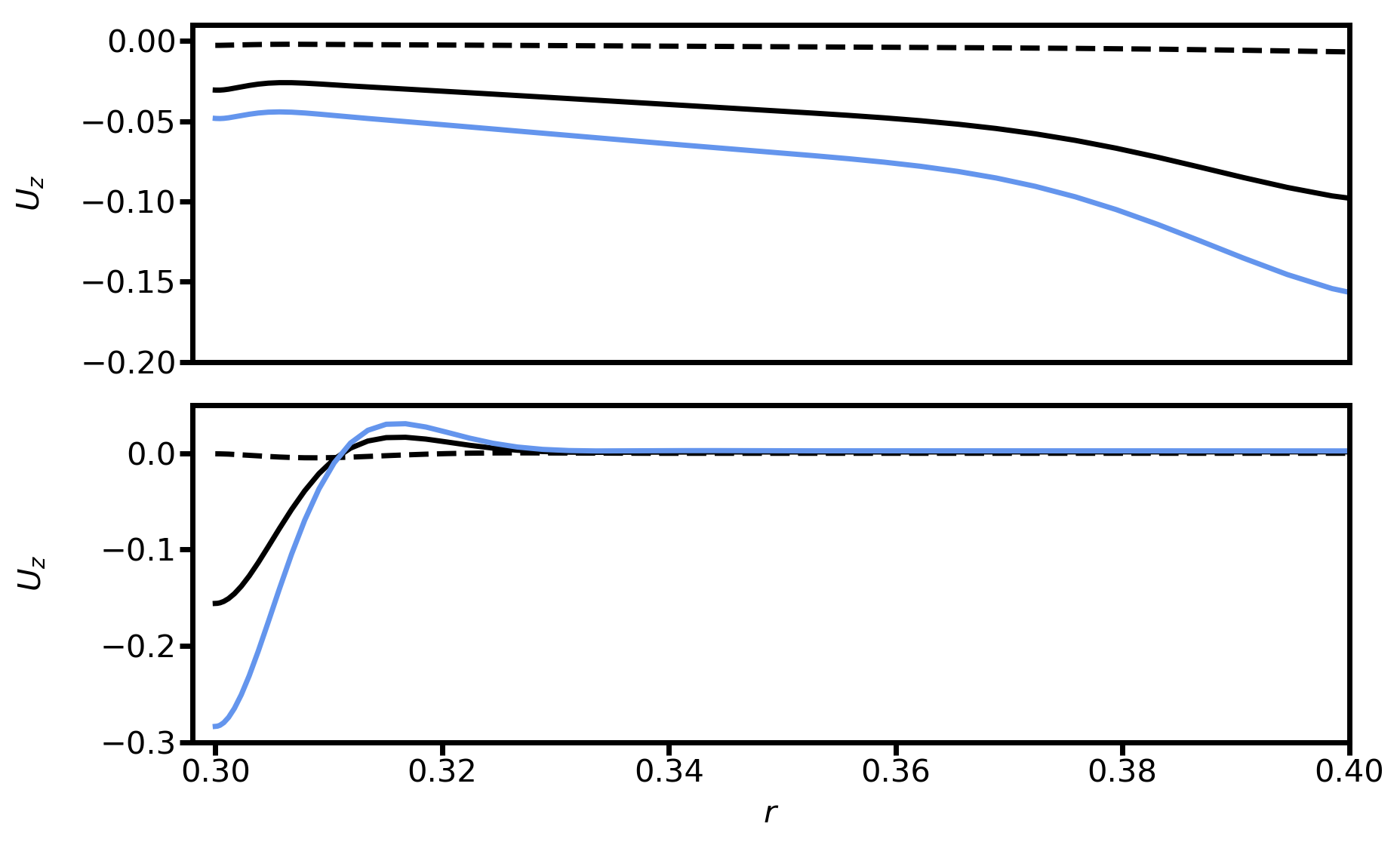}
\caption{Radial partial derivative of the ratio between the azimuthal velocity field $U_{\phi}$ and the radius $r$ (left panel), cylindrical radial velocity field $U_s = U_r \sin{\theta} + U_{\theta} \cos{\theta}$ (middle panel) and vertical velocity field $U_z = U_r \cos{\theta} - U_{\theta} \sin{\theta}$ (right panel), as a function of radius. In each panel, the top row corresponds to the latitude $\theta = \pi/4$. In the left and middle panels, the bottom row corresponds to the equator but, since the vertical velocity field $U_z$ is zero at the equator, the bottom right panel is obtained close to this location, at about $85$-degree angle with the vertical axis. In each case, $E=10^{-5}$, $P_r \left(N_0/\Omega_0\right)^2 = 10^{-4}$ and $Re_c = 10^{-2}$. In all the panels, the plain blue curve corresponds to the Boussinesq case (run $1.4$ of Table \ref{parameters_TP}) while the other curves are obtained by varying the density contrast between the inner and outer spheres. Thus, in the left panel we have $\rho_i/\rho_0 = 10.8$ in dashed lines, $20.9$ in dash-dotted lines and $60.2$ in dotted lines (runs $2.2$ to $2.4$ of Table \ref{parameters_tp_anel}) while in the middle and right panels, the two black curves are respectively obtained for $\rho_i/\rho_0 = 1.75$ (plain curve) and $\rho_i/\rho_0 = 20.9$ (dashed lines) (runs $2.1$ and $2.3$ of Table \ref{parameters_tp_anel}).}
\label{TP_Anel_dUphiRdr}
\end{center}
\end{figure*}

The marked difference between the meridional flows in the Boussinesq and anelastic cases is first due to the dependence of the imposed contraction field on the density stratification, with radial velocities 
decreasing inwards for high enough density ratios. The linear and inviscid solution of the induced meridional circulation (see Appendix \ref{demonstration_rot_diff_TP_anelastic}) has the same behaviour thus explaining the global distribution of the total velocity observed in the right panel of Fig. \ref{Rot_Diff_TP_Anel}. We also find that the equatorial boundary layer is practically suppressed at high density stratification which in turn explains the lack of a strong vertical jet near the equator. The behaviour close to the inner sphere of three different fields, the radial gradient of angular velocity $\partial \Omega(r,\theta) / \partial r $, the cylindrical and vertical velocity fields, is displayed in Fig. \ref{TP_Anel_dUphiRdr}. In the Boussinesq case, we observe that an equatorial boundary layer is present and, just like in the classical spherical Couette flow, it is associated with a strong vertical jet along the tangent cylinder. However, when the density contrast increases, the jumps of the three fields near the inner sphere are practically suppressed together with the vertical jet. 

The absence of the equatorial boundary layer could be explained by the fact that the inviscid interior solution connects in a smoother way to the boundary conditions at the inner sphere. This is indeed the case for the cylindrical radial velocity because the mismatch at the inner equator falls from $ \Delta U_s \approx 11.1$ for the Boussinesq inviscid solution to $\Delta U_s \approx 0.18$ when $\rho_i/\rho_0 = 60.2$. 

\subsection{Anelastic viscous case}
\label{anelastic_viscous_case}

Here, we study in the viscous regime the modification to the differential rotation and the meridional flow induced by the introduction of a varying density across the shell. Simulations are performed for the various density contrasts and contraction Reynolds numbers $Re_c$ listed  in Table \ref{parameters_viscous_anel}.

\begin{table}[h]
\begin{center}
\begin{tabular}{c|c|c|c|c}
\hline
\hspace*{0.01cm} \textbf{$\text{Simulation}$} \hspace*{0.01cm} & \hspace*{0.3cm} $\boldsymbol{E}$ \hspace*{0.3cm} &  $\boldsymbol{P_r \left( N_0 / \Omega_0 \right)^2}$ & \hspace*{0.1cm} $\boldsymbol{Re_c}$ \hspace*{0.1cm} & \hspace*{0.1cm} $\boldsymbol{\rho_i} \mathbf{/} \boldsymbol{\rho_0}$ \hspace*{0.1cm} \\ 
\hline
1.1 & $10^{-4}$ & $10^{4}$ & $10^{-2}$ & $1.75$\\
1.2 & $10^{-4}$ & $10^{4}$ & $10^{-2}$ & $10.8$\\
1.3 & $10^{-4}$ & $10^{4}$ & $10^{-2}$ & $20.9$\\
1.4 & $10^{-4}$ & $10^{4}$ & $10^{-2}$ & $60.2$\\
\hline
2.1 & $10^{-4}$ & $10^{4}$ & $10^{-1}$ & $1.75$\\
2.2 & $10^{-4}$ & $10^{4}$ & $10^{-1}$ & $10.8$\\
2.3 & $10^{-4}$ & $10^{4}$ & $10^{-1}$ & $20.9$\\
2.4 & $10^{-4}$ & $10^{4}$ & $10^{-1}$ & $60.2$\\
\hline
3.1 & $10^{-4}$ & $10^{4}$ & $1$ & $1.75$\\
3.2 & $10^{-4}$ & $10^{4}$ & $1$ & $10.8$\\
3.3 & $10^{-4}$ & $10^{4}$ & $1$ & $20.9$\\
3.4 & $10^{-4}$ & $10^{4}$ & $1$ & $60.2$\\
\hline
4.1 & $10^{-4}$ & $10^{4}$ & $10$ & $1.75$\\
4.2 & $10^{-4}$ & $10^{4}$ & $10$ & $10.8$\\
4.3 & $10^{-4}$ & $10^{4}$ & $10$ & $20.9$\\
\hline
5.1 & $10^{-3}$ & $10^{4}$ & $10^{-2}$ & $20.9$\\
5.2 & $10^{-5}$ & $10^{4}$ & $10^{-2}$ & $20.9$\\
\hline
6.1 & $10^{-3}$ & $10^{3}$ & $10^{-2}$ & $20.9$\\
6.2 & $10^{-4}$ & $10^{3}$ & $10^{-2}$ & $20.9$\\
6.3 & $10^{-5}$ & $10^{3}$ & $10^{-2}$ & $20.9$\\
\hline
\end{tabular}
\vspace*{0.2cm}
\caption{Parameters of the simulations in the anelastic viscous regime.}
\label{parameters_viscous_anel}
\end{center}
\end{table}

\begin{figure*}[h]
\begin{center}
\includegraphics[width=4.4cm]{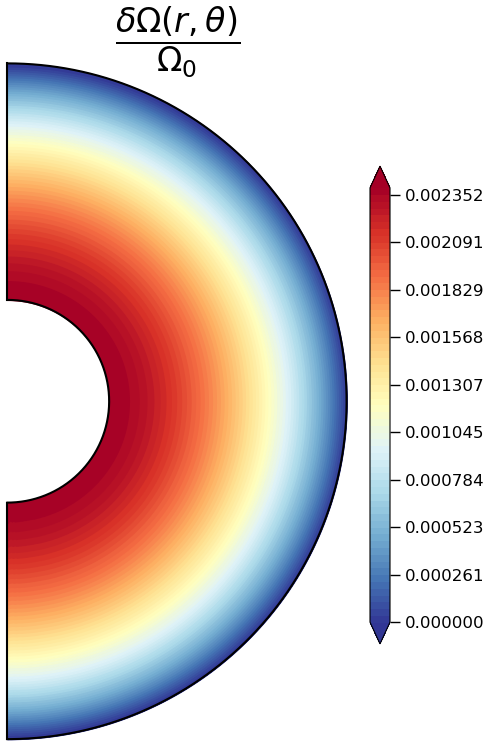}
\includegraphics[width=4.4cm]{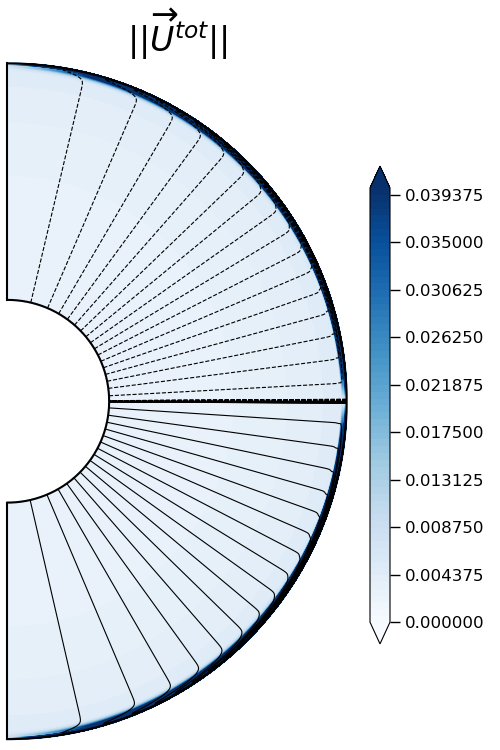}
\includegraphics[width=4.4cm]{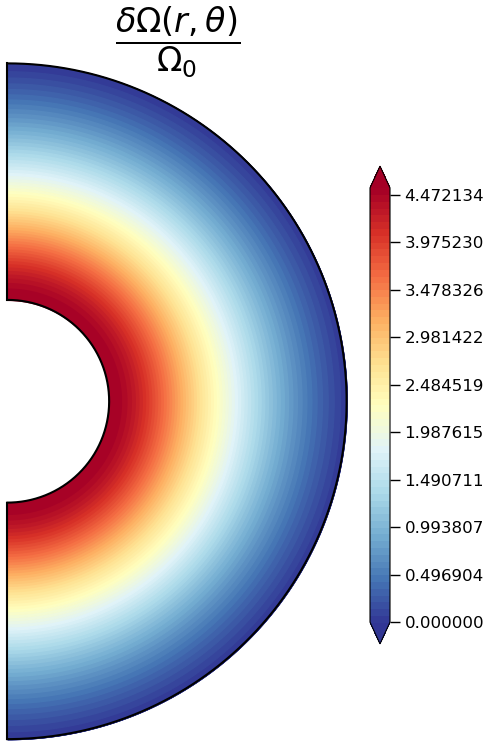}
\includegraphics[width=4.4cm]{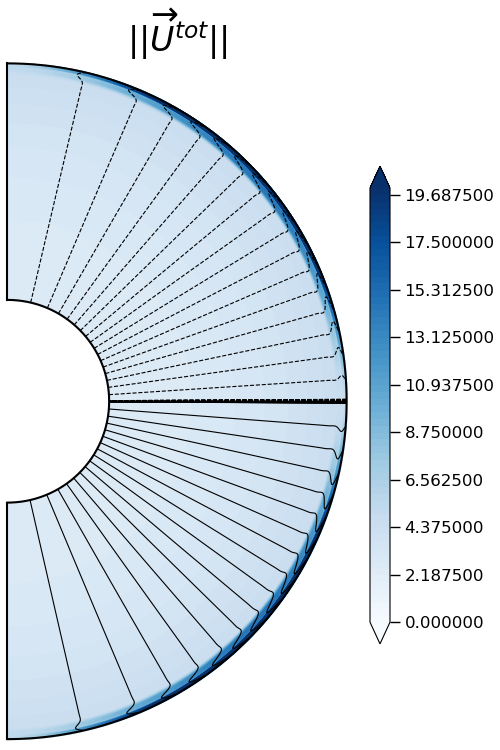}
\caption{Stationary differential rotation $\delta \Omega(r,\theta) / \Omega_0$ (first and third panels) and norm of the total meridional velocity field $\protect\vv{U}^{\hspace*{0.05cm} \text{tot}} = \left(U_r + V_f \hspace*{0.02cm} (r) \right) \protect\vv{e}_r + U_{\theta} \protect\vv{e}_{\theta}$ (second and fourth panels) with its associated streamlines in black, in the viscous anelastic regime. The parameters of the simulation shown in the two left panels are $E=10^{-5}$, $P_r \left(N_0/\Omega_0\right)^2 = 10^{\hspace*{0.02cm}4}$, $Re_c = 10^{-2}$ and $\rho_i/\rho_0 = 20.9$ (run $5.2$ of Table $\ref{parameters_viscous_anel}$), and in the two right panels $E=10^{-4}$, $P_r \left(N_0/\Omega_0\right)^2 = 10^{\hspace*{0.02cm}4}$, $Re_c = 10$ and $\rho_i/\rho_0 = 20.9$ (run $4.3$ of Table $\ref{parameters_viscous_anel}$). In the first and third panels, the coloured contours are normalised to the top value $\Omega_0$.}
\label{Rot_Diff_Viscous_Anel}
\end{center}
\end{figure*}

Figure \ref{Rot_Diff_Viscous_Anel} shows the coloured contours of the rotation rate $\Omega$ and the streamlines of the meridional circulation in the steady state for two anelastic simulations with  $\rho_i/\rho_0 = 20.9$. The other parameters are the same as the ones used in the Boussinesq cases represented in Fig. \ref{Rot_Diff_Viscous}. Similar to the Boussinesq case, the differential rotation profile is still radial and the amplitude of differential rotation still increases with $Re_c$ (first and third panels of Fig. \ref{Rot_Diff_Viscous_Anel}). However, the amplitude of the differential rotation is decreased compared to the Boussinesq case. Indeed, for $\rho_i/\rho_0 = 20.9$, the maximum value of the rotation contrast between the inner and the outer spheres is nearly four times smaller than in the Boussinesq case when $Re_c=10^{-2}$, and almost twice as small as when $Re_c=10$. The total meridional circulation, visible in the second and fourth panel of Fig. \ref{Rot_Diff_Viscous}, is dominated by the radial contraction field. As in the Boussinesq case, a circulation confined in a thin layer close to the outer boundary can be seen when the forced radial flow is subtracted. However, the circulation at the inner sphere boundary is virtually suppressed by the density stratification.

We now derive an analytical expression of the differential rotation in the linear regime $ \Delta \Omega / \Omega_0 \ll 1 $ that clearly shows how the density stratification lowers the level of differential rotation as compared to the Boussinesq case.

In the anelastic approximation, the AM equation reads:

\begin{equation}
\begin{array}{lll}
\displaystyle \frac{\partial U_{\phi}}{\partial t} + \text{NL} + \underbrace{2 \Omega_0 U_s}_{\text{Coriolis term}} -\underbrace{\nu \hspace*{0.05cm} \left \lbrack D^2 U_{\phi} + \displaystyle \frac{r}{\overline{\rho}} \displaystyle \frac{\text{d} \overline{\rho}}{\text{d} r} \displaystyle \frac{\partial}{\partial r} \left( \displaystyle \frac{U_{\phi}}{r} \right) \right \rbrack}_{\text{Viscous term}} ~ = \\\\ \underbrace{\displaystyle \frac{V_0 r{_0^2} \hspace*{0.02cm} \rho_0}{\overline{\rho} r^3} \left( \displaystyle \frac{\partial}{\partial r} \left( r U_{\phi} + r^2 \sin{\theta} ~ \Omega_0 \right) \right)}_{\text{Contraction}}
\end{array}
\label{angular_momentum_evolution_anel}
\end{equation}

\noindent where $D^2$ is the vector Laplace operator in the azimuthal direction $D^2 = \left(\vv{\nabla}^2 - 1/r^2 \sin^2 \theta\right)$. When the Coriolis and the non-linear advection terms are neglected in Eq. \eqref{angular_momentum_evolution_anel}, and that the differential rotation is sufficiently weak (i.e. $ \Delta \Omega / \Omega_0 \ll 1 $ and thus the first part of the contraction term can be neglected), the steady balance is the linear equation:

\begin{equation}
- \nu \hspace*{0.05cm} \left \lbrack D^2 U_{\phi} + \displaystyle \frac{r}{\overline{\rho}} \displaystyle \frac{\text{d} \overline{\rho}}{\text{d} r} \displaystyle \frac{\partial}{\partial r} \left( \displaystyle \frac{U_{\phi}}{r} \right) \right \rbrack = 2 \Omega_0 \sin{\theta} \displaystyle \frac{V_0 r{_0^2} \hspace*{0.02cm} \rho_0}{\overline{\rho} r^2} 
\label{balance_contraction_viscous}
\end{equation}

\noindent whose the solution is:

\begin{equation}
\displaystyle \frac{\delta \Omega_{\nu}(r)}{\Omega_0} = Re_c \displaystyle \int_{1}^{r/r_0} \displaystyle \frac{\rho_0}{\overline{\rho}(r)} \left \lbrack \displaystyle \frac{\left(r_i/r_0\right)^2 - \left(r/r_0\right)^2}{\left(r/r_0\right)^4} \right \rbrack \hspace*{0.05cm} \text{d} \left(r/r_0\right)
\label{integral_solution}
\end{equation}

It simply corresponds to the Boussinesq linear viscous solution Eq. \eqref{simple_sol} weighted by the inverse of the background density profile. 

\begin{figure*}[h]
\begin{center}
\includegraphics[width=6.02cm]{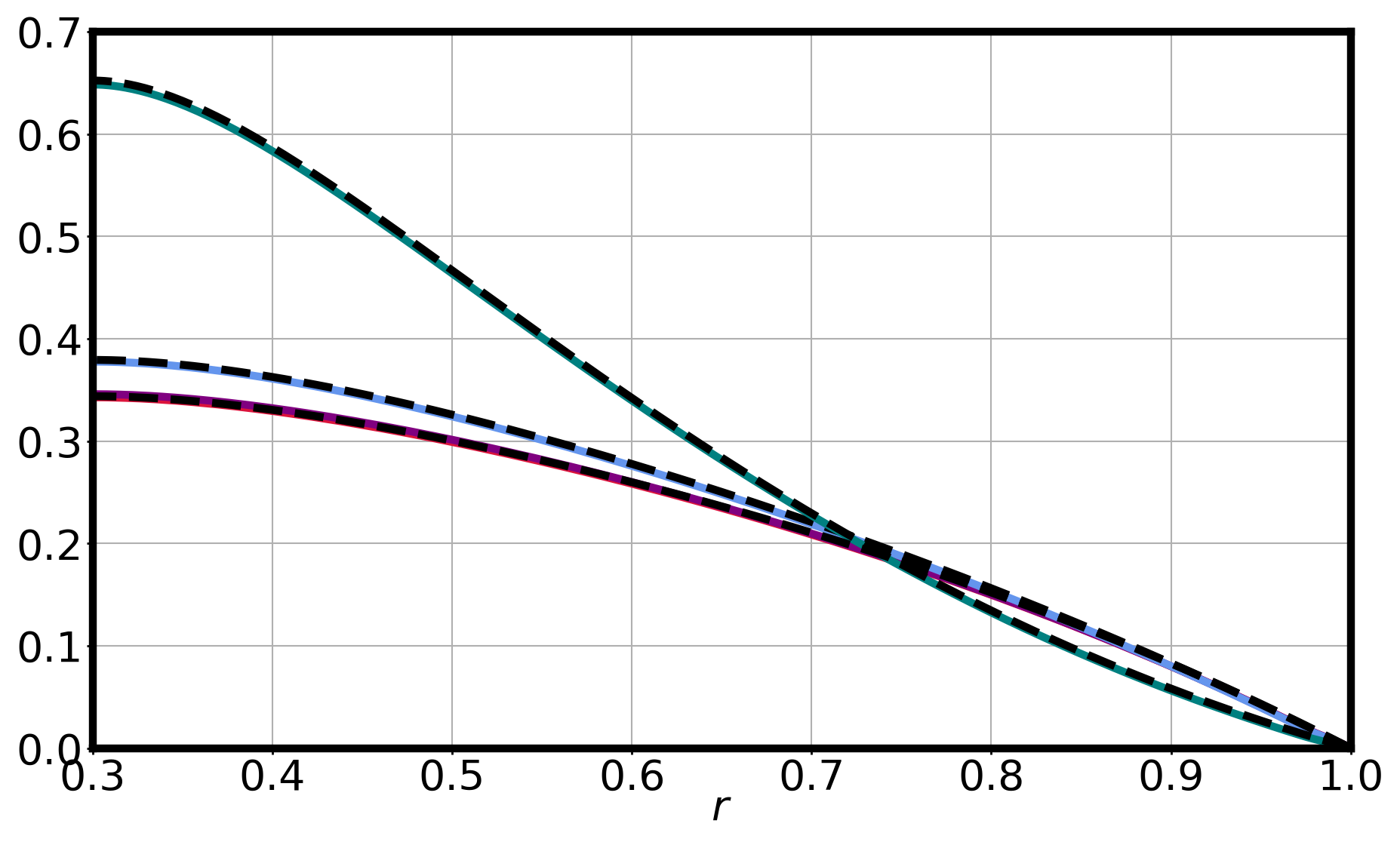}
\includegraphics[width=6.1cm]{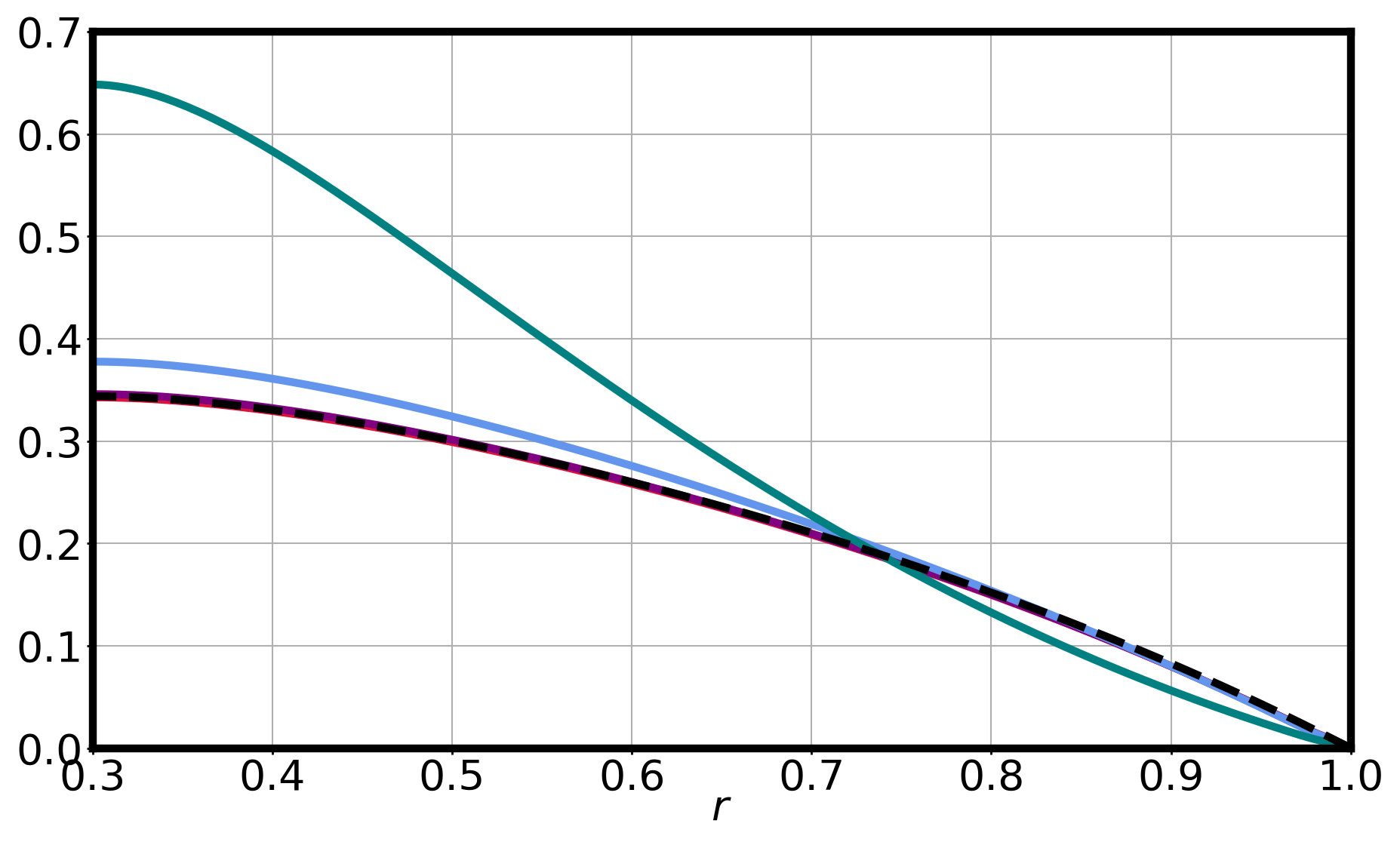}
\includegraphics[width=6.1cm]{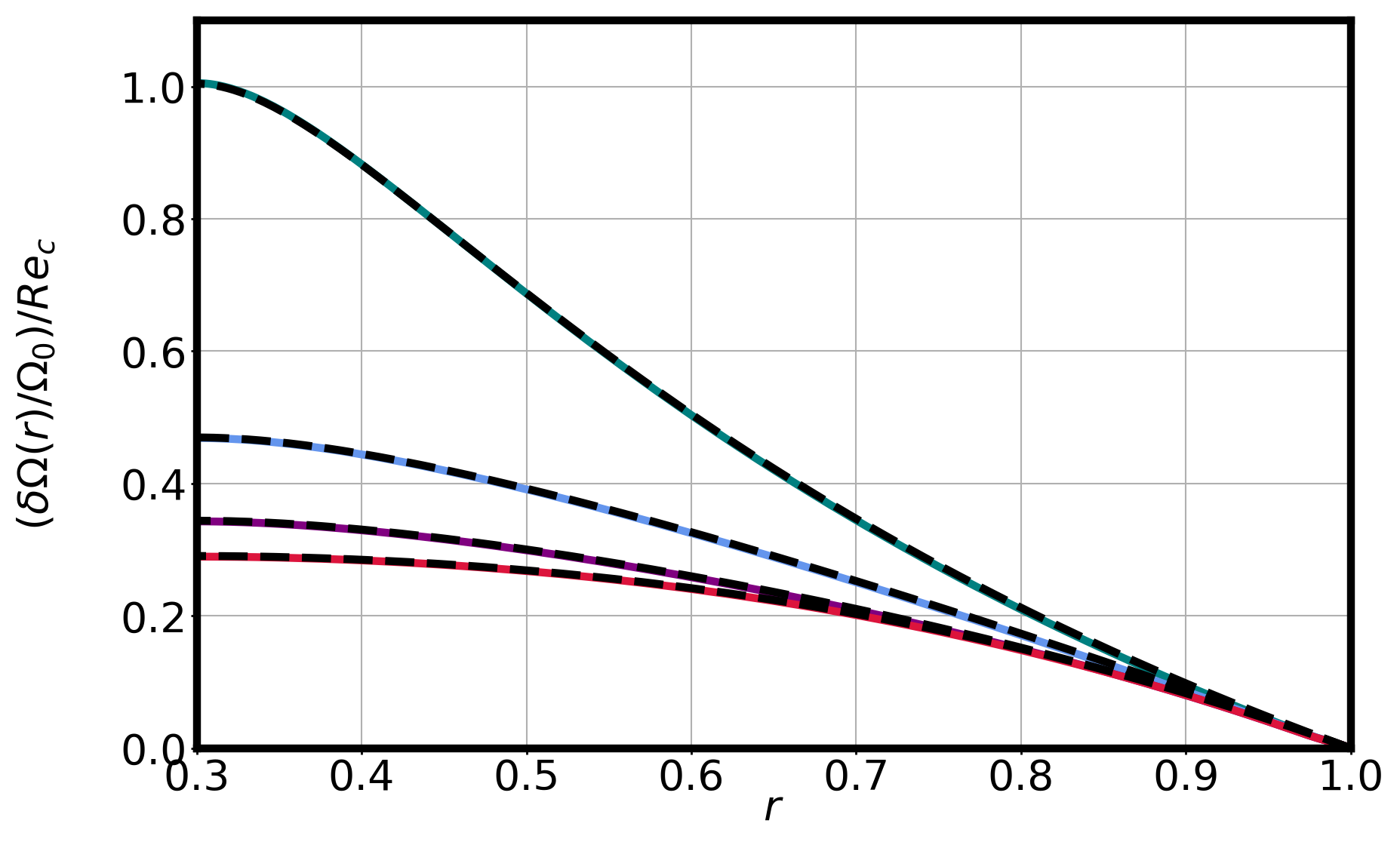}
\caption{Differential rotation normalised to the rotation rate of the outer sphere $\delta \Omega/\Omega_0$, rescaled with $Re_c$ and plotted at the latitude $\theta = \pi/8$ as a function of radius. In each case, $E = 10^{-4}$ and $P_r \left(N_0/\Omega_0\right)^2 = 10^{\hspace*{0.02cm} 4}$. Left panel: the analytical solution Eq. \eqref{integral_solution} is represented in dashed lines and numerical solutions (runs $1.1$ to $1.4$ of Table \ref{parameters_viscous_anel}) in plain coloured lines for the linear regime $Re_c = 10^{-2}$. The colours green, blue, purple and red correspond respectively to the values of the density contrasts between the inner and outer spheres $\rho_i / \rho_0 = 1.75$, $10.8$, $20.9$ and $60.2$. Middle Panel: numerical solutions are now represented for $Re_c = 10^{-2}$, $10^{-1}$, $1$ and $10$ (green, blue, purple and red respectively) for $\rho_i/\rho_0=20.9$ (runs $1.3$, $2.3$, $3.3$ and $4.3$ of Table \ref{parameters_viscous_anel}). Right Panel: same but when $Re_c \geq 1$, the solution in dashed lines is numerically estimated by taking the full contraction term in Eq. \eqref{balance_contraction_viscous} into account.}
\label{Analytical_versus_Numerical}
\end{center}
\end{figure*}

In Fig. \ref{Analytical_versus_Numerical}, we show the agreement between the analytical expression \eqref{integral_solution} and the numerical simulations. In all panels of this figure, the profiles of the differential rotation found in the numerical calculations for various values of the parameters are plotted at a particular latitude (here, $\theta = \pi/8$) as a function of radius. We also overplot in dashed lines the analytical profile found by solving Eq. \eqref{balance_contraction_viscous}. In the left panel, the linear regime (with $Re_c=10^{-2}$) is represented at different density contrasts. In the other panels, the density contrast between the inner and outer spheres is fixed to $\rho_i/\rho_0=20.9$ but $Re_c$ is increased. It is quite clear from the left panel that the agreement between the numerical and analytical solutions is perfect for the low Reynolds number cases. When the Reynolds number is increased, a departure from the analytical solution appears, as observed in the middle panel of Fig. \ref{Analytical_versus_Numerical}, where the profile of Eq. \eqref{integral_solution}, scaled with $Re_c$, is shown in dashed lines. Indeed in this case, the level of differential rotation becomes comparable to $\Omega_0$ and the first part of the contraction term $\left( V_0 r_0^2 \rho_0 / \overline{\rho} r^3\right) \partial ( r U_{\phi}) / \partial r$ in Eq. \eqref{angular_momentum_evolution_anel} is no longer negligible, leading to a correction to Eq. \eqref{integral_solution}. When this correction is applied, the agreement between the numerical solutions at higher $Re_c$ and the new analytical expression (not given here) is recovered, as shown in the last panel of Fig. \ref{Analytical_versus_Numerical}.

To summarise, we find that in the anelastic viscous regime, the presence of a density stratification leads to a weaker differential rotation between the inner and outer spheres. In the linear regime $\Delta \Omega/\Omega_0 \ll 1 $, the differential rotation is given by Eq. \eqref{integral_solution} i.e. the analytical solution previously derived in the Boussinesq case but weighted by the inverse of the background density.

\subsection{Anelastic Eddington-Sweet regime}
\label{anelastic_edd_regime}

\begin{table}[h]
\begin{center}
\begin{tabular}{c|c|c|c|c}
\hline
\hspace*{0.01cm} \textbf{$\text{Simulation}$} \hspace*{0.01cm} & \hspace*{0.3cm} $\boldsymbol{E}$ \hspace*{0.3cm} &  $\boldsymbol{P_r \left( N_0 / \Omega_0 \right)^2}$ & \hspace*{0.1cm} $\boldsymbol{Re_c}$ \hspace*{0.1cm} & \hspace*{0.1cm} $\boldsymbol{\rho_i} \mathbf{/} \boldsymbol{\rho_0}$ \hspace*{0.1cm} \\ 
\hline
1.1 & $10^{-5}$ & $10^{-1}$ & $10^{-2}$ & $1.75$\\
1.2 & $10^{-5}$ & $10^{-1}$ & $10^{-2}$ & $10.8$\\
1.3 & $10^{-5}$ & $10^{-1}$ & $10^{-2}$ & $20.9$\\
1.4 & $10^{-5}$ & $10^{-1}$ & $10^{-2}$ & $60.2$\\
\hline
2.1 & $10^{-5}$ & $10^{-1}$ & $10^{-1}$ & $1.75$\\
2.2 & $10^{-5}$ & $10^{-1}$ & $10^{-1}$ & $10.8$\\
2.3 & $10^{-5}$ & $10^{-1}$ & $10^{-1}$ & $20.9$\\
2.4 & $10^{-5}$ & $10^{-1}$ & $10^{-1}$ & $60.2$\\
\hline
3.1 & $10^{-5}$ & $10^{-1}$ & $1$ & $1.75$\\
3.2 & $10^{-5}$ & $10^{-1}$ & $1$ & $10.8$\\
3.3 & $10^{-5}$ & $10^{-1}$ & $1$ & $20.9$\\
3.4 & $10^{-5}$ & $10^{-1}$ & $1$ & $60.2$\\
\hline
4.1 & $10^{-5}$ & $10^{-1}$ & $10$ & $1.75$\\
4.2 & $10^{-5}$ & $10^{-1}$ & $10$ & $10.8$\\
4.3 & $10^{-5}$ & $10^{-1}$ & $10$ & $20.9$\\
\hline
5.1 & $10^{-5}$ & $10^{-2}$ & $10^{-2}$ & $1.75$\\
5.2 & $10^{-5}$ & $10^{-2}$ & $10^{-2}$ & $10.8$\\
5.3 & $10^{-5}$ & $10^{-2}$ & $10^{-2}$ & $20.9$\\
5.4 & $10^{-5}$ & $10^{-2}$ & $10^{-2}$ & $60.2$\\
\hline
6.1 & $10^{-5}$ & $10^{-2}$ & $10^{-1}$ & $1.75$\\
6.2 & $10^{-5}$ & $10^{-2}$ & $10^{-1}$ & $10.8$\\
6.3 & $10^{-5}$ & $10^{-2}$ & $10^{-1}$ & $20.9$\\
6.4 & $10^{-5}$ & $10^{-2}$ & $10^{-1}$ & $60.2$\\
\hline
7.1 & $10^{-5}$ & $10^{-2}$ & $1$  & $1.75$\\
7.2 & $10^{-5}$ & $10^{-2}$ & $1$  & $10.8$\\
7.3 & $10^{-5}$ & $10^{-2}$ & $1$ & $20.9$\\
7.4 & $10^{-5}$ & $10^{-2}$ & $1$ & $60.2$\\
\hline
8.1 & $10^{-5}$ & $10^{-2}$ & $10$ & $1.75$\\
8.2 & $10^{-5}$ & $10^{-2}$ & $10$ & $10.8$\\
8.3 & $10^{-5}$ & $10^{-2}$ & $10$ & $20.9$\\
8.4 & $10^{-5}$ & $10^{-2}$ & $10$ & $60.2$\\
\hline
9.1 & $10^{-6}$ & $10^{-2}$ & $10^{-2}$ & $20.9$\\
9.2 & $10^{-6}$ & $10^{-2}$ & $10^{-1}$ & $20.9$\\
9.3 & $10^{-6}$ & $10^{-2}$ & $1$ & $20.9$\\
9.4 & $10^{-6}$ & $10^{-2}$ & $10$ & $20.9$\\
\hline
10.1 & $10^{-3}$ & $10^{-1}$ & $10^{-2}$ & $20.9$\\
10.2 & $10^{-4}$ & $10^{-1}$ & $10^{-2}$ & $20.9$\\
10.3 & $10^{-6}$ & $10^{-1}$ & $10^{-2}$ & $20.9$\\
\hline
\end{tabular}
\vspace*{0.2cm}
\caption{Parameters of the simulations in the Eddington-Sweet regime under the anelastic approximation.}
\label{parameters_edd_anel}
\end{center}
\end{table}

The role of the density stratification is now studied in the regime $\sqrt{E} \ll P_r \left(  N_0/\Omega_0 \right)^2 \ll 1$ where the transport of AM is dominated by an Eddington-Sweet type meridional circulation. The density ratios $\rho_i /\rho_0$ considered together with the other parameters of the simulations, are listed in Table \ref{parameters_edd_anel}. The parameters $E$, $P_r \left(N_0/\Omega_0\right)^2$ and $Re_c$ being defined in the same way and covering the same range as in the Boussinesq case (see Table \ref{parameters_edd}), we shall be able to simply compare the anelastic simulations to the Boussinesq ones.

\begin{figure}[h]
\begin{center}
\includegraphics[width=4.4cm]{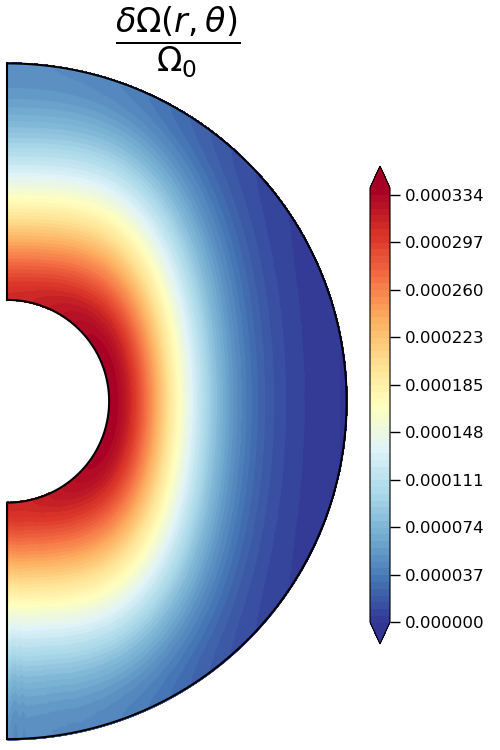}
\includegraphics[width=4.4cm]{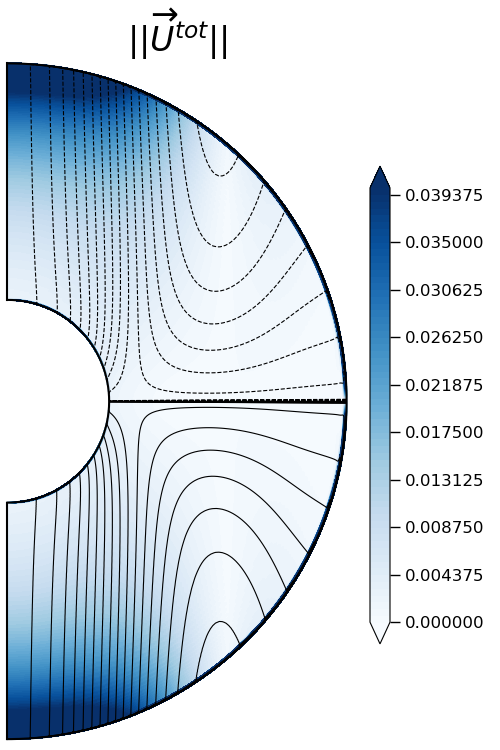}
\caption{Steady differential rotation (left panel) and norm of the total meridional velocity field $\protect\vv{U}^{\hspace*{0.05cm} \text{tot}} = \left(U_r + V_f \hspace*{0.02cm} (r) \right) \protect\vv{e}_r + U_{\theta} \protect\vv{e}_{\theta}$ with the associated streamlines in black (right panel), in the Eddington-Sweet regime obtained for $E=10^{-6}$, $P_r \left(N_0/\Omega_0\right)^2 = 10^{-1}$, $Re_c = 10^{-2}$ and $\rho_i/\rho_0 = 20.9$ (run $10.3$ of Table \ref{parameters_edd_anel}). In the left panel, the coloured contours are normalised to the rotation rate of the outer sphere $\Omega_0$. In the right panel, the streamlines are as follows: outside the tangent cylinder the dashed lines describe a counterclockwise circulation and the full ones a clockwise circulation whereas inside the tangent cylinder the dashed lines correspond to a downward meridional flow and the full lines to an inward one.}
\label{Rot_Diff_Edd_Anel}
\end{center}
\end{figure}

The meridional cut displayed in the left panel of Fig. \ref{Rot_Diff_Edd_Anel} shows that, as in the Boussinesq case, the differential rotation profile is neither radial nor cylindrical. But comparing Fig. \ref{Rot_Diff_Edd_Anel} to its Boussinesq counterpart Fig. \ref{Rot_Diff_Edd_Sweet} reveals clear differences between the two simulations that only differ by their density stratification. Firstly, the amplitude of the rotation contrast taken between the inner and outer spheres is smaller in the anelastic case. Secondly, the distribution of the differential rotation within the spherical shell is globally smoother, predominantly radial, and in particular the region of high rotation rates localised at the inner sphere close to the equator that was clearly visible in the Boussinesq case is now absent. Quantitatively, along the inner sphere, the rotation rates of the equator and the pole only differ by $11\%$ while this contrast was $67\%$ in the Boussinesq case. From the right panel of Fig. \ref{Rot_Diff_Edd_Anel}, it is obvious that the meridional flow in the anelastic case is also smoother and not focused towards the equator at the inner sphere. That comes as no surprise since the meridional flow is similar in the Eddington-Sweet and Taylor-Proudman regimes and we have already found in the anelastic Taylor-Proudman case that the circulation is weighted by the inverse of the background density profile and is no longer affected by an equatorial boundary layer.

We have computed the global differential rotation between the inner and the outer spheres for different density ratios and we propose a scaling relation to describe how it decreases with higher density ratios.  
As in Sect. \ref{eddington_sweet_regime} we estimate the dominant terms in the governing equations, but paying attention to the effects of the density stratification. In particular, in determining the contraction timescale, we take into account that the contraction velocity field is weighted by the inverse of the density. This allows us a more accurate determination of the contraction timescale:

\begin{equation}
\tau_{\text{c}}^{\hspace*{0.05cm}  \text{A}} = \displaystyle \int_{r_0}^{r_i} \displaystyle \frac{\text{d} r}{V_f \hspace*{0.02cm}(r)}  \approx  \left( \displaystyle \int_{r_i/r_0}^{1} \displaystyle \frac{\overline{\rho}}{\rho_0} ~ \text{d} \tilde{r} \right) \tau_{\text{c}}
\end{equation}

\noindent where the superscript $\text{A}$ for "Anelastic" differentiates this contraction timescale from $ \tau_{\text{c}} = r_0/V_0$ used in the Boussinesq case. The second expression is obtained using the definition of $V_f \hspace*{0.02cm} (r)$ and assuming that $r \sim r_0$.

\begin{figure}[h]
\begin{center}
\hspace*{-0.5cm}
\includegraphics[width=9.0cm]{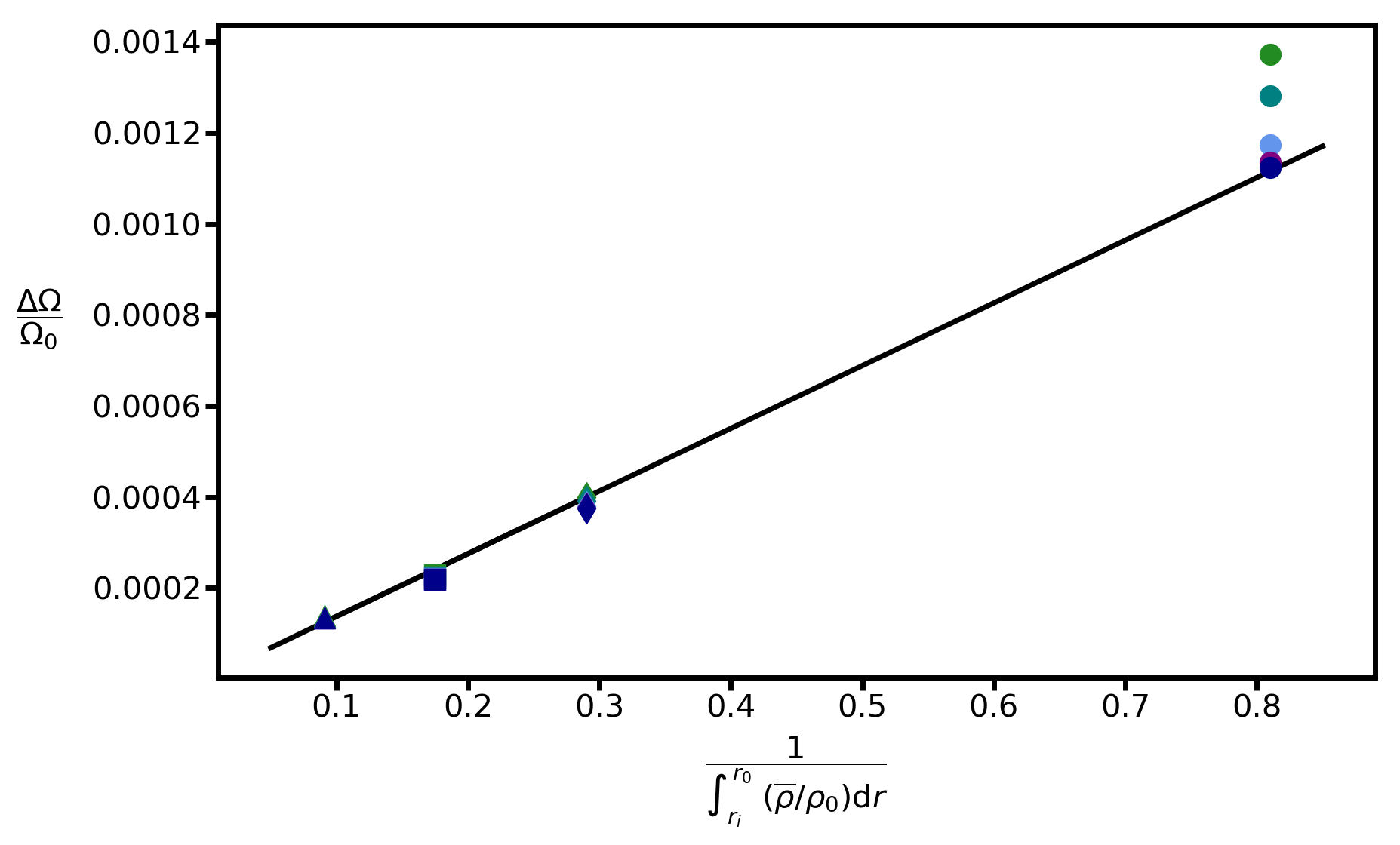}
\caption{Differential rotation between the inner and outer spheres $\Delta \Omega = \Omega(r_i) - \Omega(r_0)$ normalised to the top value $\Omega_0$ obtained after subtracting the numerical Taylor-Proudman contribution, plotted as a function of the inverse of the integral of the background density profile $\overline{\rho}(r)$ normalised to the outer sphere value $\rho_0$ for different latitudes namely, $\theta = \pi/2$, $\pi/4$, $\pi/8$, $\pi/16$ and $\pi/32$ respectively in green, cyan, light blue, purple and blue. The parameters are $E=10^{-5}$, $P_r \left(N_0/\Omega_0\right)^2 = 10^{-1}$ and $Re_c = 10^{-2}$. The symbols circle, diamond, square and triangle correspond to the different density contrasts between the inner and outer spheres $\rho_i/\rho_0 = 1.75$, $10.8$, $20.9$ and $60.2$ (runs $1.1$ to $1.4$ of Table \ref{parameters_edd_anel}).}
\label{Edd_Scaling_Anel}
\end{center}
\end{figure}

Another specificity of the anelastic equations is the presence of an additional term in the thermal balance:

\begin{equation}
U_r \displaystyle \frac{\text{d} \overline{S}}{\text{d} r} = \kappa_{\text{T}} \left \lbrack \displaystyle \left( \displaystyle \frac{\text{d} \ln{\overline{\rho}}}{\text{d} r} + \displaystyle \frac{\text{d} \ln{\overline{T}}}{\text{d} r} \right) \displaystyle \frac{\text{d} \delta S}{\text{d} r} + \vv{\nabla}^2 \delta S \right \rbrack
\end{equation}

\noindent namely the first term in the brackets. But in our simulations, we found that this term is of the same order as the second term, so that we can use only this second term to derive a scaling relation. This scaling relation takes the following form:

\begin{equation}
\displaystyle \frac{\Delta \Omega}{\Omega_0} \approx \displaystyle \frac{\tau_{\text{ED}}}{\tau_{\text{c}}^{\hspace*{0.05cm} \text{A}}} = P_r \left( \displaystyle \frac{N_0}{\Omega_0} \right)^2 \displaystyle \frac{R_o}{E} \left( \displaystyle \int_{r_i/r_0}^{1} \left( \displaystyle \frac{\overline{\rho}}{\rho_0} \right) \text{d} \tilde{r} \right)^{-1}
\label{estimate_anel_edd_rotation_rate}
\end{equation}

\noindent and provides a reasonable approximation for the differential rotation between the inner and outer spheres obtained in the anelastic Eddington-Sweet regime. However, as for the Boussinesq case, we found that the agreement is better when the contribution of the outer Ekman layer is taken away by subtracting the numerical Taylor-Proudman solution. This is done in Fig. \ref{Edd_Scaling_Anel} which shows that the global differential rotation does scale approximately with the inverse of the integral of the density between the two spheres. It also shows that the rotation contrast is practically independent of the latitude except for the lowest density ratio $\rho_i/\rho_0=1.75$.

\begin{figure*}[h]
\begin{center}
\hspace*{-0.5cm}
\includegraphics[width=6cm]{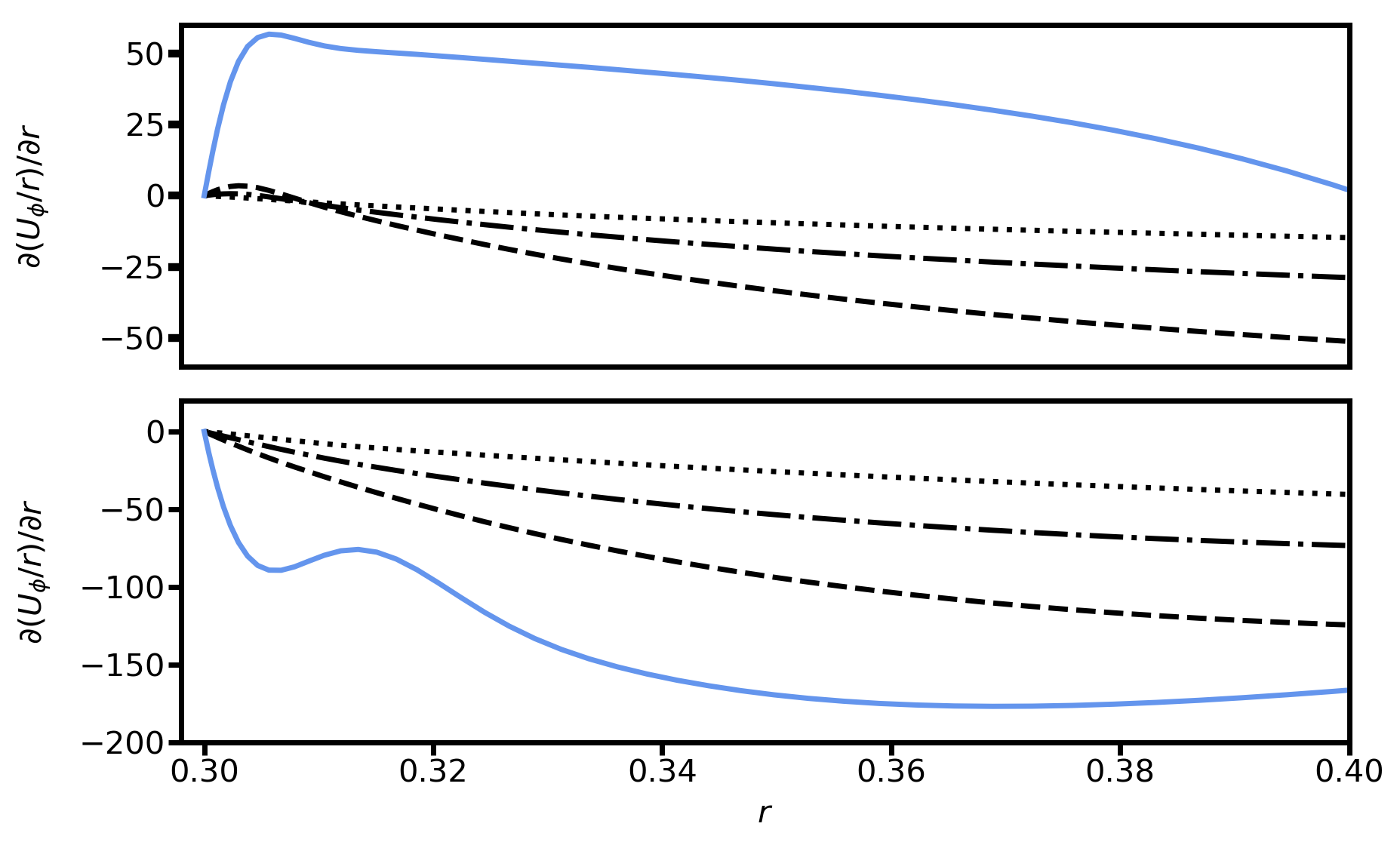}
\includegraphics[width=6cm]{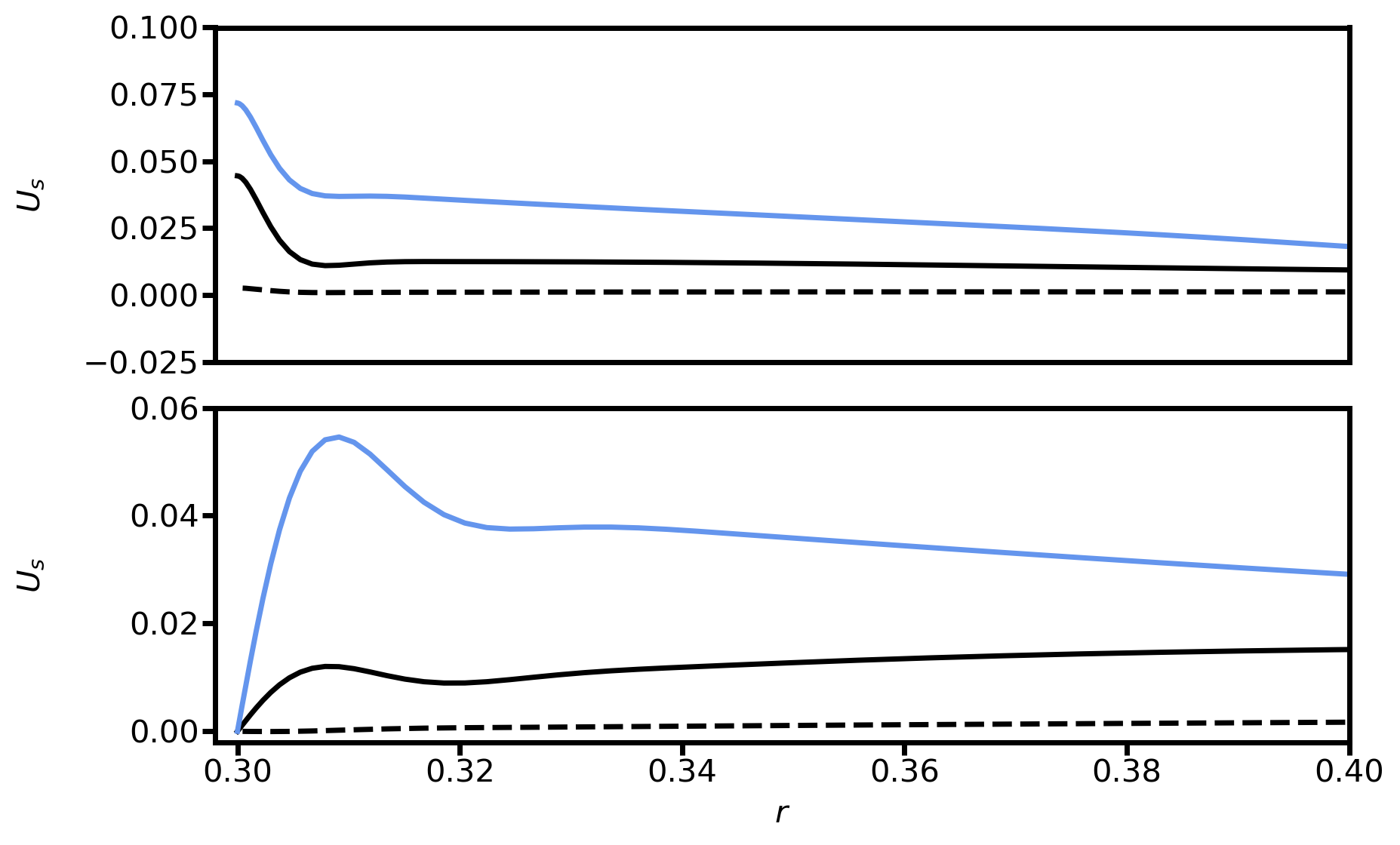}
\includegraphics[width=6cm]{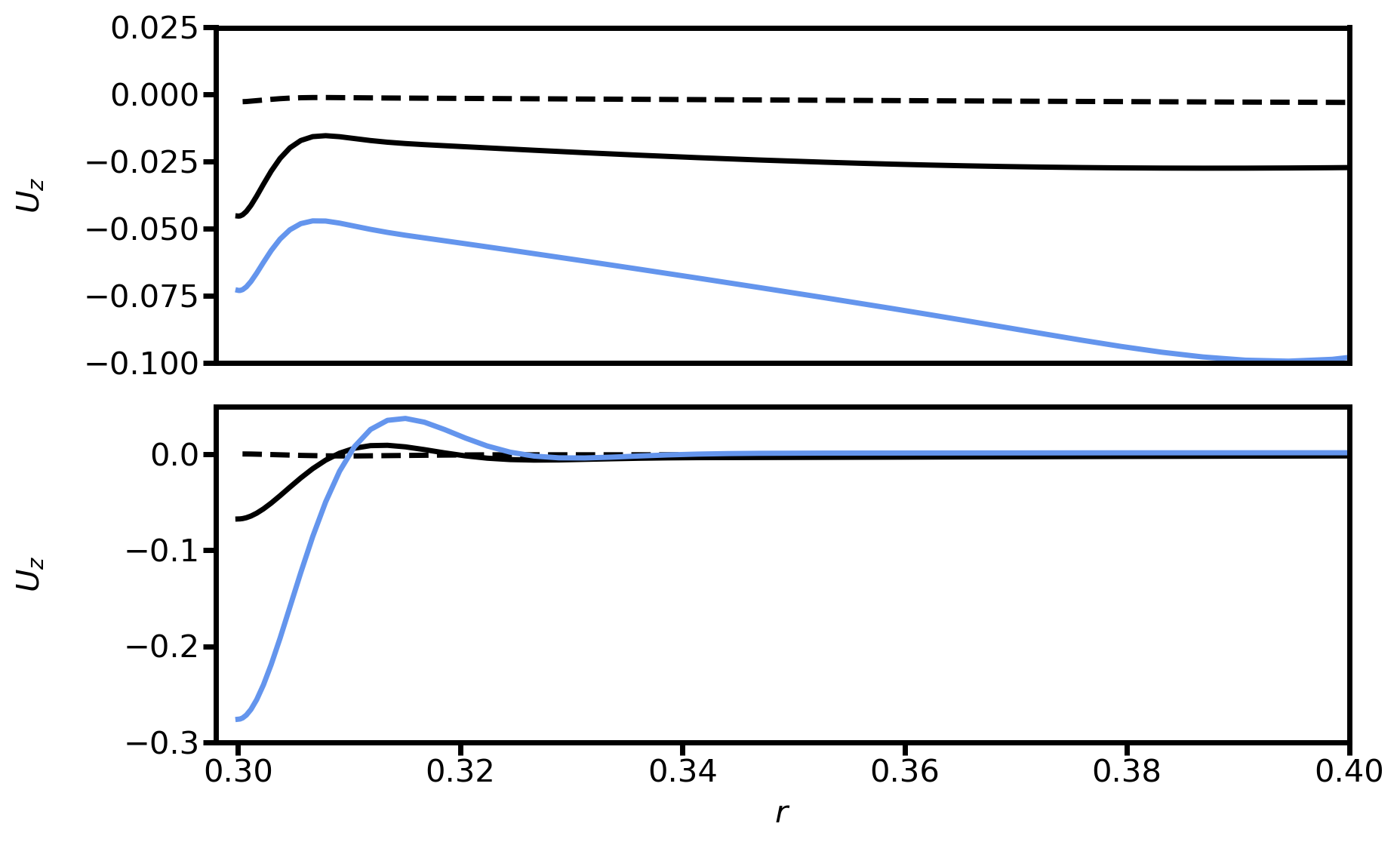}
\caption{Same as Fig. \ref{TP_Anel_dUphiRdr} but in the Eddington-Sweet regime with parameters $E=10^{-5}$, $P_r \left(N_0/\Omega_0\right)^2 = 10^{-1}$ and $Re_c = 10^{-2}$. The plain blue curve still corresponds to the Boussinesq case (run $1.3$ of Table \ref{parameters_edd}) while the other curves are obtained by varying the density contrast between the inner and outer spheres in the same way that previously but for runs $1.2$ to $1.4$ of Table \ref{parameters_edd_anel} in the left panel, and for cases $1.1$ and $1.3$ of Table \ref{parameters_edd_anel} in the middle and right panels.}
\label{Edd_Anel_dUphiRdr}
\end{center}
\end{figure*}

The latitudinal variation of the inner sphere rotation rate found at $\rho_i/\rho_0=1.75$ is also found in the Boussinesq simulations and is a manifestation of the local rotation rate maximum found near the equator in these simulations. This local maximum seems to be related to the strong vertical jet which advects AM towards the equator, this jet being in turn associated with the presence of an equatorial boundary layer. Figure \ref{Edd_Anel_dUphiRdr} shows the radial profiles of the gradient of the angular velocity together with the cylindrical radial and vertical velocity fields (respectively left, middle and right panels) near the inner sphere at $\theta = \pi/4$ and at the equator (respectively top and bottom rows), both for the Boussinesq (blue curve) and anelastic cases (black curves). As in the Taylor-Proudman regime, we observe that the strong equatorial boundary layer and the vertical jet that are present in the Boussinesq case nearly disappear in the anelastic simulations. Thus, contrary to the Boussinesq case, the meridional flow is no longer focused towards the equator. The different AM advections between the Boussinesq and anelastic cases thus appears to explain the smoother, more radial differential rotation observed in the anelastic case. 

\section{Summary and conclusions}
\label{summary_and_conclusions}


In this paper, we have presented the results of our extensive parametric study of an axisymmetric rotating spherical layer undergoing contraction, mimicking a stellar radiative zone during a rapid contraction phase. Our goal was to understand the motions induced by the contraction, in particular the amplitude and the spatial distribution of the differential rotation and of the meridional circulation. The contraction, modelled through an imposed mass-conserving radial velocity field $\vv{V_f}$, produces an inward transport of AM that is balanced, in a stationary state, by an outward viscous or advective transport of AM. The full compressible gas dynamics has been approximated using either the Boussinesq or the anelastic equations.

We show that the parameter $P_r \left(N_0/\Omega_0\right)^2$ controls the amplitude and the distribution of the differential rotation. When $P_r \left(N_0/\Omega_0\right)^2 \gg 1$, the buoyancy force inhibits the circulation and the viscosity dominates the AM transport. For a weak differential rotation $\Delta \Omega/\Omega \ll 1$, the amplitude of the differential rotation is proportional to the ratio between the viscous and the contraction timescales and its purely radial profile can be derived analytically. This regime is relevant in the core of contracting subgiants. 
If $P_r \left(N_0/\Omega_0\right)^2 \ll 1$, an Eddington-Sweet type circulation dominates the AM transport. Then, for a weak differential rotation and in the absence of density stratification, the amplitude of the differential rotation is of the order of the ratio between the Eddington-Sweet and the contraction timescales. In the anelastic simulations, this amplitude decreases approximately as the inverse of the integral of the background density across the layer. The rotation is neither cylindrical nor radial, but when the density stratification is taken into account, it tends to be predominantly radial and smoothly distributed. This regime holds for PMS stars and outside the degenerate core of subgiants. 

Although less relevant for stars, a third regime of weak stratification $P_r \left(N_0/\Omega_0\right)^2 \le \sqrt{E}$ was studied to understand the effects of the small though finite Ekman numbers in our simulations. It enabled us to understand the effects of the boundary layers formed at the inner and outer limits of the numerical domain. In particular, we found that in the Boussinesq simulations the presence of an equatorial boundary layer and of an associated vertical jet along the tangent cylinder produces a localised region of high rotation rates near the equator. This equatorial layer and the vertical jet are absent from the more realistic density stratified anelastic simulations leading to a smoother distribution of the differential rotation.  This is at odds with the conclusions of \cite{hypolite2014dynamics} where such jets are invoked as a way to connect the core and the convective envelope of the contracting subgiants.


In this study, we have been mostly concerned with the weak differential rotation regime although the limits of this linear regime $\Delta \Omega / \Omega_0 \sim 1$ have been exceeded in various cases. It would be interesting to further investigate the non-linear regime, although this may be numerically challenging even for axisymmetric studies. 
In our setup the rotation of the frame of reference has been kept constant while we expect it to increase as the star contracts. This effect could be introduced by adding an Euler term that describes the time evolution of rotating frame of reference (e.g. \cite{hypolite2014dynamics}). However, in density stratified cases we expect that this temporal variation is slow compared to the contraction time in most parts of the domain. 

We also want to stress two important points about the anelastic approximation. Firstly, our formulation of the momentum equation Eq. \eqref{anel_momentum_adim} neglects the $\left(P^{'} / \overline{\rho} \hspace*{0.05cm} C_p \right) \text{d} \overline{S} / \text{d} r$ term: this is the LBR approximation (\cite{lantz1992dynamical}, \cite{braginsky1995equations} and \cite{lantz1999anelastic}). Physically, it means that the density scale height is smaller than the potential temperature scale height, which is for example approximately satisfied in the earth's atmosphere \citep{vallis2017atmospheric}. This hypothesis is ensured in our simulations by keeping a low deviation from isentropy ($\epsilon_s \ll 1$). The legitimate question is, what is the validity of this approximation for a stellar interior? For a convective zone, this assumption is obviously correct since $\text{d} \overline{S} / \text{d} r \approx 0$. For a radiative interior, a model of subgiant computed with the stellar evolution code MESA shows that outside the degenerate core, the ratio $h_{\theta}/h_{\rho}$ between the potential temperature and density scale heights is only four during the evolution of the subgiant up to the red giant branch (RGB). Inside the core, the condition $h_{\rho} \ll h_{\theta}$ can be satisfied at the terminal age main sequence (TAMS) as well as at the base of the RGB, but never during the subgiant phase where $h_{\theta}$ and $h_{\rho}$ are of the same order. Nevertheless, using a sound-proof approach other than the LBR approximation would lead to problems associated with the non-conservation of energy \citep{brown2012energy}. 

Secondly, in the thermal balance equation, we used the entropy diffusion term instead of the temperature diffusion because temperature diffusion leads to numerical difficulties with the code MagIC. In principle, these numerical difficulties could be relaxed by ensuring a strong coupling between the pressure and velocity fields, either by using staggered grids \citep{harlow1965numerical} or through an interpolation method \citep{rhie1983numerical}.

In this work, we present the Eddington-Sweet regime as a possible axisymmetric steady state in a contracting stellar radiative zone. This might seem at odds with previous studies that questioned the existence of an Eddington-Sweet meridional circulation \citep{busse1981eddington, busse1982problem} but as we discuss now there is actually no contradiction between these studies and the present one. \cite{busse1981eddington} showed that for an inviscid baroclinic stably stratified rotating fluid, the only possible steady state is a purely zonal flow and that any meridional motions should be only transients. A stationary meridional flow would indeed produce an uncompensated advection in the AM equation. Consequently, if an Eddington-Sweet circulation initially existed, it would be modified on a rotation timescale \citep{busse1982problem} because of the lack of AM conservation. Such a system would then evolve thermally so that the divergence of the heat flux balances the temperature advection leading, after an Eddington-Sweet timescale, to a steady state where the meridional circulation as well as the divergence of the heat flux are both zero. Our work does not disagree with these results since we introduce a contraction term in the AM equation which compensates for the advection by a steady meridional flow, thus ensuring AM conservation. We then get a meridional flow whose amplitude is of the order of the forcing amplitude and whose timescale corresponds to the so-called Eddington-Sweet timescale.

The existence of the axisymmetric states described in this paper can also be debated against potential hydrodynamical instabilities. Axisymmetric instabilities like the centrifugal instability \citep{zahn1974rotational} or the Goldreich-Schubert-Fricke instability \citep{goldreich1967differential} are unlikely because they should have been observed in our numerical simulations. But the differentially rotating solutions might still be subject to non-axisymmetric instabilities either barotropic instabilities driven by the radial \citep{lignieres1999shear} or the horizontal shear \citep{Deloncle2007}, or baroclinic instabilities which rest on the non-coincidence of the isobaric and isentropic surfaces \citep{Spruit1984}.


Besides the hydrodynamical stability of the axisymmetric steady states described in this work, contracting radiative zones could also potentially trigger magneto-hydrodynamical instabilities \citep{spruit1999differential}. The AM transport associated with the development of such instabilities tends to decrease the level of differential rotation imposed by the contraction such as required by subgiant seismic data (\cite{deheuvels2012seismic},
\cite{deheuvels2014seismic}). For example \cite{fuller2019slowing} propose a revised mechanism for the saturation of the Tayler instability that reproduces fairly well the core rotation rates along the red giant branch, during the red clump and up to the white dwarf phase. The magneto-rotational instability (MRI) (\cite{rudiger2015angular}, \cite{jouve2020interplay}) could also be at play in contracting radiative zones.

In intermediate-mass PMS stars, such instabilities have also been invoked as a way to explain the observed dichotomy between the magnetism of Ap/Bp stars and the weak complex magnetic fields discovered on bright A and Am stars \citep{lignieres2013dichotomy,gaurat2015evolution}.

\begin{acknowledgements}

The authors acknowledge the developers of MagIC (\url{https://github.com/magic-sph/magic}) for the open-source code thanks to which the simulations were performed. This work was granted access to the HPC resources of CALMIP supercomputing center under the allocation P$1118$. The authors also wish to thank Sébastien Deheuvels for providing us with stellar parameters from the stellar evolution code MESA. LJ acknowledges funding by the Institut Universitaire de France.

\end{acknowledgements}

\bibliographystyle{Config/aa}
\bibliography{Bibliography/biblio.bib}

\appendix

\section{Hydrostatic state forced by the contraction}
\label{second_hydrostatic_state}

In order to derive the different timescales of AM transport in Sect. \ref{timescales_physical_processes}, we will subtract the hydrostatic state forced by the contraction velocity field from our equations. For this, we first decompose the temperature fluctuations as the sum of a spherical contribution and a deviation to the spherical symmetry:

\begin{equation}
\Theta^{'}(r,\theta) = T^{'}(r) + \delta \Theta(r,\theta)
\label{decomposition_temperature_fluctuations}
\end{equation}

\noindent where:

\begin{equation}
T^{'}(r) = \displaystyle \frac{1}{2} \displaystyle \int_0^{\pi} \Theta^{'}(r,\theta) \sin{\theta} ~ \text{d} \theta
\label{spherical_contribution_definition}
\end{equation}

\noindent and,

\begin{equation}
\displaystyle \frac{1}{2} \displaystyle \int_0^{\pi} \delta \Theta(r,\theta) \sin{\theta} ~ \text{d} \theta = 0
\label{no_spherical_contribution_definition}
\end{equation}

We rewrite Eq. \eqref{bouss_entropie_adim} by using the decomposition of Eq. \eqref{decomposition_temperature_fluctuations}, wherein $T^{'}(r)$ is adimensionalised in the same way as $\overline{T}(r)$ and $\delta \Theta(r,\theta)$ is non-dimensionalised like $\Theta^{'}(r,\theta)$:

\begin{equation}
\begin{array}{lll}
Pe_c \hspace*{0.05cm} P_r \hspace*{0.05cm} R_0 \left \lbrack \displaystyle \frac{\partial \delta \tilde{\Theta}}{\partial t} + \left( \left( \vv{\tilde{U}} + \vv{\tilde{V}_f} \right) \cdot \vv{\nabla} \right) \delta \tilde{\Theta} \right \rbrack ~ + \\\\ Pe_c \hspace*{0.05cm} P_r \left( \displaystyle \frac{N_0}{\Omega_0} \right)^2 \left( \tilde{U}_r + \tilde{V}_f \right) \left( \displaystyle \frac{\text{d} \tilde{\overline{T}}}{\text{d} r} + \displaystyle \frac{\text{d} \tilde{T}^{'}}{\text{d} r} \right) = Pe_c \hspace*{0.05cm} E ~ \vv{\nabla}^2 \delta \tilde{\Theta} ~ + \\\\ P_r \left( \displaystyle \frac{N_0}{\Omega_0} \right)^2 ~ \vv{\nabla}^2 \tilde{T}^{'}
\end{array}
\label{entropy_equation_with_decomposition}
\end{equation}

If $R_o \ll \left(N_0/\Omega_0\right)^2$, the steady case of the foregoing equation reads:

\begin{equation}
\begin{array}{lll}
Pe_c \hspace*{0.05cm} P_r \left( \displaystyle \frac{N_0}{\Omega_0} \right)^2 \left( \tilde{U}_r + \tilde{V}_f \right) \left( \displaystyle \frac{\text{d} \tilde{\overline{T}}}{\text{d} r} + \displaystyle \frac{\text{d} \tilde{T}^{'}}{\text{d} r} \right) = Pe_c \hspace*{0.05cm} E ~ \vv{\nabla}^2 \delta \tilde{\Theta} ~ + \\\\ P_r \left( \displaystyle \frac{N_0}{\Omega_0} \right)^2 ~ \vv{\nabla}^2 \tilde{T}^{'}
\end{array}
\label{steady_entropy_equation_adim}
\end{equation}

Using both the Ostrogradsky's theorem and the continuity equation, we can then write:

\begin{equation}
\oiint_{\mathcal{S}} \pm ~ \vv{\tilde{U}} \cdot \text{d} \mathcal{S} ~ \vv{e_r} = \iiint_{\mathcal{V}} \left( \vv{\nabla} \cdot \vv{\tilde{U}} \right) \text{d} \mathcal{V} = 0
\end{equation}

\noindent which then allows us to deduce:

\begin{equation}
\displaystyle \frac{1}{2} \int_{0}^{\pi} \tilde{U}_r \sin{\theta} ~ \text{d} \theta = 0
\end{equation}

Furthermore,

\begin{equation}
\hspace*{-0.18cm}
\begin{array}{lll}
\displaystyle \frac{1}{2} \int_{0}^{\pi} \vv{\nabla}^2  \delta \tilde{\Theta}  \sin{\theta} ~ \text{d} \theta = \displaystyle \frac{1}{\tilde{r}^2} \displaystyle \frac{\partial}{\partial r} \left \lbrack \tilde{r}^2 \displaystyle \frac{\partial}{\partial r} \left( \displaystyle \frac{1}{2} \displaystyle \int_{0}^{\pi} \delta \tilde{\Theta} \sin{\theta} ~ \text{d} \theta \right) \right \rbrack \\\\ \hspace*{3.0cm} + ~ \displaystyle \frac{1}{\tilde{r}^2} \left \lbrack \displaystyle \frac{1}{2} \displaystyle \int_0^{\pi} \displaystyle \frac{\partial}{\partial \theta} \left( \sin{\theta} \displaystyle \frac{\partial \delta \tilde{\Theta}}{\partial \theta} \right) \text{d} \theta  \right \rbrack
\end{array}
\end{equation}

According to Eq. \eqref{no_spherical_contribution_definition} the first integral in brackets is zero. Since the last integral is also zero, we conclude:

\begin{equation}
\displaystyle \frac{1}{2} \int_{0}^{\pi} \vv{\nabla}^2 \left( \delta \tilde{\Theta} \right) \sin{\theta} ~ \text{d} \theta = 0
\end{equation}

Consequently, the latitudinal mean of Eq. \eqref{steady_entropy_equation_adim} gives:

\begin{equation}
Pe_c \tilde{V}_f \left( \displaystyle \frac{\text{d} \tilde{\overline{T}}}{\text{d} r} + \displaystyle \frac{\text{d} \tilde{T}^{'}}{\text{d} r} \right) = \vv{\nabla}^2 \tilde{T}^{'}
\label{equation_for_t_prime}
\end{equation}

The radial derivative of $\tilde{\overline{T}}$ is given by solving $\vv{\nabla}^{2} \tilde{\overline{T}} = 0$ and leads to:

\begin{equation}
\displaystyle \frac{\text{d} \tilde{\overline{T}}}{\text{d} r} = \displaystyle \frac{- \tilde{r}_i}{\tilde{r}^2 \left(\tilde{r}_i - 1 \right)}
\end{equation}

Finally, $\tilde{T}^{'}(r)$ is derived with the solving of

\begin{equation}
\displaystyle \frac{\text{d}^2 \tilde{T}^{'}}{\text{d} r^2} + \displaystyle \frac{1}{\tilde{r}} \left(\displaystyle \frac{Pe_c}{\tilde{r}} + 2 \right) \displaystyle \frac{\text{d} \tilde{T}^{'}}{\text{d} r} = \displaystyle \frac{Pe_c \tilde{r}_i}{\tilde{r}^4 \left( \tilde{r}_i - 1 \right)}
\end{equation}

\noindent using the boundary conditions $\tilde{T}^{'}(1) = \tilde{T}^{'}(\tilde{r}_i) = 0$. The analytical solution is readily:

\begin{equation}
\tilde{T}^{'}(r) = \displaystyle \frac{\tilde{r}_i}{\tilde{r}}\left(\displaystyle \frac{\tilde{r}-1}{\tilde{r}_i-1}\right) - \left \lbrack \displaystyle \frac{\exp{\left(Pe_c \left( \displaystyle \frac{1}{\tilde{r}} -1 \right)\right)}-1}{\exp{\left(Pe_c \left( \displaystyle \frac{1}{\tilde{r}_i} -1 \right)\right)}-1}\right \rbrack
\end{equation}

We thus define:

\begin{equation}
\tilde{T}_{\text{m}}(r) = \tilde{\overline{T}}(r) + \tilde{T}^{'}(r) = \tilde{\overline{T}}(1) - \left \lbrack \displaystyle \frac{\exp{\left(Pe_c \left( \displaystyle \frac{1}{\tilde{r}} -1 \right)\right)}-1}{\exp{\left(Pe_c \left( \displaystyle \frac{1}{\tilde{r}_i} -1 \right)\right)}-1}\right \rbrack
\end{equation}

\noindent and likewise:

\begin{equation}
\displaystyle \frac{\text{d} \tilde{T}_{\text{m}}}{\text{d} r} = \displaystyle \frac{\text{d} \left(\tilde{\overline{T}}+\tilde{T}^{'}\right)}{\text{d}r} = \displaystyle \frac{- Pe_c}{\tilde{r}^2} \left \lbrack \displaystyle \frac{\exp{\left(Pe_c \left( \displaystyle \frac{1}{\tilde{r}} - 1 \right)\right)}}{\exp{\left(Pe_c \left( \displaystyle \frac{1}{\tilde{r}_i} -1 \right)\right)}-1} \right \rbrack 
\end{equation}

After subtracting the spherically symmetric temperature field $\tilde{T}^{'}(r)$ induced by the contraction velocity field, the evolution equation of temperature fluctuations finally reads:

\begin{equation}
\begin{array}{lll}
P_r \hspace*{0.05cm} R_0 \left \lbrack \displaystyle \frac{\partial \delta \tilde{\Theta}}{\partial t} + \left( \left( \vv{\tilde{U}} + \vv{\tilde{V}_f} \right) \cdot \vv{\nabla} \right) \delta \tilde{\Theta} \right \rbrack + P_r \left( \displaystyle \frac{N_0}{\Omega_0} \right)^2 \tilde{U}_r \displaystyle \frac{\text{d} \tilde{T}_{\text{m}}}{\text{d} r} ~ = \\\\ E ~ \vv{\nabla}^2 \delta \tilde{\Theta} 
\end{array}
\label{final_form_of_entropy_equation}
\end{equation}

\noindent while the momentum equation Eq. \eqref{bouss_momentum_adim} becomes:

\begin{equation}
\begin{array}{lll}
R_o \left \lbrack \displaystyle \frac{\partial \vv{\tilde{U}}}{\partial t} + \left( \left( \vv{\tilde{U}} + \vv{\tilde{V}_f} \right) \cdot \vv{\nabla} \right) \left( \vv{\tilde{U}} + \vv{\tilde{V}_f} \right)\right \rbrack + 2 \vv{e}_z \times \left( \vv{\tilde{U}} + \vv{\tilde{V}_f} \right) \\\\ = - \vv{\nabla} \delta \tilde{\Pi} + \delta \tilde{\Theta} \hspace*{0.02cm} \vv{\tilde{r}} ~ + E ~ \vv{\nabla}^2 \left( \vv{\tilde{U}} + \vv{\tilde{V}_f} \right)
 \label{bouss_momentum_adim_without_contraction_pressure}
\end{array}
\end{equation}

\noindent in which we have subtracted the hydrostatic state induced by the radial contraction:

\begin{equation}
\displaystyle \frac{\text{d} \tilde{P}_{\text{c}}(r)}{\text{d} r} = \displaystyle \frac{\tilde{T}^{'}(r)}{\tilde{\overline{T}}(r)} ~ \tilde{r}
\end{equation}

\noindent with $\delta \tilde{\Pi} = \tilde{\Pi}^{'} - \tilde{P}_{\text{c}} $ and where $\tilde{P}_{\text{c}}$ is the contraction-induced hydrostatic pressure non-dimensionalised with the hydrostatic pressure scale $r_0 \hspace*{0.05cm} \rho_0 \hspace*{0.05cm} g_0$.

\section{Analytical solution for the differential rotation in the Taylor-Proudman regime (Boussinesq case)}
\label{demonstration_rot_diff_TP}

In this appendix, we seek an analytical solution for the differential rotation in the Taylor-Proudman regime under the Boussinesq approximation.

The linear steady inviscid AM equation Eq. \eqref{angular_momentum_evolution} provides a balance between the Coriolis and contraction terms. This balance allows us to derive an analytical solution for the cylindrical radial velocity field:

\begin{equation}
\tilde{U}_s = \displaystyle \frac{\sin{\theta}}{\tilde{r}^2} = \displaystyle \frac{\tilde{s}}{\left(\tilde{s}^2 + \tilde{z}^2 \right)^{3/2}}
\label{Us}
\end{equation}

\noindent where $\tilde{s} = \tilde{r} \sin{\theta}$ and $\tilde{z} = \tilde{r} \cos{\theta}$. In cylindrical coordinates, the steady axisymmetric velocity field reads:

\begin{equation}
\vv{\tilde{U}}(r,\theta) = \left( \displaystyle \frac{-1}{\tilde{s}} \displaystyle \frac{\partial \tilde{\Psi}(s,z)}{\partial z}, \tilde{s} \hspace*{0.05cm} \delta \tilde{\Omega}(s,z), \displaystyle \frac{1}{\tilde{s}} \displaystyle \frac{\partial \tilde{\Psi}(s,z)}{\partial s} \right)
\label{axisymmetric_cylindrical_flow}
\end{equation}
 
\noindent where $\tilde{\Psi}$ is the dimensionless stream function. By using Eq. \eqref{Us} we can then derive an analytical solution for $\tilde{\Psi}$:

\begin{equation}
\tilde{\Psi}(s,z) = \displaystyle \frac{-\tilde{z}}{\sqrt{\tilde{s}^2 + \tilde{z}^2}} + \tilde{f}(s)
\end{equation}

\noindent while the use of Eq. \eqref{axisymmetric_cylindrical_flow} allows us to determine the vertical velocity field:

\begin{equation}
\tilde{U}_z = \displaystyle \frac{\cos{\theta}}{\tilde{r}^2} + \displaystyle \frac{\tilde{f}^{'}(s)}{\tilde{s}} = \displaystyle \frac{\tilde{z}}{\left(\tilde{s}^2 + \tilde{z}^2 \right)^{3/2}} + \displaystyle \frac{\tilde{f}^{'}(s)}{\tilde{s}}
\label{Uz}
\end{equation}

By using the classical relations between the cylindrical and spherical coordinates we deduce:

\begin{equation}
\tilde{U}_r = \displaystyle \frac{1}{\tilde{r}^2} + \displaystyle \frac{\tilde{f}^{'}(s)}{\tilde{s}} \cos{\theta}
\end{equation}

\begin{equation}
\tilde{U}_{\theta} = - \displaystyle \frac{\tilde{f}^{'}(s)}{\tilde{s}} \sin{\theta} 
\end{equation}

The no-slip condition at the inner sphere $\tilde{U}_r(r=\tilde{r}_i,\theta) = 0$ leads to:

\begin{equation}
\left. \displaystyle \frac{\tilde{f}^{'}(s)}{\tilde{s}} \right|_{\theta \hspace*{0.05cm} \in \hspace*{0.05cm} [0 \hspace*{0.03cm} ; \hspace*{0.03cm} \pi/2]} = \displaystyle \frac{-1}{\tilde{r}_i^2 \cos{\theta_i}} \quad \text{;} \quad \left. \displaystyle \frac{\tilde{f}^{'}(s)}{\tilde{s}} \right|_{\theta \hspace*{0.05cm} \in \hspace*{0.05cm} [\pi/2 \hspace*{0.03cm} ; \hspace*{0.03cm} \pi]} = \displaystyle \frac{1}{\tilde{r}_i^2 \cos{\theta_i}}
\label{fs}
\end{equation}

By symmetry, $\tilde{U}_z = 0$ at the equator, which means that when $\tilde{r}>\tilde{r}_i$, $\tilde{f}^{'}(s)/\tilde{s} = 0$. This allows us to find the analytical solutions of the meridional velocity fields:

\begin{equation}
\begin{array}{lll}
\left. \tilde{U}_r(s \leq r_i,z) \hspace*{0.05cm} \right|_{\theta \hspace*{0.05cm} \in \hspace*{0.05cm} [0 \hspace*{0.03cm} ; \hspace*{0.03cm} \pi/2]} = \displaystyle \frac{1}{\tilde{r}^2} - \displaystyle \frac{\cos \theta}{\tilde{r}_i^2 \cos \theta_i} \\\\
\left. \tilde{U}_r(s \leq r_i,z) \hspace*{0.05cm} \right|_{\theta \hspace*{0.05cm} \in \hspace*{0.05cm} [\pi/2 \hspace*{0.03cm} ; \hspace*{0.03cm} \pi]} = \displaystyle \frac{1}{\tilde{r}^2} + \displaystyle \frac{\cos \theta}{\tilde{r}_i^2 \cos \theta_i} \\\\
\tilde{U}_r(s>r_i,z) = \displaystyle \frac{1}{\tilde{r}^2}
\end{array}
\label{radial_Edd}
\end{equation}

\begin{equation}
\begin{array}{lll}
\left. \tilde{U}_{\theta}\hspace*{0.05cm}(s \leq r_i,z) \hspace*{0.05cm} \right|_{\theta \hspace*{0.05cm} \in \hspace*{0.05cm} [0 \hspace*{0.03cm} ; \hspace*{0.03cm} \pi/2]} = \displaystyle \frac{\sin \theta}{\tilde{r}_i^2 \cos \theta_i} \\\\
\left. \tilde{U}_{\theta}\hspace*{0.05cm}(s \leq r_i,z) \hspace*{0.05cm} \right|_{\theta \hspace*{0.05cm} \in \hspace*{0.05cm} [\pi/2 \hspace*{0.03cm} ; \hspace*{0.03cm} \pi]} = -\displaystyle \frac{\sin \theta}{\tilde{r}_i^2 \cos \theta_i} \\\\
\tilde{U}_{\theta}\hspace*{0.05cm}(s>r_i,z) = 0
\end{array}
\label{latitudinal_Edd}
\end{equation}

By integrating Eq. \eqref{fs}, we deduce the stream function (up to a constant):

\begin{equation}
\begin{array}{lll}
\left. \tilde{\Psi}(r,\theta) \right|_{\theta \hspace*{0.05cm} \in \hspace*{0.05cm} [0 \hspace*{0.03cm} ; \hspace*{0.03cm} \pi/2]} = - \cos{\theta} + \cos{\theta_i} + \text{cte} \\\\
\left. \tilde{\Psi}(r,\theta) \right|_{\theta \hspace*{0.05cm} \in \hspace*{0.05cm} [\pi/2 \hspace*{0.03cm} ; \hspace*{0.03cm} \pi]} = - \cos{\theta} - \cos{\theta_i} + \text{cte}
\end{array}
\end{equation}

\noindent and since the stream function $\tilde{\Psi}$ is zero at the poles, we deduce that $\text{cte} = 0$. 

The mass conservation leads to the following relationship:

\begin{equation}
\iiint_{\mathcal{V}} \left( \vv{\nabla} \cdot \vv{\tilde{U}} \right) \text{d} V = \oiint_{\mathcal{S}} \vv{\tilde{U}} \cdot \text{d} \vv{S} = 0
\label{bouss_mass_conservation}
\end{equation}

\noindent which gives in spherical coordinates:

\begin{equation}
\oiint  \tilde{U}_r ~ \tilde{r} ~ \text{d} \theta ~ \tilde{r} \sin{\theta} ~ \text{d} \phi + \oiint \tilde{U}_{\theta} ~ \text{d} r ~ \tilde{r} \sin{\theta} ~ \text{d} \phi = 0
\label{bilan}
\end{equation}

From the analysis of the Ekman layer located at the outer sphere, we obtain the non-dimensional form of the Ekman condition:

\begin{equation}
\tilde{U}_r(r=1,\theta) = \displaystyle \frac{- \sqrt{E}}{\sin \theta} \displaystyle \frac{\partial}{\partial \theta} \left( \displaystyle \frac{\text{sgn} \left(\cos{\theta}\right) \hspace*{0.05cm} \sin \theta ~ \tilde{U}_{\phi}^{I}\left(r=1,\theta\right)}{2 \sqrt{\left | \cos \theta \right |}} \right)
\label{ekman_pumping_dimensionless}
\end{equation}

We now rewrite Eq. \eqref{ekman_pumping_dimensionless} in cylindrical coordinates using the Taylor-Proudman constraint and by noting that $\cos{\theta} = \sqrt{1 - \left(\tilde{s}/\tilde{r}\right)^2}$ which leads to $\cos{\theta_0} = \sqrt{1 - \tilde{s}^2}$ at the outer sphere: 

\begin{equation}
\tilde{U}_r = \displaystyle \frac{- \sqrt{E}}{\tilde{s}} \cos{\theta} \displaystyle \frac{\partial}{\partial s} \left( \displaystyle \frac{\text{sgn} \left(\cos{\theta}\right) \hspace*{0.05cm} \tilde{s}^2 ~ \delta \tilde{\Omega}(s)}{2 \sqrt{\left| \cos \theta_0 \right|}} \right)
\label{ur_integrated}
\end{equation}

In addition,

\begin{equation}
\tilde{U}_{\theta} = \displaystyle \frac{-1}{\tilde{r} \sin{\theta}} \displaystyle \frac{\partial \tilde{\Psi}}{\partial r} = \displaystyle \frac{- \cos{\theta}}{\tilde{s}} \displaystyle \frac{\partial \tilde{\Psi}}{\partial z}
\end{equation}

\noindent thus, with the following substitutions $\tilde{r} \cos{\theta} \hspace*{0.05cm} \text{d}\theta = \text{d} s$ and $\text{d} z = \cos{\theta} \hspace*{0.05cm} \text{d} r $, the relationship Eq. \eqref{bilan} becomes:

\begin{equation}
- \oiint \sqrt{E} \displaystyle \frac{\partial}{\partial s} \left( \displaystyle \frac{\text{sgn} \left(\cos{\theta}\right) \hspace*{0.05cm} \tilde{s}^2 ~ \delta \tilde{\Omega}(s)}{2 \sqrt{\left | \cos \theta_0 \right |}} \right) \text{d} s \text{d} \phi = - \oiint \displaystyle \frac{\partial \tilde{\Psi}}{\partial z} \text{d} z \text{d} \phi
\label{bilan_cylindrical}
\end{equation}

\begin{figure}[h]
\begin{center}
\includegraphics[width=8.5cm]{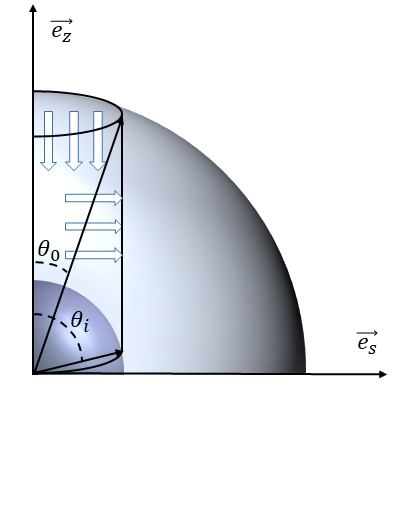}
\vspace*{-2.8cm}
\caption{Sketch of a mass-flux balance inside a cylinder of radius $s = r \sin{\theta}$ \hspace*{0.01cm} between $z_0= r_0 \sqrt{1 - \left(s/r_0\right)^2}$ and $z_i = r_i \sqrt{1 - \left(s/r_i\right)^2}$.}
\label{mass_flux_balance}
\end{center}
\end{figure}

The choice of the closed surface is a cylinder of radius $s = r \sin{\theta} $ between $z_0 = r_0 \sqrt{1 - \left(s/r_0\right)^2}$ and $z_i = r_i \sqrt{1 - \left(s/r_i\right)^2}$ as represented in Fig. \ref{mass_flux_balance}.

By noting $\tilde{\mathcal{Q}}_s$ the dimensionless outward mass-flux crossing this cylinder, we have:

\begin{equation}
\tilde{\mathcal{Q}}_s = - 2 \pi \displaystyle \int_{\tilde{z}_i}^{\tilde{z}_0} \hspace*{0.02cm} \displaystyle \frac{\partial \tilde{\Psi}}{\partial z} \hspace*{0.05cm} \text{d} z = 2 \pi \left( \cos{\theta_0} - \cos{\theta_i} \right)
\end{equation}

Likewise, with $\tilde{\mathcal{Q}}_z$ the inward dimensionless mass-flux crossing the cylinder, we obtain:

\begin{equation}
\begin{array}{lll}
\tilde{\mathcal{Q}}_z = - 2 \pi \sqrt{E} \displaystyle \int_{0}^{\tilde{s}} \frac{\partial}{\partial s} \left( \displaystyle \frac{\text{sgn} \left(\cos{\theta}\right) \hspace*{0.05cm} \tilde{s}^2 \delta \tilde{\Omega}(s)}{2 \sqrt{\left|\cos{\theta_0}\right|}} \right) \hspace*{0.05cm} \text{d} s \\\\ \hspace*{0.5cm}
= - \hspace*{0.05cm} 2 \pi \sqrt{E} \hspace*{0.05cm} \displaystyle \frac{\tilde{s}^2 ~ \delta \tilde{\Omega}(s)}{2 \sqrt{|\cos{\theta_0}|}}
\end{array}
\end{equation}

Finally, from Eq. \eqref{bilan_cylindrical} we get: 

\begin{equation}
2 \pi \left( \cos{\theta_0} - \cos{\theta_i} \right) = 2 \pi \sqrt{E} \hspace*{0.05cm} \cdot \displaystyle \frac{\tilde{s}^2 ~ \delta \tilde{\Omega}(s)}{2 \sqrt{|\cos{\theta_0}|}}
\end{equation}

\noindent whence we conclude:

\begin{equation}
\begin{array}{lll}
\delta \tilde{\Omega}_{\text{TP}}(s \leq r_i) = \displaystyle \frac{1}{\sqrt{E}} \left \lbrack \displaystyle \frac{2}{\tilde{s}^2} \left( \cos{\theta_0} - \cos{\theta_i} \right) \sqrt{\left| \cos{\theta_0} \right|}\right \rbrack \\\\
\delta \tilde{\Omega}_{\text{TP}}(s >r_i) = \displaystyle \frac{1}{\sqrt{E}} \left \lbrack \displaystyle \frac{2}{\tilde{s}^2} \cos{\theta_0}  \sqrt{\left| \cos{\theta_0} \right|}\right \rbrack
\end{array}
\label{analytical_solution_dimensionless_rot_diff_tp}
\end{equation}

\noindent which can then be rewritten under dimensional form to lead to the solution Eq. \eqref{differential_rotation_TP}.

\section{Analytical solution for the differential rotation in the Taylor-Proudman regime (anelastic case)}
\label{demonstration_rot_diff_TP_anelastic}

We will now show that the analytical solution previously derived, namely Eq. \eqref{analytical_solution_dimensionless_rot_diff_tp}, is still valid independently of the amplitude of the density contrast applied between the inner and outer spheres.

In the anelastic approximation, the dimensionless balance between the Coriolis and contraction terms now provides:

\begin{equation}
\tilde{U}_s = \displaystyle \frac{\sin{\theta}}{\tilde{\overline{\rho}} \hspace*{0.02cm} \tilde{r}^2} = \displaystyle \frac{\tilde{s}}{\tilde{\overline{\rho}} \left(\tilde{s}^2 + \tilde{z}^2 \right)^{3/2}}
\label{Us_anel}
\end{equation}

The poloidal-toroidal decomposition now relates the stream function to the meridional velocity fields as follows:

\begin{equation}
\tilde{\overline{\rho}} \tilde{U}_s = \displaystyle \frac{-1}{\tilde{s}} \displaystyle \frac{\partial \tilde{\Psi}(s,z)}{\partial z} \hspace*{0.1cm} \text{;} \quad \tilde{\overline{\rho}} \tilde{U}_z = \displaystyle \frac{1}{\tilde{s}} \displaystyle \frac{\partial \tilde{\Psi}(s,z)}{\partial s}
\end{equation}

\noindent or equivalent to,

\begin{equation}
\tilde{\overline{\rho}} \tilde{U}_r = \displaystyle \frac{1}{\tilde{r}^2 \sin{\theta}} \displaystyle \frac{\partial \tilde{\Psi}(r,\theta)}{\partial \theta} \hspace*{0.1cm} \text{;} \quad 
\tilde{\overline{\rho}} \tilde{U}_\theta = \displaystyle \frac{-1}{\tilde{r} \sin{\theta}} \displaystyle \frac{\partial \tilde{\Psi}(r,\theta)}{\partial r}
\end{equation}

From these relationships, it is obvious that all the analytical solutions derived for the meridional fields will be identical to the Boussinesq case except that they will now be weighted by the inverse of the background density profile:

\begin{equation}
\begin{array}{lll}
\left. \tilde{U}_z(s \leq r_i,z) \hspace*{0.05cm} \right|_{\theta \hspace*{0.05cm} \in \hspace*{0.05cm} [0 \hspace*{0.03cm} ; \hspace*{0.03cm} \pi/2]} = \displaystyle \frac{1}{\tilde{\overline{\rho}}} \left( \displaystyle \frac{\cos{\theta}}{\tilde{r}^2} - \displaystyle \frac{1}{\tilde{r}_i^2 \cos{\theta_i}} \right) \\\\
\left. \tilde{U}_z(s \leq r_i,z) \hspace*{0.05cm} \right|_{\theta \hspace*{0.05cm} \in \hspace*{0.05cm} [\pi/2 \hspace*{0.03cm} ; \hspace*{0.03cm} \pi]}= \displaystyle \frac{1}{\tilde{\overline{\rho}}} \left( \displaystyle \frac{\cos{\theta}}{\tilde{r}^2} + \displaystyle \frac{1}{\tilde{r}_i^2 \cos{\theta_i}} \right) \\\\
\tilde{U}_z(s>r_i,z) = \displaystyle \frac{\cos{\theta}}{\tilde{\overline{\rho}} \hspace*{0.02cm} \tilde{r}^2}
\end{array}
\label{vertical_Edd_anel}
\end{equation}

\begin{equation}
\begin{array}{lll}
\left. \tilde{U}_r(s \leq r_i,z) \hspace*{0.05cm} \right|_{\theta \hspace*{0.05cm} \in \hspace*{0.05cm} [0 \hspace*{0.03cm} ; \hspace*{0.03cm} \pi/2]} = \displaystyle \frac{1}{\tilde{\overline{\rho}}} \left( \displaystyle \frac{1}{\tilde{r}^2} - \displaystyle \frac{\cos \theta}{\tilde{r}_i^2 \cos \theta_i} \right) \\\\
\left. \tilde{U}_r(s \leq r_i,z) \hspace*{0.05cm} \right|_{\theta \hspace*{0.05cm} \in \hspace*{0.05cm} [\pi/2 \hspace*{0.03cm} ; \hspace*{0.03cm} \pi]} = \displaystyle \frac{1}{\tilde{\overline{\rho}}} \left( \displaystyle \frac{1}{\tilde{r}^2} + \displaystyle \frac{\cos \theta}{\tilde{r}_i^2 \cos \theta_i} \right) \\\\
\tilde{U}_r(s>r_i,z) = \displaystyle \frac{1}{\tilde{\overline{\rho}} \hspace*{0.02cm} \tilde{r}^2}
\end{array}
\label{radial_Edd_anel}
\end{equation}

\begin{equation}
\begin{array}{lll}
\left. \tilde{U}{_{\theta}}\hspace*{0.05cm}(\tilde{s} \leq \tilde{r}_i,\tilde{z}) \hspace*{0.05cm} \right|_{\theta \hspace*{0.05cm} \in \hspace*{0.05cm} [0 \hspace*{0.03cm} ; \hspace*{0.03cm} \pi/2]} = \displaystyle \frac{\sin \theta}{\tilde{\overline{\rho}} \hspace*{0.05cm} \tilde{r}_i^2 \cos \theta_i} \\\\
\left. \tilde{U}{_{\theta}}\hspace*{0.05cm}(\tilde{s} \leq \tilde{r}_i,\tilde{z}) \hspace*{0.05cm} \right|_{\theta \hspace*{0.05cm} \in \hspace*{0.05cm} [\pi/2 \hspace*{0.03cm} ; \hspace*{0.03cm} \pi]} = \displaystyle \frac{- \sin \theta}{\tilde{\overline{\rho}} \hspace*{0.05cm} \tilde{r}_i^2 \cos \theta_i} \\\\
\tilde{U}{_{\theta}}\hspace*{0.05cm}(\tilde{s}>\tilde{r}_i,\tilde{z}) = 0
\end{array}
\label{latitudinal_Edd_anel}
\end{equation}

\noindent while the previous analytical solution derived for the stream function is unmodified:  

\begin{equation}
\begin{array}{lll}
\left. \tilde{\Psi}(r,\theta) \right|_{\theta \hspace*{0.05cm} \in \hspace*{0.05cm} [0 \hspace*{0.03cm} ; \hspace*{0.03cm} \pi/2]} = - \cos{\theta} + \cos{\theta_i}\\\\
\left. \tilde{\Psi}(r,\theta) \right|_{\theta \hspace*{0.05cm} \in \hspace*{0.05cm} [\pi/2 \hspace*{0.03cm} ; \hspace*{0.03cm} \pi]} = - \cos{\theta} - \cos{\theta_i}
\end{array}
\end{equation}

In the anelastic approximation, Eq. \eqref{bouss_mass_conservation} is rewritten:

\begin{equation}
\iiint_{\mathcal{V}} \left( \vv{\nabla} \cdot \left( \tilde{\overline{\rho}} \vv{\tilde{U}} \right) \right) \text{d} V = \oiint_{\mathcal{S}} \left( \tilde{\overline{\rho}} \vv{\tilde{U}} \right) \cdot \text{d} \vv{S} = 0
\end{equation}

Since the Ekman condition Eq. \eqref{ekman_pumping_dimensionless} is given in $\tilde{r} = 1$ and that the background density profile at this location is such as $\tilde{\overline{\rho}}(r = 1) = 1$, we obtain the following relationship: 

\begin{equation}
- \oiint \sqrt{E} \displaystyle \frac{\partial}{\partial s} \left( \displaystyle \frac{\text{sgn} \left(\cos{\theta}\right) \hspace*{0.05cm} \tilde{s}^2 ~ \delta \tilde{\Omega}(s)}{2 \sqrt{\left | \cos \theta_0 \right |}} \right) \text{d} s \text{d} \phi = - \oiint \displaystyle \frac{\partial \tilde{\Psi}}{\partial z} \text{d} z \text{d} \phi
\label{bilan_cylindrical_anel}
\end{equation}

\noindent in other words, the same dimensional analytical solution for the differential rotation than in the Boussinesq case, namely Eq. \eqref{analytical_solution_dimensionless_rot_diff_tp} (or under dimensional form Eq. \eqref{differential_rotation_TP}).

\section{Solution in the Boussinesq Eddington-Sweet regime}
\label{superposition}

In this appendix, we show that in the Eddington-Sweet regime, a solution 
for the differential rotation can be expressed as the sum of an Eddington-Sweet type solution plus a Taylor-Proudman solution. The second term comes from the Ekman boundary layer at the outer sphere and can not be entirely neglected near this boundary even at very small Ekman numbers. 

The dimensionless linear axisymmetric steady equations of momentum, temperature and continuity read:

\begin{equation}
\begin{array}{lll}
- 2 \sin{\theta} ~ \tilde{U}_{\phi} = - \displaystyle \frac{\partial \delta \tilde{\Pi}}{\partial r} + \delta \tilde{\Theta} ~ \tilde{r} + \left. E ~ \vv{\nabla}^2 \vv{\tilde{U}} \right|_{\vv{e}_r} \\\\
- 2 \cos{\theta} ~ \tilde{U}_{\phi} = - \displaystyle \frac{1}{\tilde{r}} \displaystyle \frac{\partial \delta \tilde{\Pi}}{\partial \theta} + \left. E ~ \vv{\nabla}^2 \vv{\tilde{U}} \right|_{\vv{e}_{\theta}} \\\\
2 \cos{\theta} ~ \tilde{U}_{\theta} + 2 \sin{\theta} ~ \tilde{U}_r - \displaystyle \frac{2 \sin{\theta}}{\tilde{r}^2} = E \left( \vv{\nabla}^2 - \displaystyle \frac{1}{\tilde{r}^2 \sin^2 \theta} \right) \tilde{U}_{\phi} \\\\
P_r \left( \displaystyle \frac{N_0}{\Omega_0} \right)^2 \tilde{U}_r ~ \displaystyle \frac{\text{d} \tilde{\overline{T}}_{\text{m}}}{\text{d} r} = E ~ \vv{\nabla}^2 \delta \tilde{\Theta} \\\\
\displaystyle \frac{1}{\tilde{r}^2} \displaystyle \frac{\partial}{\partial r} \left( \tilde{r}^2 \tilde{U}_r \right) + \displaystyle \frac{1}{\tilde{r} \sin{\theta}} \displaystyle \frac{\partial}{\partial \theta} \left( \sin \theta ~ \tilde{U}_{\theta} \right) = 0
\end{array}
\label{eq_axi}
\end{equation}

For small Ekman numbers, solutions are sought as the sum of an interior inviscid solution and a boundary layer solution, that is $\vv{\tilde{U}}(r,\theta) = \vv{\tilde{U}}_{I}(r,\theta) + \vv{\tilde{U}}_E(\xi,\theta)$ where by construction the boundary layer solution must vanish in the interior. For an $E^{-1/2}$ wide boundary layer at the outer sphere, a dimensionless stretched coordinate $\tilde{\xi} = E^{-1/2} \left(1 - \tilde{r}\right)$ is introduced and the previous condition reads $\lim\limits_{\tilde{\xi} \to \infty} \vv{\tilde{U}}_{E}(\xi,\theta) = 0$. 

For the interior solution, we assume the following expansion:

\begin{equation}
\begin{array}{lll}
\tilde{U}_{\phi_I} = \displaystyle \frac{P_r \left(N_0/\Omega_0\right)^2}{E} \hspace*{0.05cm} \tilde{U}_{\phi_I}^{(0)} + \displaystyle \frac{1}{\sqrt{E}} \hspace*{0.05cm} \tilde{U}_{\phi_I}^{(1)} \\\\
\tilde{U}_{r_I} = \tilde{U}_{r_I}^{(0)} \\\\
\tilde{U}_{\theta_I} = \tilde{U}_{\theta_I}^{(0)} \\\\
\delta \tilde{\Theta}_I = \displaystyle \frac{P_r \left(N_0/\Omega_0\right)^2}{E} \hspace*{0.05cm} \delta \tilde{\Theta}_{I}^{(0)}
\end{array}
\label{expansion_interior_flow}
\end{equation}
\noindent where $E^{-1/2} \ll P_r \left(N_0/\Omega_0\right)^2 / E$.

Substitution of this expansion in Sys. \eqref{eq_axi} yields at the dominant order the following system:

\begin{equation}
 \begin{array}{lll}
 2 \displaystyle \frac{\partial \tilde{U}_{\phi_I}^{(0)}}{\partial z} = \displaystyle \frac{\partial \hspace*{0.02cm} \delta \tilde{\Theta}_I^{(0)}}{\partial \theta} \\\\
\tilde{U}_{s_I}^{(0)} = \displaystyle \frac{\sin{\theta}}{\tilde{r}^2} \\\\
 \tilde{U}_{r_I}^{(0)} \displaystyle \frac{\text{d} \tilde{\overline{T}}_{\text{m}}}{\text{d} r} = \vv{\nabla}^2 \delta \tilde{\Theta}_I^{(0)} \\\\
 \vv{\nabla} \cdot \vv{\tilde{U}_I}^{(0)} = 0
 \end{array}
\end{equation}

From the expression of $\tilde{U}_{s_I}^{(0)}$, the radial velocity field $ \tilde{U}_{r_I}^{(0)}$ can be determined using mass conservation and the boundary condition $ \tilde{U}_{r_I}^{(0)}(r = \tilde{r_i},\theta) = 0$. Then, we assume that the system of equations for the remaining unknowns $\tilde{U}_{\phi_I}^{(0)}$ and $\delta \tilde{\Theta}_{I}^{(0)}$ admits a solution for the full boundary conditions $\left(\tilde{U}_{\phi_I}^{(0)}(r = 1,\theta) = 0 \right.$ , $\partial \left({U}_{\phi_I}^{(0)} / \tilde{r} \right) / \partial r = 0$ in $r=\tilde{r}_i$, and  $\left.\delta \tilde{\Theta}_{I}^{(0)}(r = 1,\theta) = \delta \tilde{\Theta}_{I}^{(0)}(r = \tilde{r}_i,\theta) = 0 \right)$. 

However, the meridional flow $\tilde{U}_{r_I}^{(0)}, \tilde{U}_{\theta_I}^{(0)}$ does not satisfy the no-slip condition at the outer sphere. An Ekman layer is therefore needed to correct the $\mathcal{O}(1)$ jump on this field and this will induced the $E^{-1/2} \tilde{U}_{\phi_I}^{(1)}$ term. 
Inserting the expansion Sys. \eqref{expansion_interior_flow} in Sys. \eqref{eq_axi}, we see at $\mathcal{O}\left( E^{-1/2} \right)$ order that such a term must verify:

\begin{equation}
\begin{array}{lll}
\displaystyle \frac{\partial \tilde{U}_{\phi_I}^{(1)}}{\partial z} = 0 \\\\
\label{edd_sys}
\end{array}
\end{equation}

Meanwhile, within the Ekman layer, the boundary layer solution must be of the following order to compensate the  $\mathcal{O}\left( 1 \right)$ meridional velocity jump:

\begin{equation}
\begin{array}{lll}
\tilde{U}_{\phi_E} = \displaystyle \frac{1}{\sqrt{E}} \hspace*{0.05cm} \tilde{U}_{\phi_E}^{(0)} \\\\
\tilde{U}_{r_E} = \tilde{U}_{r_E}^{(1)} \\\\
\tilde{U}_{\theta_E} = \displaystyle \frac{1}{\sqrt{E}} \hspace*{0.05cm} \tilde{U}_{\theta_E}^{(0)}
\label{ekman_expansion}
\end{array}
\end{equation}

By rewriting Sys. \eqref{eq_axi} with the stretched coordinate $\xi$ and by inserting the previous expressions, the boundary layer equations read:

\begin{equation}
\begin{array}{lll}
- 2 \cos{\theta} ~ \tilde{U}_{\phi_E}^{(0)} = \displaystyle \frac{\partial^2 \tilde{U}_{\theta_E}^{(0)}}{\partial \xi^2} \\\\
2 \cos{\theta} ~ \tilde{U}_{\theta_E}^{(0)} = \displaystyle \frac{\partial^2 \tilde{U}_{\phi_E}^{(0)}}{\partial \xi^2} \\\\
- \displaystyle \frac{\partial \tilde{U}_{r_E}^{(1)}}{\partial \xi} + \displaystyle \frac{1}{\sin{\theta}} \displaystyle \frac{\partial}{\partial \theta} \left( \sin \theta ~ \tilde{U}_{\theta_E}^{(0)} \right) = 0
\end{array}
\label{System4}
\end{equation}

The boundary condition $\tilde{U}_{\phi}=0$ leads to $\tilde{U}_{\phi_I}^{(1)}(r = 1,\theta) + \tilde{U}_{\phi_E}^{(0)}(\xi=0,\theta) = 0$ at the order $E^{-1/2}$. Together with the other boundary condition $\tilde{U}_{\theta_E}^{(0)}(\xi=0,\theta) = 0$ and the evanescent condition when $\tilde{\xi} \to \infty$, the first two equations are solved to give:

\begin{equation}
\begin{array}{lll}
\tilde{U}_{{\theta}_E}^{(0)}(\xi,\theta) = -\text{sgn}\left(\cos{\theta}\right) \tilde{U}_{\phi_I}^{(1)}\left(1,\theta\right) \exp\left({- \sqrt{\left| \cos{\theta} \right |}\hspace*{0.05cm} \tilde{\xi}}\right) ~ \cdot \\\\ \hspace*{2.cm} \sin\left({\sqrt{\left| \cos{\theta} \right |}\hspace*{0.05cm} \tilde{\xi}}\right) \\\\
\tilde{U}_{{\phi}_E}^{(0)}(\xi,\theta) = - \text{sgn}\left(\cos{\theta}\right) \tilde{U}_{\phi_I}^{(1)}\left(1,\theta\right) \exp\left({- \sqrt{\left| \cos{\theta} \right |}\hspace*{0.05cm} \tilde{\xi}}\right) ~ \cdot \\\\ \hspace*{2.cm} \cos\left({\sqrt{\left| \cos{\theta} \right |}\hspace*{0.05cm} \tilde{\xi}}\right)
\end{array}
\end{equation}

Then, integrating the last equation of Sys. \eqref{System4} and using both the evanescent condition and the boundary condition $\tilde{U}_{r_I}^{(0)}(r = 1,\theta) + \tilde{U}_{r_E}^{(1)}(\xi=0,\theta) = 0$ lead to the Ekman pumping condition Eq. \eqref{ekman_pumping_dimensionless} in its dimensionless form. From this condition on the interior flow, one can show as in Appendix \ref{demonstration_rot_diff_TP}, that $\tilde{U}_{\phi_I}^{(1)}$ is the Taylor-Proudman solution. 

In the interior, its contribution to the total differential rotation decreases as $P_r \left(N_0/\Omega_0\right)^2 / \sqrt{E}$ increases. But near the outer sphere its contribution can never be neglected because the Eddington-Sweet term vanishes there $\left(\tilde{U}_{\phi_I}^{(0)}(r=1, \theta)=0\right)$. 

\end{document}